# On Distribution of Superconductivity in Metal Hydrides

*Dmitrii V. Semenok,[1,*] Ivan A. Kruglov,[2,3] Igor A. Savkin,[4] Alexander G. Kvashnin,[1,2,*] and Artem R. Oganov[1,2,5]*

[1] Skolkovo Institute of Science and Technology, Skolkovo Innovation Center, 3 Nobel Street, Moscow 121205, Russia

[2] Moscow Institute of Physics and Technology, 9 Institutsky Lane, Dolgoprudny 141700, Russia

[3] Dukhov Research Institute of Automatics (VNIIA), Moscow 127055, Russia

[4] Research Computer Center of Lomonosov Moscow State University, Moscow, Russia

[5] International Center for Materials Discovery, Northwestern Polytechnical University, Xi'an 710072, China

**Corresponding authors**

*Alexander G. Kvashnin: A.Kvashnin@skoltech.ru

*Dmitrii V. Semenok: dmitrii.semenok@skoltech.ru

**ABSTRACT.** Using the data on the superconducting critical temperature ($T_C$) for a number of metal hydrides, we found a rule that makes it possible to predict the maximum $T_C$ based only on the information about the electronic structure of metal atoms. Using this guiding principle, we explored the hydride systems for which no reliable information existed, predicted new higher hydrides in the K-H, Zr-H, Hf-H, Ti-H, Mg-H, Sr-H, Ba-H, Cs-H, and Rb-H systems at high pressures, and calculated their $T_C$. Results of the study of actinides and lanthanides show that they form highly symmetric superhydrides $XH_7$-$XH_9$. However, actinide hydrides do not exhibit high-temperature superconductivity (except Th-H system) and might not be considered as promising materials for experimental studies, as well as all $d^m$-elements with m > 4, including hydrides of the noble elements. Designed neural network allowing the prediction of $T_C$ of various hydrides shows high accuracy and was used to estimate upper limit for $T_C$ of the materials with absence of the data. The developed rule, based on regular behavior of the maximum achievable critical temperature as a function of number of $d$+$f$ electrons, enables targeted predictions about the existence of new high-$T_C$ superconductors.

## Introduction

Recent progress in experimental synthesis and measurement of superconducting properties of $H_3S$,[1] $H_3Se$,[2,3] $Si_2H_6$,[4] $LaH_{10}$,[5-7] $FeH_5$,[8,9] $UH_7$,[10] $CeH_9$,[11] $ThH_9$, $ThH_{10}$,[12] and $PrH_9$,[13] together with theoretical predictions including the above compounds and $LiH_8$,[14] $CaH_6$,[15] $BaH_6$,[2] $TaH_6$,[29] $ScH_9$,[18] $VH_8$,[19] $SrH_{10}$,[16] $YH_{10}$,[20] $LaH_{10}$,[20] $ThH_{10}$,[21] and $AcH_{10}$,[22] emphasizes the need for a law governing the distribution of superconducting properties of metal hydrides over Mendeleev's Periodic Table. We have suggested earlier[22] that the majority of high-$T_C$ metal hydrides are formed by metals at the border between *s*/*p* and *s*/*d* blocks. The possible reason[22] is the anomalously strong electron-phonon interaction caused by a decrease in the energy difference between the *s*- and *d*-orbitals of metals atoms under pressure.

Here, using this idea of "orbital lability" to predict new superconducting hydrides that have not yet been studied. We considered the maximum value of $T_C$ achieved among all stable compounds in each M-H system at all pressures found in literature as a target value for this study, i.e. $F = \max T_C(P, M_nH_m)$. To describe the behavior of $f$ as a function of pressure $P$, hydrogen content $N_H$, and the total number of $d$- and $f$-electrons $N_{d+f}$, we analyzed the $\max T_C$-$N_{d+f}$, $\max T_C$-$P$, and $\max T_C$-$N_H$ diagrams. Those systems for which we could not find reliable data on stable polyhydrides and their $T_C$ were carefully investigated.

## Methods

A powerful tool for predicting thermodynamically stable compounds at given pressure, the evolutionary algorithm USPEX,[23–25] was used to perform variable-composition searches of the $M_xH_y$ phases (M = K, Ca, Zr, Hf, Pr, Nd, Ho, Er, Tm, Lu, Pa, Np, Am, Cm) at pressures of 50, 150, 250, and 300 GPa. The first generation of 120 structures was created using a random symmetric and random topological[26] generators, while all subsequent generations contained 20% of random structures and 80% of structures created using heredity, softmutation, and transmutation operators. The evolutionary searches were combined with structure relaxations using density functional theory (DFT)[27, 28] within the generalized gradient approximation (the Perdew-Burke-Ernzerhof functional)[29] and the projector augmented wave method[30, 31] as implemented in the VASP code.[32–34] The plane wave kinetic energy cutoff was set to 600 eV and the Brillouin zone was sampled using the Γ-centered *k*-points meshes with the resolution of $2\pi\times0.05$ Å$^{-1}$.

Phonon frequencies and electron-phonon coupling (EPC) coefficients were computed using density functional perturbation theory,[35] employing the plane-wave pseudopotential method with the norm-conserving Troullier-Martins pseudopotentials and the Perdew-Burke-Ernzerhof exchange-correlation functional[29] as implemented in QUANTUM ESPRESSO (QE) package.[36] The $3s^23p^64s^1$, $3s^23p^64s^2$, $3d^24s^2$, $4d^25s^2$, $4f^{14}5d^26s^2$, $6s^24f^3$, $6s^24f^4$, $6s^24f^{11}$, $6s^24f^{12}$, $6s^24f^{13}$, $6s^24f^{14}5d^1$, $7s^26d^15f^2$, $7s^26d^15f^4$, $7s^15f^7$, $7s^26d^15f^7$ electrons of K, Ca, Zr, Hf, Pr, Nd, Ho, Er, Tm, Lu, Pa, Np, Am, Cm were treated explicitly. Convergence tests showed that for all elements 120 Ry is a suitable kinetic energy cutoff for the plane wave basis set. In our ab initio calculations of the electron-phonon coupling (EPC) parameter λ, the first Brillouin zone was sampled using 2×2×2 or 4×4×4 *q*-points meshes and denser 16×16×16 and 24×24×24 *k*-points meshes. The superconducting transition temperature $T_C$ was estimated using the Allen-Dynes formula (see Supporting Information for details).

A regression model was created on the basis of a fully connected (each next layer is connected with all outputs of the previous one) neural network to predict max$T_C$ = Y(A,B) for ternary systems A-B-H with a lack of data. It was found empirically that the optimal topology consists of 11 layers of 12 neurons in a layer. The number of *s*-, *p*-, *d*-, and *f*-electrons in an atom and the atomic number were used as input for this network. The neural network was trained on a set of 180 reference max$T_C$ values of binary (A-A-H, A-H-H) and ternary hydrides using the RMSProp (Root Mean Square Propagation) method in 50000 steps. Boundary (Y(A,1) = Y(1,A) = Y(A,A)) and symmetry (Y(A,B) = Y(B,A)) conditions were included in the training set. The *tensorflow* library was applied to create the model. After the training, the network was validated on 20% of randomly selected max$T_C$ points. The average error for binary systems did not exceed 20 K.

## Results and Discussion

### Period 4. K–Ca–Sc–Ti–V–Cr

The main results for period 4 of Mendeleev's Periodic Table are presented in Table 1 and Figure 1. The highest $T_C$ reaches 235 K for $CaH_6$.[15] The highest-$T_C$ potassium hydride is *C*2/*c*-$KH_6$ with the maximum $T_C$ of 70 K at 166 GPa, predicted by Zhou et al.[37] (Table 1). We expect that an increase in the number of *d*-electrons in the Sc-Ti-V-Cr row should lead to a smooth drop in $T_C$: max$T_C$(Sc-H) > max$T_C$(Ti-H) > max$T_C$(V-H) > max$T_C$(Cr-H). No reliable data on titanium polyhydrides were available (Table 1), except $TiH_2$ having a low $T_C$ = 6.7 K. Predicted $T_C$ for $VH_8$ [19] may be an overestimate (Figure **1**). We decided to perform additional computational searches for new polyhydrides in the K-H, Ca-H, and Ti-H systems.

| Hydride (pressure, GPa) | $\lambda$ | $\omega_{log}$, K | max$T_C$, K |
|---|---|---|---|
| $KH_6$ (166) | 0.9 | 1165 | 70[37] |
| $CaH_6$ (150) | 2.7 | - | 235[15] |
| $ScH_9$ (300) | 1.94 | 1156 | 163[18,38] |
| $TiH_2$ (0) | 0.84 | 127 | 6.7[39] |
| $VH_8$ (200) | 1.13 | 876 | 71.4[19] |
| $CrH_3$ (81) | 0.95 | 568 | 37.1[40] |
| Newly predicted compounds | | | |
| $KH_{10}$ (150) | 1.34 | 1301 | 148 |
| $TiH_{14}$ (200) | 0.81 | 1063 | 54 |

**Table 1.** Previously predicted or synthesized binary hydrides with maximum $T_C$, period 4.

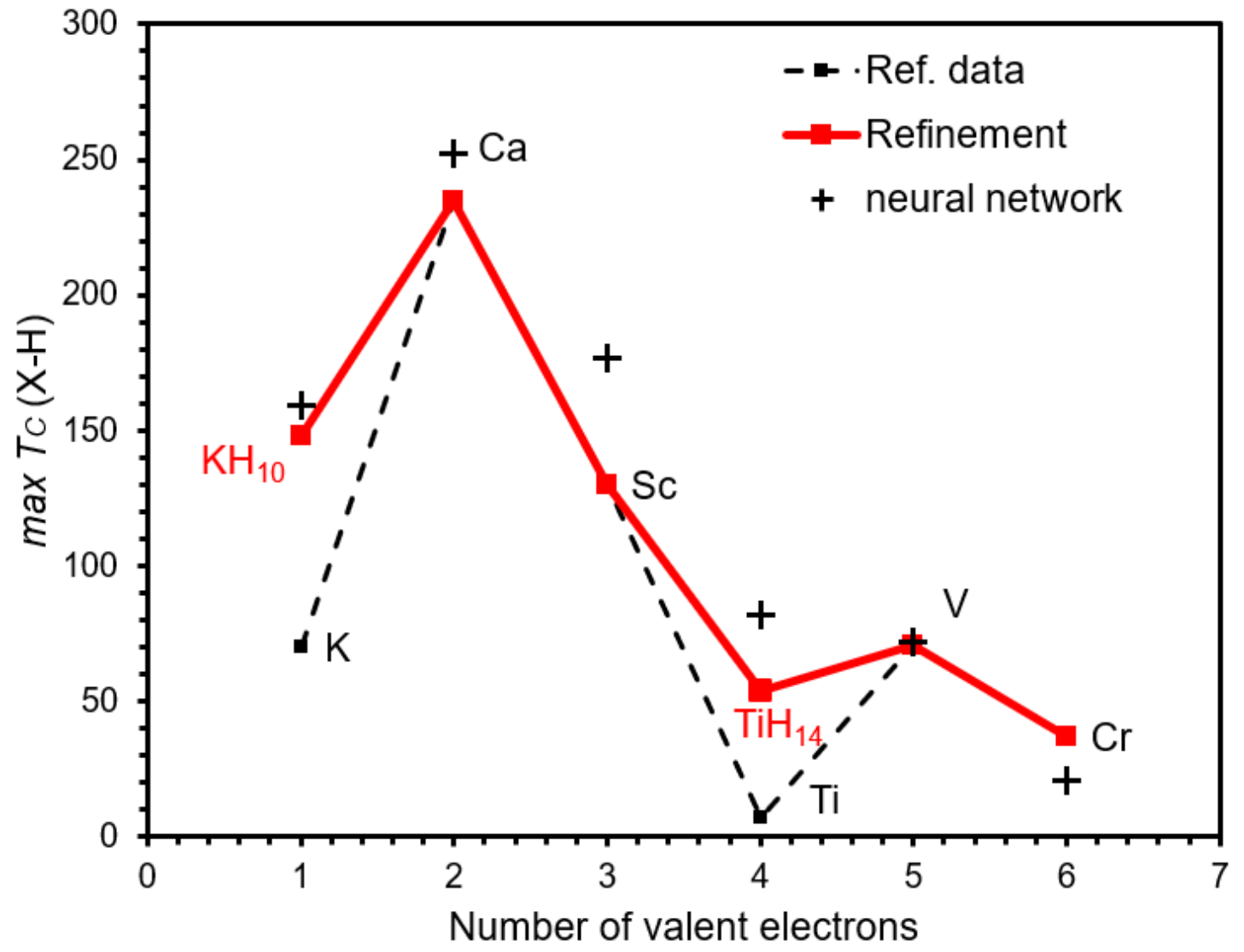

**Figure 1.** Maximum $T_C$ of hydrides of metals from 4th period.

Because the 3*d*-shell of the considered row of metals begins to be filled with electrons starting from Sc, we presumed that the energy difference between the *s*- and *d*-orbitals will decrease as the pressure grows. By analogy with *d*-elements, higher hydrides in the potassium and calcium systems, $KH_n$ and $CaH_n$, where $n > 6$ (e.g. $n = 10, 12 \ldots$), may also be formed. To test this idea, we performed the variable-composition evolutionary search for stable hydrides in the K-H and Ca-H systems at various pressures. The predicted thermodynamically stable compounds in these systems are shown in Table 2 and Table 3. The convex hull for the K-H system at 150 GPa is shown in Figure 2, while the convex hull at 50 GPa and those for the Ca-H system are shown in Supporting Information Figures S1 and S2, respectively.

| Pressure, GPa | Stable phases |
|---|---|
| 50 | *Imm2*-$KH_{11}$<br>*Pm-3m*-KH<br>*Cm*-$KH_9$<br>*Cmcm*-$KH_5$ |
| 150 | *Immm*-$KH_{10}$<br>*Immm*-$KH_{12}$<br>*Pm-3m*-KH<br>*Cmcm*-$KH_5$<br>*C2/m*-$KH_2$ |

**Table 2.** Predicted stable phases in the K-H system.

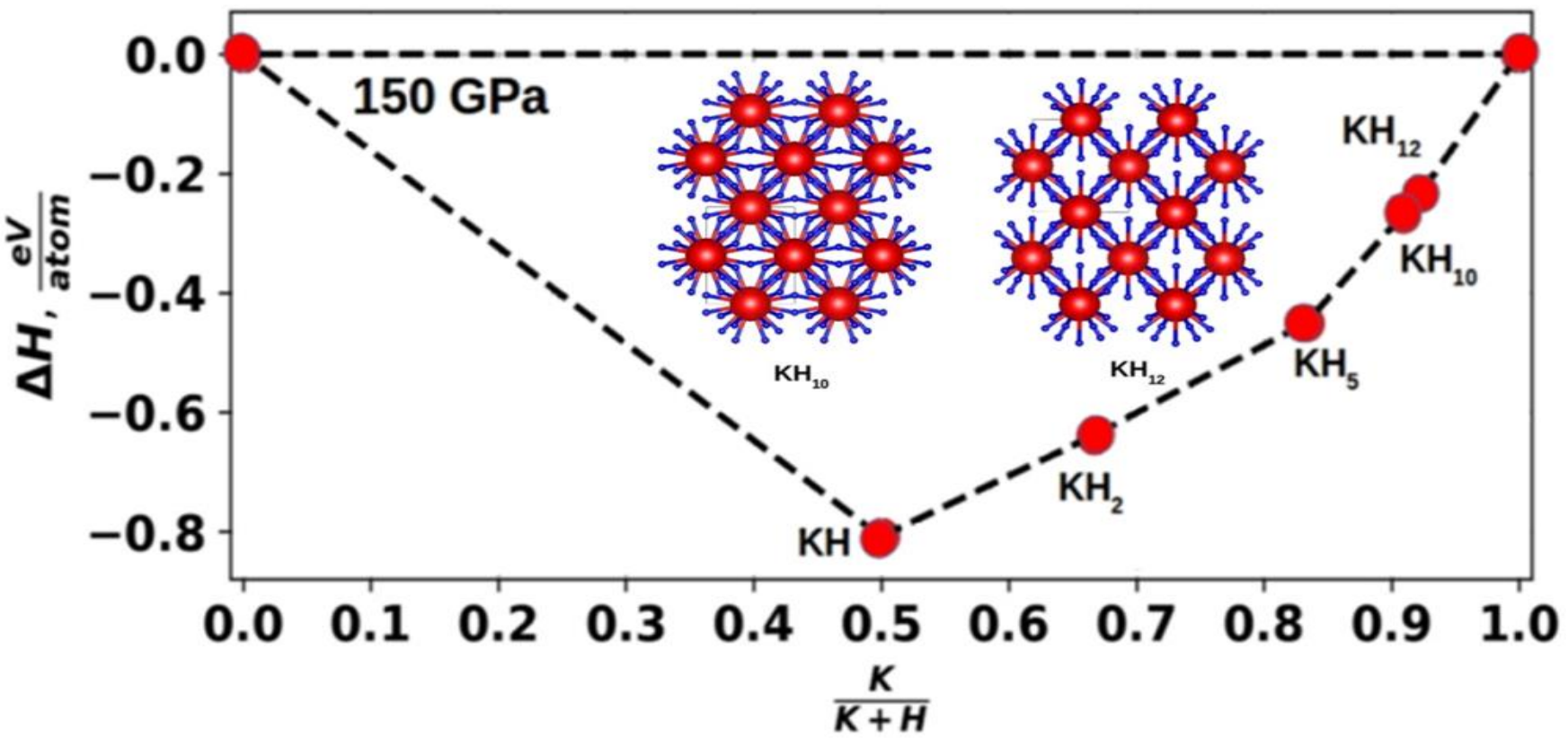

**Figure 2.** Convex hull for the K-H system at 150 GPa.

In the K-H system we predicted new thermodynamically stable higher hydrides at various pressures which have not been considered before.[2,41] We showed that the *Cmcm*-$KH_5$[42] would remove $KH_6$ from the convex hull at 150 GPa (Figure 2). $KH_5$ as well as the predicted *Imm2*-$KH_{11}$ are semiconductors with a DFT band gaps of ~1 eV and 2.03 eV at 100 and 50 GPa, respectively (the systematic underestimation of the band gap by the DFT-PBE method should be taken into account). Superconducting properties were studied only for the metallic *Immm*-$KH_{10}$ and *Immm*-$KH_{12}$. The newly predicted *Immm*-$KH_{10}$ displays the highest $T_C = 148$ K among all the K-H phases (Supporting Information Figure S1), while for *Immm*-$KH_{12}$ $T_C$ is 116 K.

The comparison of our results with those obtained in Ref.[15,41] allows us to state that new stable superhydrides, *C2/m*-$CaH_{12}$ and *C2*-$CaH_{18}$, are discovered among the already known phases in the pressure range of 0 to 200 GPa (Table 3). *C2/m*-$CaH_{12}$ has $T_C$ = 206 K and a high EPC coefficient λ = 2.16 (Supporting Information Figure S2), while *C2*-$CaH_{18}$ has a lower $T_C$ = 88 K at 150 GPa. These predicted phases have a lower $T_C$ than that of $Im\bar{3}m$-$CaH_6$.[15]

Table 3. Predicted stable phases in the Ca-H system.

| **Pressure, GPa** | **Stable phases** | **Pressure, GPa** | **Stable phases** |
|---|---|---|---|
| 0 | *Pnma*-$CaH_2$ | 150 | $Im\bar{3}m$-$CaH_6$ |
| | | | *P1*-$CaH_6$ |
| | | | *C2*-$CaH_{18}$ |
| 50 | *I4/mmm*-$CaH_4$ | | *C2/m*-$CaH_{12}$ |
| | $P6_3/mmc$-$CaH_2$ | | *I4/mmm*-$CaH_4$ |
| | $R\bar{3}c$-$Ca_3H$ | | $P6_3/mmc$-$CaH_2$ |
| | $I\bar{4}3d$-$Ca_3H_4$ | | *C2/m*-$Ca_2H$ |
| | | | *Pm*-CaH |
| 100 | *Pm*-$CaH_2$ | 200 | *C2*-$CaH_{18}$ |
| | *I4/mmm*-$CaH_4$ | | $Im\bar{3}m$-$CaH_6$ |
| | *C2/m*-$CaH_8$ | | *Cmcm*-$CaH_4$ |
| | $P\bar{1}$-$CaH_{14}$ | | *Imma*-$CaH_2$ |
| | $I\bar{4}3d$-$Ca_3H_4$ | | *P6/mmm*-$CaH_2$ |
| | $P\bar{1}$-$Ca_{12}H$ | | $Pm\bar{3}m$-CaH |

In the Ti-H system, we predicted one new stable higher hydride, *C2/m*-$TiH_{14}$, which is thermodynamically stable at 200 GPa (Supporting Information Table S1). This hydride features superconductivity with $T_C$ = 54.2 K (Supporting Information Figure S23). This additional data eliminates the abrupt minimum at $N_d$ = 4, making the max$T_C(N_d)$ function decrease more monotonically from Ca to Cr (red dotted line in Figure 1).

Thus, additional variable-composition evolutionary searches for stable phases in the K-H, Ca-H, and Ti-H systems reveal the existence of new polyhydrides with $T_C$ exceeding the values known to date, supporting our rule of maximum $T_C$ for electronically labile "boundary" elements, at the boundary between s|d block.

## Period 5. Rb–Sr–Y–Zr–Nb–Mo–Tc

Some of the metal hydrides from period 5 have been studied earlier[17, 43, 44, 16, 20] (Table 4). However, no reliable information is available about the superconducting properties of the Rb-H and Mo-H compounds. Hooper et al.[46] studied the Rb-H system ($RbH_9$-$RbH_{14}$) only at 250 GPa, but superconductivity was not investigated.

At 300 GPa, the artificially constructed $SrH_{10}$[16] displays high $T_C$ = 259 K and large $\lambda$ = 3.08, indicating the possibility to find high-$T_C$ superconductors in the Sr-H system. Our USPEX search at 50, 100, and 150 GPa (Supporting Information Figures S3 and S4) showed the presence of many stable and metastable higher strontium polyhydrides. At 50 GPa *I4/mmm*-$SrH_4$ is stable, isostructural to the previously found $CaH_4$.[43, 47] However, there are many low-symmetry higher hydrides in close proximity to the convex hull. At 100 and 150 GPa, $R\bar{3}m$-$SrH_6$ and molecular *C2/m*-$SrH_{10}$ are found to be stable and located on the convex hull (see Figure S28 for superconducting properties). $SrH_6$ becomes stable at 100 GPa, i.e. at much lower pressure than its analog $Im\bar{3}m$-$CaH_6$[15] (200 GPa). These two compounds possess pronounced superconducting properties, and at 200 GPa $T_C$ of $R\bar{3}m$-$SrH_6$ may exceed 189 K (Supporting Information Figure S29). In our calculations at 50 to 300 GPa we could not find any stable high-symmetry phases of $SrH_{10}$.

The values of max$T_C$ for the rest of the metal hydrides from period 5 drop from 323 K for $YH_{10}$ at 250 GPa[20] to 11 K for $TcH_2$ at 200 GPa[44] (Figure 3). The max$T_C$ function is not monotonically decreasing because the only stable compound in the Zr-H system is ZrH displaying a low $T_C$ of 10 K. In Ref. [48], the compound $P2_1/c$-$ZrH_6$ having $T_C$ = 153 K at 295 GPa was predicted using the random structure search method (green point in Figure 3). Taking into account a low number of valence *d*-electrons, we undertook a variable-composition evolutionary search for new stable zirconium hydrides and studied in detail the superconducting properties. At various pressures, we predicted new higher hydrides, differing from the prediction made in Ref. [48] (Supporting Information Figure S5 and Table S2). The newly predicted molecular $P\bar{1}$-$ZrH_{16}$, stable at the pressures over 150 GPa, is the most promising among all stable phases because of the high hydrogen concentration (Table S2), having a structure close to that of titanium hydride $P2_1/c$-$TiH_{14}$. The calculated value of $T_C$ of $P\bar{1}$-$ZrH_{16}$ is 88 K. With the new $ZrH_{16}$, the max$T_C$ function for period 5 becomes monotonically decreasing (red solid line in Figure 3), which is consistent with the general idea of decreasing of maximum $T_C$ with the growing number of outer electrons in metal atoms.

| Hydride (Pressure, GPa) | $\lambda$ | $\omega_{log}$, K | max$T_C$, K |
|---|---|---|---|
| $SrH_{10}$ (300) | 3.08 | 767 | 259 [16] |
| $YH_{10}$ (300) | 2.6 | 1282 | 323 [20] |
| ZrH (120) | 0.71 | 295 | 10 [45] |
| $ZrH_6$ (295) | 1.7 | 914 | 153 [48] |
| $NbH_4$ (300) | 0.82 | 938 | 38 [17] |
| $TcH_2$ (200) | 0.52 | 736 | 11 [44] |
| Newly predicted compounds | | | |
| $ZrH_{16}$ (200) | 1.19 | 852 | 88 |
| $SrH_6$ (100) | 1.65 | 1316 | 189 |

**Table 4.** Predicted or synthesized binary hydrides and their maximum critical temperatures, period 5.

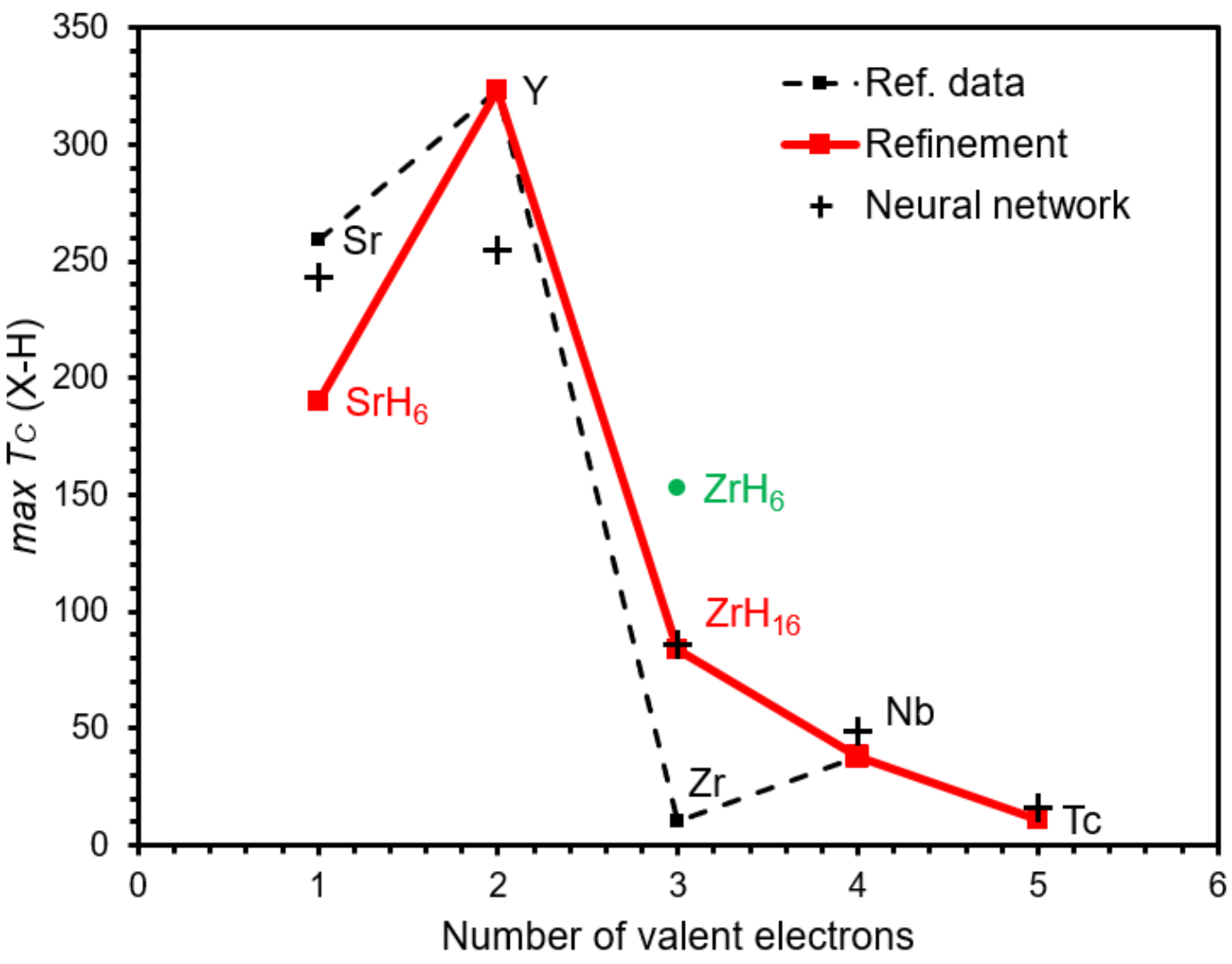

**Figure 3.** Maximum $T_C$ of hydrides of metals from 5th period.

### Period 6. Cs–Ba–La–Hf–Ta

Available data on superconducting properties of the predicted or synthesized metal hydrides from period 6 of Mendeleev's table are shown in Figure 4 and Table 5. Shamp et al.[49] predicted several Cs-H compounds, namely $CsH_7$ and $CsH_{16}$, presumably stable at 150 GPa, but their superconducting properties were not studied. In the Ba-H system, $BaH_6$ displays the maximum value of $T_C$ = 31-38 K at 100 GPa,[2] whereas $LaH_{10}$ displays $T_C$ = 286 K at 210 GPa.[20] In the Hf-H system, only $HfH_2$ is known with $T_C$ of 13 K at 260 GPa[50] (Figure 4). In the Ta-H system, $Fdd2$-$TaH_6$ was predicted to have $T_C$ = 136 K at 300 GPa.[51] A sharp drop in max$T_C$ at Hf (Figure 4), similar to Zr (Figure 3), indicates the lack of data about the possible superconducting polyhydrides in the Hf-H system.

The data from our evolutionary searches for Hf-H phases at different pressures are shown in Supporting Information Figure S6. The obtained data show the existence of novel higher hafnium hydrides, $Amm2$-$Hf_3H_{13}$ at 100 GPa with $T_C$ = 43 K, and $C2/m$-$HfH_{14}$ at 300 GPa with $T_C$ = 76 K (Supporting Information Figure S25).

| Hydride (Pressure, GPa) | $\lambda$ | $\omega_{log}$, K | max$T_C$, K |
|---|---|---|---|
| $BaH_6$ (100) | 0.77 | 878 | 38[41] |
| $LaH_{10}$ (210) | 3.41 | 848 | 286[20] |
| $LaH_{10}$ (170) | - | - | 250-260[6,7] |
| $HfH_2$ (260) | 0.87 | - | 13[50] |
| $TaH_6$ (300) | 1.56 | 1151 | 136[51] |
| New predicted compounds | | | |
| $HfH_{14}$ (300) | 0.93 | 1138 | 76 |
| $BaH_{12}$ (135) | 2.64 | 927 | 214 |

**Table 5.** Predicted binary hydrides and their maximum critical temperatures, period 6.

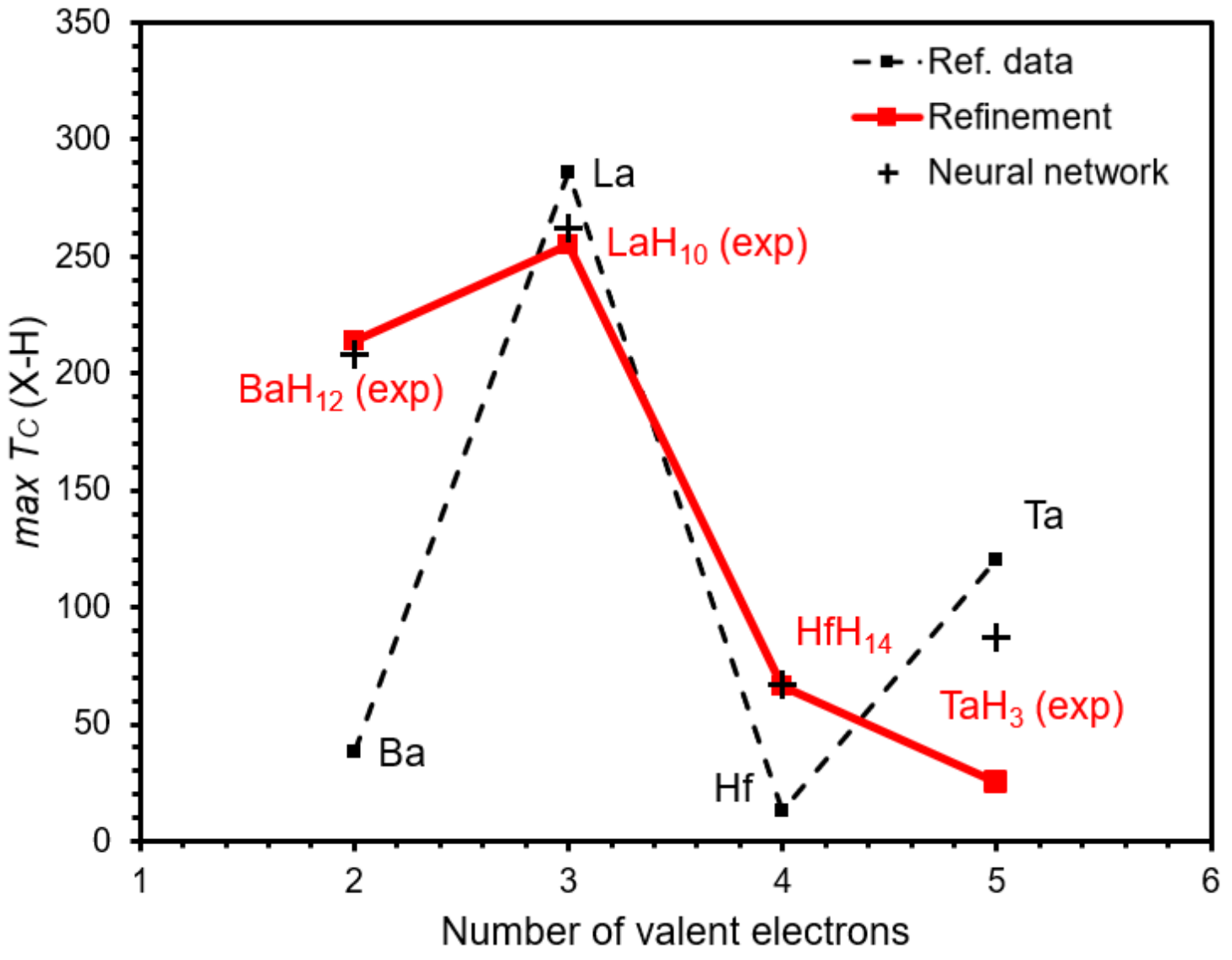

**Figure 4.** Maximum $T_C$ of hydrides of metals from 6th period. For $LaH_{10}$ and $TaH_3$, the recent experimental results from Refs. [6, 7, 51] are presented.

The results for *P4/mmm*-$BaH_6$ obtained in Ref. [41] show an incredibly low $T_C$. The Ba-H system should contain hydrides with a higher $T_C$, such as $BaH_{10}$ and $BaH_{12}$, as it is for the other alkaline-earth elements. Barium is the closest neighbor of the La-H system with the record high-$T_C$ superconductor $LaH_{10}$, [6, 7] and it is reasonable to expect remarkable superconducting properties in barium hydrides. Indeed, preliminary experimental results [53] confirm the existence of the tetragonal $BaH_{12}$, stable at pressures over 120 GPa, with the predicted $T_C$ over 214 K (Figure 4).

The stability of the *Fdd*2-$TaH_6$ phase (max$T_C$ = 136 K at 300 GPa) is also questionable. Zhuang et al.[51] performed calculations using ELocR code and a limited integer set of possible $TaH_n$ compositions with n = 1-6. However, such a high pressure may cause *Fdd2*-$TaH_6$ to be expelled from the convex hull by other compounds. Actually the comparison with the neighboring systems V-H, Nb-H, and Hf-H shows that max$T_C$(Ta-H) should lie in the range of 40 to 90 K. The recent experimental investigation of the Ta-H system did not show the presence of any higher tantalum hydrides, only $TaH_{\sim3}$ and $TaH_{\sim2}$ were found. [51]

## Lanthanides and Actinides

Higher hydrides of lanthanides are of particular interest because the theoretically predicted $LaH_{10}$[20] is the highest-$T_C$ superconductor confirmed experimentally.[6, 7] Peng et al.[54] conducted a theoretical study of lanthanide hydrides, looking in particular at the formation of clathrate-like shells composed of the H atoms. Assuming $REH_n$ ($n$ = 8-10) to be thermodynamically stable, $T_C$ was estimated at ~ 56 K at 100-200 GPa for Ce and Pr.[54] Our calculations yielded similar results for *P6₃mc*-$CeH_6$, with $T_C$ not exceeding 51 K. Recent theoretical predictions and experimental synthesis of $CeH_9$ by Salke et al.[11] show that *P6₃/mmc*-$CeH_9$ should have $\lambda = 2.30$, $\omega_{log} = 740$ K, and $T_C = 117$ K at 200 GPa (assuming $\mu^* = 0.1$ as usual). The preliminary measurements of superconducting properties in the Ce-H system confirmed the predicted $T_C$ of $CeH_9$ to be 100-110 K in the pressure range of 110 to 140 GPa.[53]

To gain a more precise insight into the superconductivity of lanthanide hydrides, we performed variable-composition evolutionary searches for thermodynamically stable compounds and crystal structures in almost all these systems. The obtained data are presented in Supporting Information Tables S1, S3-5.

According to the calculated results, the lanthanide hydrides generally have a low EPC parameter (λ) and are weak superconductors compared to other metal hydrides. The critical temperatures display a monotonic decrease followed by the *f*-shell getting filled, with the maximum $T_C$ displayed in lanthanum and cerium hydrides (Figure **5**). Moving from the light to heavy lanthanides, we observe vanishing of superconductivity for hydrides of metals with half-filled *d*- and *f*-shells (Mn - $d^5$, Re – $d^5$, Eu – $f^7$, Am – $f^7$), and then gradually increase in a "secondary wave" of superconductivity upon further filling the *d*- and *f*-shells. We think this is because of the known stability of half-filled *d*- and *f*-shells, making these atoms exactly opposite of the electronically labile atoms amenable to high-$T_C$ superconductivity.

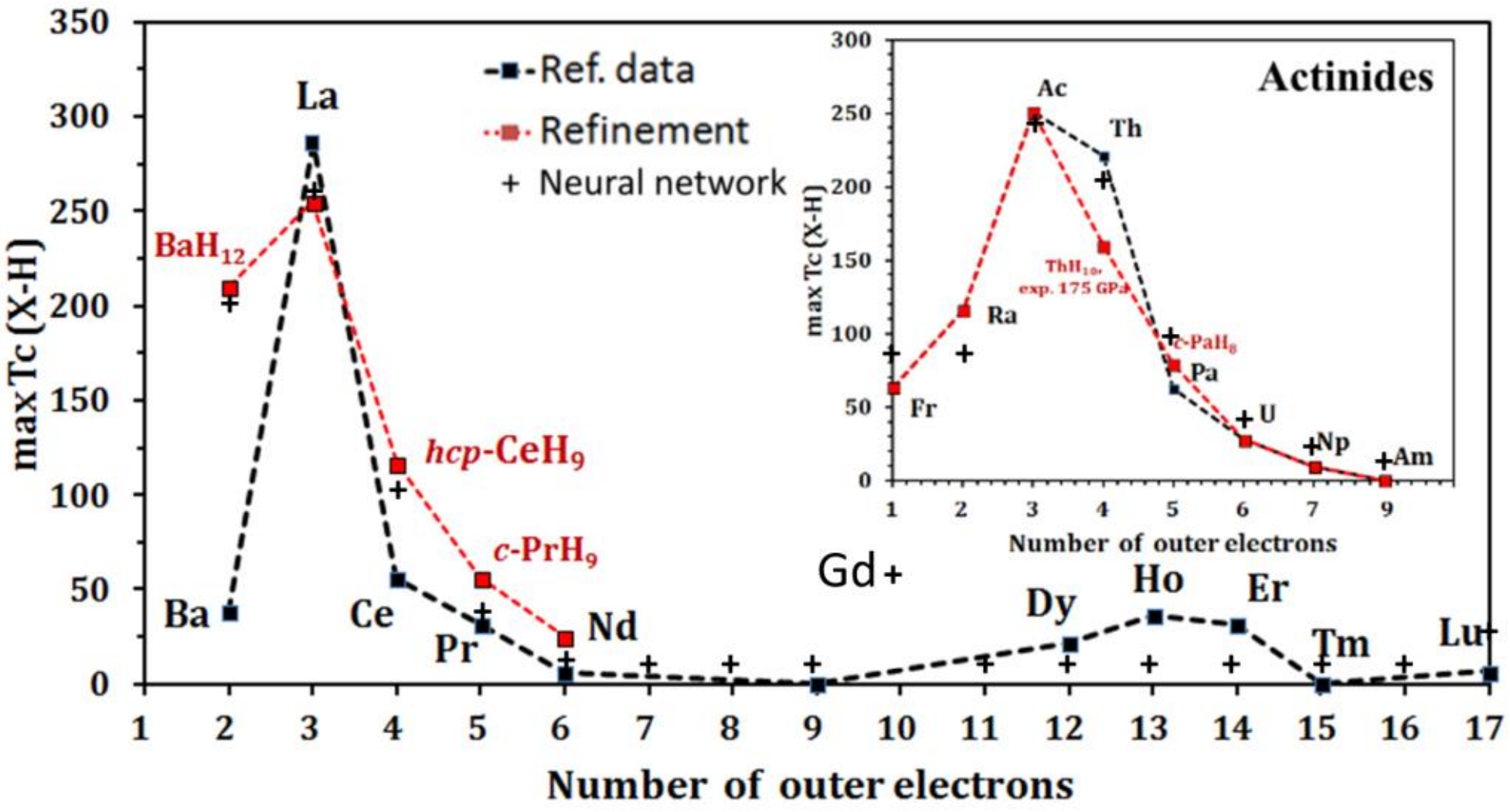

**Figure 5.** Maximum critical temperature (max$T_C$) in lanthanide and actinide hydrides as a function of the number of outer electrons in the metal atom. One can see "secondary wave" of superconductivity in heavy lanthanides. Experimental data of $T_C$ for $ThH_{10}$ was taken from Ref. [12].

The significant difference between the structures of hydrides of *d*- and *f*-elements is shown in Supporting Information Tables S4 and S5. The "pure" *d*-elements (Y, La, Ac, and Th) tend to form higher hydrides with cubic crystal structure, such as $XH_{10}$, with $XH_{32}$ clathrate cages. Adding two or more *f*-electrons leads to changes in the crystal structure: the cubic $XH_8$ motif becomes predominant for the hydrides at 150 GPa. As long as the *f*-shell is filled sequentially within a certain range, the situation does not change much, as illustrated by the series of hydrides $Fm\bar{3}m$-$PaH_8$, $Fm\bar{3}m$-$UH_8$, $Fm\bar{3}m$-$NpH_8$, and $Fm\bar{3}m$-$AmH_8$.

The new effect - the formation of a unique layered structure with hydrogen nonagons (9-angle polygons) - was observed in erbium hydride $P\bar{6}m2$-$ErH_{15}$ (Supporting Information Table S5). The closest analog with a similar crystal structure is $P\bar{6}m2$-$AcH_{16}$.[22] The hydrogen nonagons have not been observed in any other superhydrides considered above. Lutetium, despite having one *d*-electron and a completely filled *f*-shell, is not an analog of the $d^1$-elements Sc-Y-La-Ac because the *d*- and *f*-shells are still similar in energy. $LuH_{13}$ contains molecular fragments and is a semiconductor like many other hydrides of heavy lanthanides and actinides.

An increase in the number of *f*-electrons in lanthanides has an adverse effect on their superconductivity. We investigated the Pa-H, Np-H, Am-H, and Cm-H systems in addition to the already studied Ac-H,[22] Th-H,[21] and U-H[10] systems (Table 6) to confirm that max$T_C$ decreases as the number of *f*-electrons grows. The dependence of max$T_C$ on the number of outer electrons is shown in the inset in Figure 5. Pr ($4f^3 6s^2$) and Pa ($5f^2 6d^1 7s^2$) have the same total number of *d*- and *f*-electrons, but $T_C$(Pa-H) > $T_C$(Pr-H)[13, 57], because more localized *f*-electrons suppress superconductivity more than *d*-electrons (see details in the statistics section of Supporting Information).

The hydrides of actinides allow us to suggest that the critical temperature decreases as the total number of *d*- and *f*-electrons grows. The highest critical temperature among the hydrides of actinides, 251 K, belongs to the Ac-H system. It starts the sequence of critical temperatures decreasing monotonically in the Ac-Th-Pa-U-Np-Am series as the total number of *d*- and *f*-electrons increases. The data obtained for the newly predicted metal hydrides are summarized in Table 6.

**Table 6.** Parameters of superconductivity of the newly predicted metal hydrides. The critical temperatures were calculated using the Allen-Dynes formula [56] at $\mu^* = 0.1$.

| Phases | $P$, GPa | λ | $\omega_{log}$, K | $T_C$, K | $N(E_F)$, $eV^{-1}f.u.^{-1}$ | $\mu_0 H_C$, T | $\Delta$, meV |
|---|---|---|---|---|---|---|---|
| $C2/m$-$TiH_{14}$ | 200 | 0.81 | 1063 | 54 | 0.87 | 12 | 8.8 |
| $Immm$-$KH_{10}$ | 150 | 1.34 | 1301 | 148 | 0.40 | 27 | 27.9 |
| $C2/m$-$CaH_{12}$ | 150 | 2.16 | 1074 | 206 | 0.62 | 55.7 | 45 |
| $P2_1/c$-$ZrH_{16}$ | 200 | 1.19 | 852 | 88 | 0.75 | 41.7 | 16.2 |
| $Amm2$-$Hf_3H_{13}$ | 100 | 1.10 | 497 | 43 | 0.33 | 6.4 | 7.6 |
| $C2/m$-$HfH_{14}$ | 300 | 0.93 | 1138 | 76 | 0.37 | 11.3 | 12.8 |
| $P4/nmm$-$FrH_7$* | 100 | 1.08 | 745 | 64 | 0.15 | 6.4 | 11.3 |
| $C2/m$-$RaH_{12}$* | 200 | 1.36 | 998 | 116 | 0.51 | 22 | 23 |
| $I4/mmm$-$BaH_{12}$ | 135 | 2.66 | 927 | 214 | 0.30 | 43 | 49 |

* see Supporting Information for details

All lanthanides were combined in a single group because of their similar physical and chemical properties (electronic structure, coordination numbers, atomic radii, electronegativity, etc.). For light lanthanides, the crystal structures and compositions of pressure-stabilized hydrides are also similar (Figure 6). The unexpected result is the extremely strong dependence of the superconducting critical temperature on the number of $d$+$f$ electrons. Despite the same structure and seemingly the same stabilization pressure of the metallic hydrogen sublattice of $Fm\bar{3}m$-$PrH_8$, $Fm\bar{3}m$-$NdH_8$, and $Fm\bar{3}m$-$TmH_8$, they have completely different λ, $T_C$ and $N_F$ (Supporting Information Tables S7 and S8). This clearly shows the crucial importance of the lanthanide atom in superconductivity – i.e. that lanthanide hydrides should not be considered as analogs of metallic hydrogen.

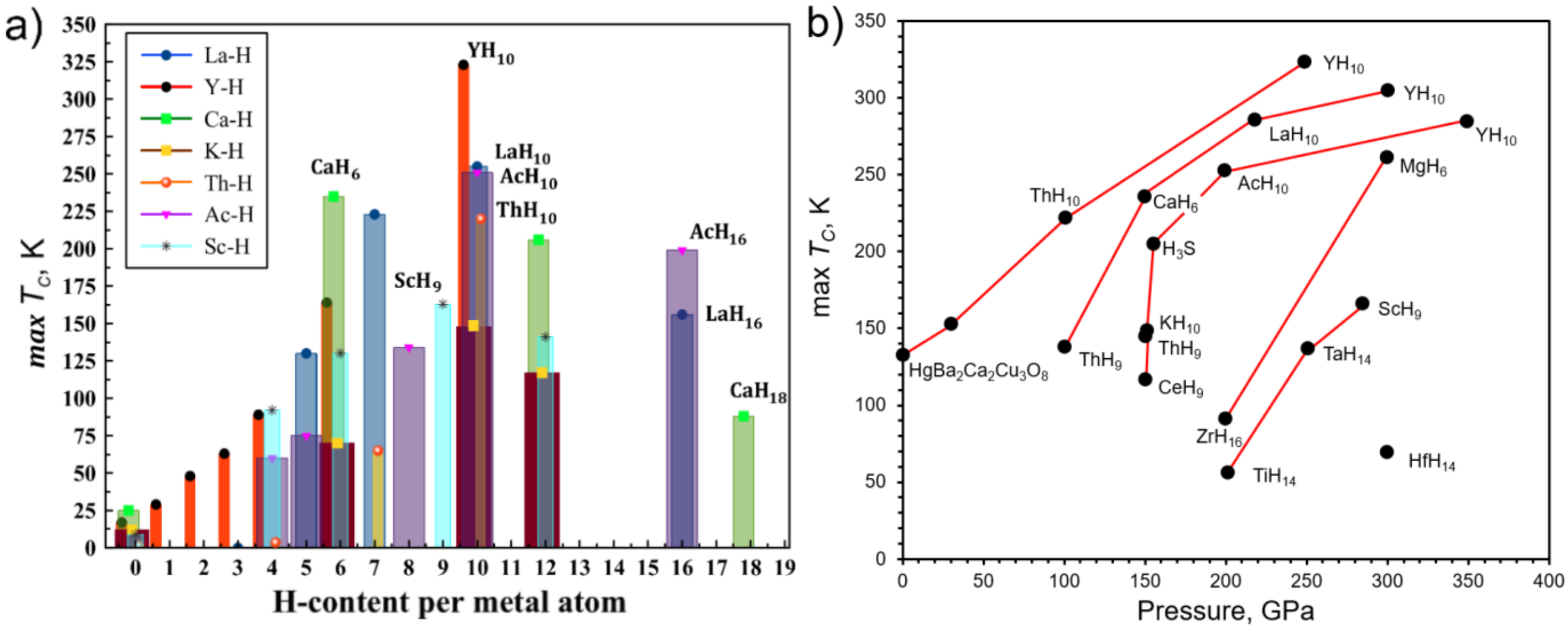

**Figure 6.** a) Maximum $T_C$ for a given hydrogen content per metal atom in the $XH_n$ hydrides, calculated using the Allen-Dynes formula. b) Ashby plot of the maximum $T_C$ versus pressure for the best-known calculated metal hydrides.

Using the data from Refs. [15, 18, 21, 22, 37, 38, 62–66] for the La-H, Y-H, Ca-H, K-H, Ac-H, Th-H, and Sc-H systems, the dependence of max$T_C$ on the number of the hydrogen atoms $n$ in the $XH_n$ compositions was determined (Figure 6a). The highest max$T_C$ values were achieved in compositions with $n = 6…12$ and particularly $n = 10$ (Y, La, Th, Ac). Further growth of $n$ (> 12) often leads to the formation of molecular hydrides with low $T_C$. As molecular hydrogen transforms to monoatomic at ≥ 500 GPa, such stoichiometries can also be very-$T_C$ superconductors at pressures above molecular dissociation.

The Ashby plot of max$T_C$ versus pressure (Figure 6b) helps to identify the superconducting metal hydrides with the highest $T_C$ and lowest stabilization pressure. From these two criteria, the best superconductors are $HgBa_2Ca_2Cu_3O_8$ [62], $ThH_{10}$, $YH_{10}$. The expected critical temperature increases with the external pressure, probably reaches the maximum at 200-250 GPa, and then declines. Presumably, the pressure range of 100 to 250 GPa is required to achieve $T_C$ > 200 K in metal hydrides.

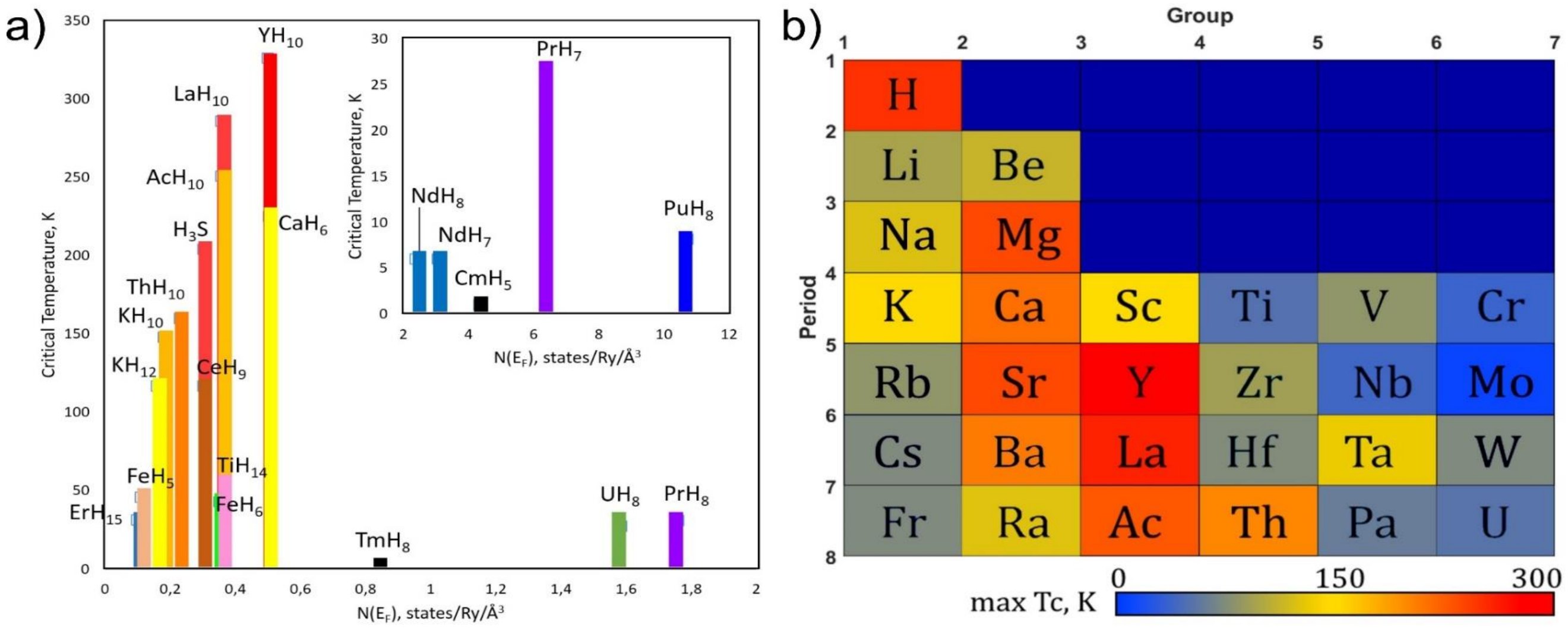

**Figure 7.** a) Maximum $T_C$ versus the electronic density of states at Fermi level for the studied higher metal hydrides. b) The distribution of max$T_C$ of metal hydrides in the left part of Mendeleev's table showing the diagonal rule.

The highest $T_C$ values usually correspond to transfer of ~0.3 electron per H atom. [54] More precisely, the number of electrons per H atom is 0.33$e$ for Ca ($s^2$), and 0.28-0.33$e$ for Ti-Zr-Hf ($s^2d^2$).[54] In other words, metal atoms serve as donors of electron density for the metallic hydrogen sublattice in polyhydrides. The optimal level of the electron doping can be estimated by calculating the electronic density of states of each compound. For the $YH_{10}$-$LaH_{10}$-$AcH_{10}$ series, the electronic density of states at the Fermi level ($N(E_F)$) is 10-12 states/Ry/cell, or 0.3-0.5 states/Ry/Å$^3$, or ~ 1 state/Ry/H-atom at 200-250 GPa. A deviation from this value results in a sharp decrease in the superconducting properties (Figure 7a and Supporting Information Table S7).

The geometric profile of the distribution of superconducting properties of metal hydrides in a part of Mendeleev's table is shown in Figure 7b. We called this a "diagonal effect," i.e. the deviation from the vertical of the superconducting $T_C$ of metal hydrides from the $d$-belt. A similar effect is also seen among the hydrides of $p$-elements.

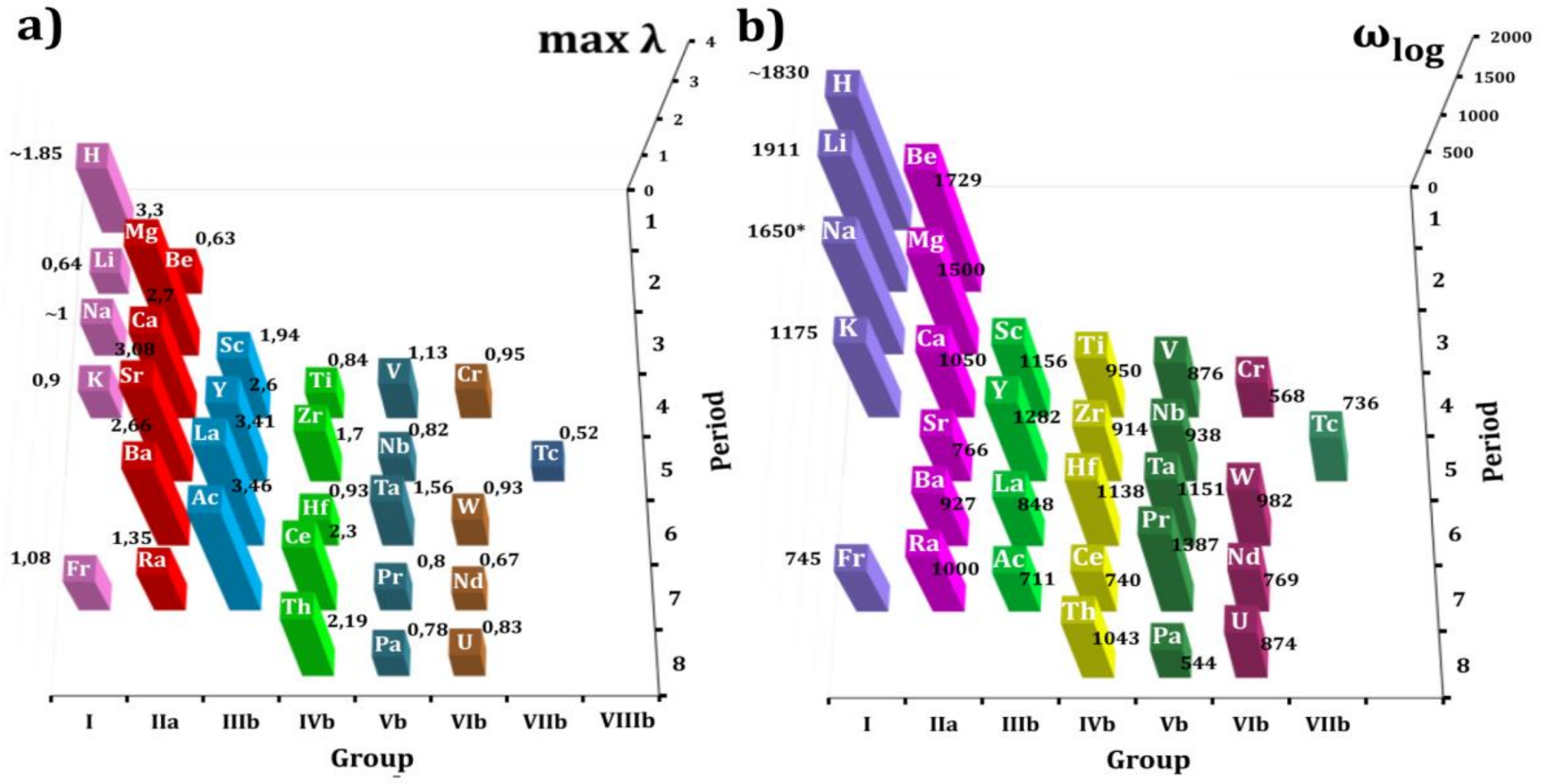

**Figure 8.** Distribution of a) the EPC coefficient λ and b) logarithmic frequency $\omega_{log}$ corresponding to the max$T_C$ points over Mendeleev's table. The data for hydrogen were taken from Refs. [69, 70]; the data for Na-H were estimated using our neural network.

The distribution of the EPC coefficients λ and logarithmic frequencies $\omega_{log}$ over Mendeleev's Periodic Table, corresponding to max$T_C$ of the studied hydrides, is shown in Figure 8. While the highest value of $\omega_{log}$ are predictably concentrated in the area of light elements, distribution of λ coincides with that of max$T_C$: the maximum values belong to the hydrides from the $d^0$ and $d^1$ belts (Figure 8a). This is not surprising because the record superconductivity is almost always associated with anomalously high λ values. The distribution of $\omega_{log}$ (Figure 8b) reaching the maximum for hydrides of light elements (H-Be-Li-Mg). High values of $\omega_{log}$ are associated with high pressures (> 200-300 GPa) which are required to stabilize and metalize molecular polyhydrides of light elements. The values of $\omega_{log}$, predicted using the neural network, may be indirectly used to estimate the minimal pressure necessary for stabilization of hydrides.

According to the obtained data and proposed rules, the most pronounced superconducting properties could be expected for hydrides of the elements lying in transition zones at the boundary between blocks of the Periodic Table (*s-p, s-d, p-d, d-f*) and are further enhanced by incipient dynamical instabilities ($H_3S$ and $LaH_{10}$).

## Author Contributions:

D.V.S., A.G.K., and I.A.K. performed calculations. A.G.K. and D.V.S. prepared the theoretical analysis. I.A.S. and D.V.S. developed neural network. D.V.S., A.G.K. and I.A.K. contributed to the interpretation of the results. D.V.S., A.G.K., A.R.O. wrote the manuscript. All authors provided critical feedback and helped shape the research, analysis and manuscript.

## Conclusions

Detailed analysis of both obtained and available reference data enabled us to propose the rules of distribution of the superconducting metal hydrides over Mendeleev's table. Our main conclusions are:

- Most of the high-temperature superconducting metal hydrides are concentrated in the $d^0$ and $d^1$ belts of Mendeleev's Table. The metals forming the high-$T_C$ hydrides are: Sc-Y-La-Ac ($d^1$ belt), Mg-Ca-Sr-Ba-Ra ($d^0$ belt), and Th ($s^2d^2$). These elements are electronically labile, i.e. their orbital populations should be sensitive to atomic environment, giving rise to strong electron-phonon coupling;
- Maximum values of $T_C$ are achieved when the number of electrons transferred from metal atoms is ~0.3*e* per hydrogen atom. In other words, the electronic density of states at the Fermi level $N(E_F)$ should be ~ 1 state/Ry/H-atom.
- The superconducting properties of hydrides become less and less pronounced as the total number of *d*- and *f*-electrons increases. In particular, the critical temperatures of lanthanide and actinide hydrides decrease almost monotonically with an increase in the number of *f*-electrons.
- Optimal hydrogen content in superconductors corresponds to $XH_{10\pm2}$ composition which can be formed at pressures of about 150 to 250 GPa. If the composition or/and the level of electron doping deviate from the optimal values, the superconductivity parameters decrease.
- The pressure required for the stabilization of hydrides of lanthanides and actinides decreases going down the Periodic Table. For actinides, the pressure of 50 GPa is enough to stabilize superconducting polyhydrides, like $UH_7$ and $PaH_8$.
- Using our rule of electronic lability, we questioned and successfully corrected previous results on K-H, Zr-H, Hf-H and Ti-H systems. New hydrides were also predicted in Mg-H, Sr-H, Ba-H, Cs-H and Rb-H systems.

## Supporting Information

The crystal data, convex hulls, Eliashberg functions, forecast list, and statistical data for the studied metal polyhydrides are included in Supporting Information.

## Acknowledgments

A.R.O. acknowledges the Russian Science Foundation (№19-72-30043). The calculations were performed on Rurik supercomputer at the MIPT. The authors thank Dr. Mahdi Davari Esfahani for the study of stable phases in the Cm-H system. A.G.K. thanks the Foundation for Assistance to Small Innovative Enterprises (FASIE) for the financial support within the UMNIK grant №13408GU/2018 and Russian Foundation for Basic Research (RFBR) project №19-03-00100.

## Data availability

The authors declare that all other data supporting the findings of this study are available within the paper and its supplementary information files

## References

(1) Drozdov, A. P.; Eremets, M. I.; Troyan, I. A.; Ksenofontov, V.; Shylin, S. I. Conventional Superconductivity at 203 Kelvin at High Pressures in the Sulfur Hydride System. *Nature* **2015**, *525* (7567), 73–76. https://doi.org/10.1038/nature14964.

(2) Duan, D.; Liu, Y.; Ma, Y.; Shao, Z.; Liu, B.; Cui, T. Structure and Superconductivity of Hydrides at High Pressures. *Natl. Sci. Rev.* **2017**, *4* (1), 121–135. https://doi.org/10.1093/nsr/nww029.

(3) Mishra, A. K.; Somayazulu, M.; Ahart, M.; Karandikar, A.; Hemley, R. J.; Struzhkin, V. V. Novel Synthesis Route and Observation of Superconductivity in the Se-H System at Extreme Conditions. In *Bulletin of the American Physical Society*; 2018.

(4) Kong, P. P.; Drozdov, A. P.; Eroke, E.; Eremets, M. I. Pressure-Induced Superconductivity above 79 K in Si2H6. In *Book of abstracts of AIRAPT 26 joint with ACHPR 8 & CHPC 19*; Biijing, China, 2017; p 347.

(5) Geballe, Z. M.; Liu, H.; Mishra, A. K.; Ahart, M.; Somayazulu, M.; Meng, Y.; Baldini, M.; Hemley, R. J. Synthesis and Stability of Lanthanum Superhydrides. *Angew. Chem. Int. Ed.* **2017**, *57* (3), 688–692. https://doi.org/10.1002/anie.201709970.

(6) Drozdov, A. P.; Kong, P. P.; Minkov, V. S.; Besedin, S. P.; Kuzovnikov, M. A.; Mozaffari, S.; Balicas, L.; Balakirev, F. F.; Graf, D. E.; Prakapenka, V. B.; et al. Superconductivity at 250 K in Lanthanum Hydride under High Pressures. *Nature* **2019**, *569* (7757), 528. https://doi.org/10.1038/s41586-019-1201-8.

(7) Somayazulu, M.; Ahart, M.; Mishra, A. K.; Geballe, Z. M.; Baldini, M.; Meng, Y.; Struzhkin, V. V.; Hemley, R. J. Evidence for Superconductivity above 260 K in Lanthanum Superhydride at Megabar Pressures. *Phys. Rev. Lett.* **2019**, *122* (2), 027001. https://doi.org/10.1103/PhysRevLett.122.027001.

(8) Pépin, C. M.; Geneste, G.; Dewaele, A.; Mezouar, M.; Loubeyre, P. Synthesis of FeH5: A Layered Structure with Atomic Hydrogen Slabs. *Science* **2017**, *357* (6349), 382–385. https://doi.org/10.1126/science.aan0961.

(9) Majumdar, A.; Tse, J. S.; Wu, M.; Yao, Y. Superconductivity in FeH5. *Phys. Rev. B* **2017**, *96* (20), 201107. https://doi.org/10.1103/PhysRevB.96.201107.

(10) Kruglov, I. A.; Kvashnin, A. G.; Goncharov, A. F.; Oganov, A. R.; Lobanov, S. S.; Holtgrewe, N.; Jiang, Sh.; Prakapenka, V. B.; Greenberg, E.; Yanilkin, A. V. Uranium Polyhydrides at Moderate Pressures: Prediction, Synthesis, and Expected Superconductivity. *Sci Adv* **2018**, *4* (10), eaat9776. https://doi.org/10.1126/sciadv.aat9776.

(11) Salke, N. P.; Esfahani, M. M. D.; Zhang, Y.; Kruglov, I. A.; Zhou, J.; Wang, Y.; Greenberg, E.; Prakapenka, V. B.; Oganov, A. R.; Lin, J.-F. Synthesis of Clathrate Cerium Superhydride CeH9 at 80 GPa with Anomalously Short H-H Distance. *arXiv:1805.02060* **2018**, https://arxiv.org/ftp/arxiv/papers/1805/1805.02060.pdf.

(12) Semenok, D. V.; Kvashnin, A. G.; Ivanova, A. G.; Svitlyk, V.; Troyan, I. A.; Oganov, A. R. Synthesis of ThH4, ThH6, ThH9 and ThH10: A Route to Room-Temperature Superconductivity. *arXiv* **2019**, No. 1902.10206.

(13) Zhou, D.; Semenok, D.; Duan, D.; Xie, H.; Huang, X.; Chen, W.; Li, X.; Liu, B.; Oganov, A. R.; Cui, T. Superconducting Praseodymium Superhydrides. *ArXiv190406643 Cond-Mat* **2019**.

(14) Xe, Y.; Li, Q.; Oganov, A. R.; Wang, H. Superconductivity of Lithium-Doped Hydrogen under High Pressure. *Acta Cryst C* **2014**, *70*, 104–111.
(15) Wang, H.; Tse, J. S.; Tanaka, K.; Iitaka, T.; Ma, Y. Superconductive Sodalite-like Clathrate Calcium Hydride at High Pressures. *Proc. Natl. Acad. Sci.* **2012**, *109* (17), 6463–6466. https://doi.org/10.1073/pnas.1118168109.
(16) Tanaka, K.; Tse, J. S.; Liu, H. Electron-Phonon Coupling Mechanisms for Hydrogen-Rich Metals at High Pressure. *Phys. Rev. B* **2017**, *96* (10), 100502. https://doi.org/10.1103/PhysRevB.96.100502.
(17) Gao, G.; Hoffmann, R.; Ashcroft, N. W.; Liu, H.; Bergara, A.; Ma, Y. Theoretical Study of the Ground-State Structures and Properties of Niobium Hydrides under Pressure. *Phys. Rev. B* **2013**, *88* (18), 184104. https://doi.org/10.1103/PhysRevB.88.184104.
(18) Abe, K. Hydrogen-Rich Scandium Compounds at High Pressures. *Phys. Rev. B* **2017**, *96* (14), 144108. https://doi.org/10.1103/PhysRevB.96.144108.
(19) Li, X.; Peng, F. Superconductivity of Pressure-Stabilized Vanadium Hydrides. *Inorg. Chem.* **2017**, *56* (22), 13759–13765. https://doi.org/10.1021/acs.inorgchem.7b01686.
(20) Liu, H.; Naumov, I. I.; Hoffmann, R.; Ashcroft, N. W.; Hemley, R. J. Potential High-Tc Superconducting Lanthanum and Yttrium Hydrides at High Pressure. *Proc. Natl. Acad. Sci.* **2017**, *114*, 6990–6995. https://doi.org/10.1073/pnas.1704505114.
(21) Kvashnin, A. G.; Semenok, D. V.; Kruglov, I. A.; Wrona, I. A.; Oganov, A. R. High-Temperature Superconductivity in Th-H System at Pressure Conditions. *ACS Appl. Mater. Interfaces* **2018**, *10* (50), 43809–43816.
(22) Semenok, D. V.; Kvashnin, A. G.; Kruglov, I. A.; Oganov, A. R. Actinium Hydrides AcH10, AcH12, and AcH16 as High-Temperature Conventional Superconductors. *J. Phys. Chem. Lett.* **2018**, *9* (8), 1920–1926. https://doi.org/10.1021/acs.jpclett.8b00615.
(23) Oganov, A. R.; Glass, C. W. Crystal Structure Prediction Using Ab Initio Evolutionary Techniques: Principles and Applications. *J Chem Phys* **2006**, *124*, 244704. https://doi.org/10.1063/1.2210932.
(24) Oganov, A. R.; Lyakhov, A. O.; Valle, M. How Evolutionary Crystal Structure Prediction Works—and Why. *Acc. Chem. Res.* **2011**, *44*, 227–237. https://doi.org/10.1021/ar1001318.
(25) Lyakhov, A. O.; Oganov, A. R.; Stokes, H. T.; Zhu, Q. New Developments in Evolutionary Structure Prediction Algorithm USPEX. *Comput. Phys. Commun.* **2013**, *184*, 1172–1182. https://doi.org/10.1016/j.cpc.2012.12.009.
(26) Bushlanov, P. V.; Blatov, V. A.; Oganov, A. R. Topology-Based Crystal Structure Generator. *Comput. Phys. Commun.* **2019**, *236*, 1–7. https://doi.org/10.1016/j.cpc.2018.09.016.
(27) Hohenberg, P.; Kohn, W. Inhomogeneous Electron Gas. *Phys Rev* **1964**, *136* (3B), B864–B871.
(28) Kohn, W.; Sham, L. J. Self-Consistent Equations Including Exchange and Correlation Effects. *Phys Rev* **1965**, *140* (4), A1133–A1138.
(29) Perdew, J. P.; Burke, K.; Ernzerhof, M. Generalized Gradient Approximation Made Simple. *Phys. Rev. Lett.* **1996**, *77* (18), 3865–3868.
(30) Blöchl, P. E. Projector Augmented-Wave Method. *Phys. Rev. B* **1994**, *50* (24), 17953–17979. https://doi.org/10.1103/PhysRevB.50.17953.
(31) Kresse, G.; Joubert, D. From Ultrasoft Pseudopotentials to the Projector Augmented-Wave Method. *Phys. Rev. B* **1999**, *59* (3), 1758–1775. https://doi.org/10.1103/PhysRevB.59.1758.
(32) Kresse, G.; Furthmüller, J. Efficient Iterative Schemes for Ab Initio Total-Energy Calculations Using a Plane-Wave Basis Set. *Phys. Rev. B* **1996**, *54*, 11169–11186. https://doi.org/10.1103/PhysRevB.54.11169.
(33) Kresse, G.; Hafner, J. Ab Initio Molecular Dynamics for Liquid Metals. *Phys. Rev. B* **1993**, *47*, 558–561. https://doi.org/10.1103/PhysRevB.47.558.
(34) Kresse, G.; Hafner, J. Ab Initio Molecular-Dynamics Simulation of the Liquid-Metal Amorphous-Semiconductor Transition in Germanium. *Phys. Rev. B* **1994**, *49*, 14251–14269. https://doi.org/10.1103/PhysRevB.49.14251.
(35) Baroni, S.; de Gironcoli, S.; Dal Corso, A.; Giannozzi, P. Phonons and Related Crystal Properties from Density-Functional Perturbation Theory. *Rev. Mod. Phys.* **2001**, *73* (2), 515–562.
(36) Giannozzi, P.; Baroni, S.; Bonini, N.; Calandra, M.; Car, R.; Cavazzoni, C.; Ceresoli, D.; Chiarotti, G. L.; Cococcioni, M.; Dabo, I.; et al. QUANTUM ESPRESSO: A Modular and Open-Source Software Project for Quantum Simulations of Materials. *J. Phys. Condens. Matter* **2009**, *21*, 395502. https://doi.org/10.1088/0953-8984/21/39/395502.

(37) Zhou, D.; Jin, X.; Meng, X.; Bao, G.; Ma, Y.; Liu, B.; Cui, T. Ab Initio Study Revealing a Layered Structure in Hydrogen-Rich KH${}_{6}$ under High Pressure. *Phys. Rev. B* **2012**, *86* (1), 014118. https://doi.org/10.1103/PhysRevB.86.014118.
(38) Ye, X.; Zarifi, N.; Zurek, E.; Hoffmann, R.; Ashcroft, N. W. High Hydrides of Scandium under Pressure: Potential Superconductors. *J. Phys. Chem. C* **2018**, *122* (11), 6298–6309. https://doi.org/10.1021/acs.jpcc.7b12124.
(39) Shanavas, K. V.; Lindsay, L.; Parker, D. S. Electronic Structure and Electron-Phonon Coupling in $TiH_2$. *Sci. Rep.* **2016**, *6*, 28102. https://doi.org/10.1038/srep28102.
(40) Yu, S.; Jia, X.; Frapper, G.; Li, D.; Oganov, A. R.; Zeng, Q.; Zhang, L. Pressure-Driven Formation and Stabilization of Superconductive Chromium Hydrides. *Sci. Rep.* **2015**, *5*, 17764. https://doi.org/10.1038/srep17764.
(41) Hooper, J.; Altintas, B.; Shamp, A.; Zurek, E. Polyhydrides of the Alkaline Earth Metals: A Look at the Extremes under Pressure. *J. Phys. Chem. C* **2013**, *117* (6), 2982–2992. https://doi.org/10.1021/jp311571n.
(42) Hooper, J.; Zurek, E. High Pressure Potassium Polyhydrides: A Chemical Perspective. *J. Phys. Chem. C* **2012**, *116* (24), 13322–13328. https://doi.org/10.1021/jp303024h.
(43) Mishra, A. K.; Muramatsu, T. S.; Liu, H.; Geballe, Z. M.; Somayazulu, M.; Ahart, M.; Baldini, M.; Meng, Y.; Zurek, E.; Hemley, R. J. New Calcium Hydrides With Mixed Atomic and Molecular Hydrogen. *J. Phys. Chem. C* **2018**, *122* (34), 19370–19378. https://doi.org/10.1021/acs.jpcc.8b05030.
(44) Li, X.; Liu, H.; Peng, F. Crystal Structures and Superconductivity of Technetium Hydrides under Pressure. *Phys. Chem. Chem. Phys.* **2016**, *18* (41), 28791–28796. https://doi.org/10.1039/C6CP05702K.
(45) Li, X.-F.; Hu, Z.-Y.; Huang, B. Phase Diagram and Superconductivity of Compressed Zirconium Hydrides. *Phys. Chem. Chem. Phys.* **2017**, *19* (5), 3538–3543. https://doi.org/10.1039/C6CP08036G.
(46) Hooper, J.; Zurek, E. Rubidium Polyhydrides Under Pressure: Emergence of the Linear H3− Species - Hooper - 2012 - Chemistry – A European Journal - Wiley Online Library. *Chem Eur J* **2012**, *18*, 5013–5021.
(47) Wu, G.; Huang, X.; Xie, H.; Li, X.; Liu, M.; Liang, Y.; Huang, Y.; Duan, D.; Li, F.; Liu, B.; et al. Unexpected Calcium Polyhydride CaH4: A Possible Route to Dissociation of Hydrogen Molecules. *J. Chem. Phys.* **2019**, *150* (4), 044507. https://doi.org/10.1063/1.5053650.
(48) Abe, K. High-Pressure Properties of Dense Metallic Zirconium Hydrides Studied by Ab Initio Calculations. *Phys. Rev. B* **2018**, *98* (13), 134103. https://doi.org/10.1103/PhysRevB.98.134103.
(49) Shamp, A.; Hooper, J.; Zurek, E. Compressed Cesium Polyhydrides: Cs+ Sublattices and H3– Three-Connected Nets - Inorganic Chemistry (ACS Publications). *Inorg Chem* **2012**, *51* (17), 9333–9342.
(50) Duda, A. M.; Szewczyk, K. A.; Jarosik, M. W.; Szcześniak, K. M.; Sowińska, M. A.; Szcześniak, D. Characterization of the Superconducting State in Hafnium Hydride under High Pressure. *Phys. B Condens. Matter* **2018**, *536*, 275–279. https://doi.org/10.1016/j.physb.2017.10.107.
(51) Zhuang, Q.; Jin, X.; Cui, T.; Ma, Y.; Lv, Q.; Li, Y.; Zhang, H.; Meng, X.; Bao, K. Pressure-Stabilized Superconductive Ionic Tantalum Hydrides. *Inorg. Chem.* **2017**, *56* (7), 3901–3908. https://doi.org/10.1021/acs.inorgchem.6b02822.
(52) Ying, J.; Li, X.; Greenberg, E.; Prakapenka, V. B.; Liu, H.; Struzhkin, V. V. Synthesis and Stability of Tantalum Hydride at High Pressures. *Phys. Rev. B* **2019**, *99* (22), 224504. https://doi.org/10.1103/PhysRevB.99.224504.
(53) Personal Information from Huang's Group, Jilin University.
(54) Peng, F.; Sun, Y.; Pickard, C. J.; Needs, R. J.; Wu, Q.; Ma, Y. Hydrogen Clathrate Structures in Rare Earth Hydrides at High Pressures: Possible Route to Room-Temperature Superconductivity. *Phys. Rev. Lett.* **2017**, *119*, 107001–107007. https://doi.org/10.1103/PhysRevLett.119.107001.
(55) Xiao, X.; Duan, D.; Xie, H.; Shao, Z.; Li, D.; Tian, F.; Song, H.; Yu, H.; Bao, K.; Cui, T. Structure and Superconductivity of Protactinium Hydrides under High Pressure. *J. Phys. Condens. Matter* **2019**, *31* (31), 315403. https://doi.org/10.1088/1361-648X/ab1d03.
(56) Allen, P. B.; Dynes, R. C. Transition Temperature of Strong-Coupled Superconductors Reanalyzed. *Phys. Rev. B* **1975**, *12* (3), 905–922. https://doi.org/10.1103/PhysRevB.12.905.
(57) Kruglov, I. A.; Semenok, D. V.; Song, H.; Szczęśniak, R.; Wrona, I. A.; Akashi, R.; Esfahani, M. M. D.; Duan, D.; Cui, T.; Kvashnin, A. G.; et al. Superconductivity of LaH10 and LaH16 Polyhydrides. *arXiv* **2018**, No. 1810.01113v3.

(58) Hamlin, J. J.; Tissen, V. G.; Schilling, J. S. Superconductivity at $17\phantom{\rule{0.3em}{0ex}}\mathrm{K}$ in Yttrium Metal under Nearly Hydrostatic Pressures up to $89\phantom{\rule{0.3em}{0ex}}\mathrm{GPa}$. *Phys. Rev. B* **2006**, *73* (9), 094522. https://doi.org/10.1103/PhysRevB.73.094522.

(59) Yabuuchi, T.; Matsuoka, T.; Nakamoto, Y.; Shimizu, K. Superconductivity of Ca Exceeding 25 K at Megabar Pressures. *J. Phys. Soc. Jpn.* **2006**, *75* (8), 083703. https://doi.org/10.1143/JPSJ.75.083703.

(60) Profeta, G.; Franchini, C.; Lathiotakis, N. N.; Floris, A.; Sanna, A.; Marques, M. A. L.; Lüders, M.; Massidda, S.; Gross, E. K. U.; Continenza, A. Superconductivity in Lithium, Potassium, and Aluminum under Extreme Pressure: A First-Principles Study. *Phys. Rev. Lett.* **2006**, *96* (4), 047003. https://doi.org/10.1103/PhysRevLett.96.047003.

(61) Nixon, L. W.; Papaconstantopoulos, D. A.; Mehl, M. J. Calculations of the Superconducting Properties of Scandium under High Pressure. *Phys. Rev. B* **2007**, *76* (13), 134512. https://doi.org/10.1103/PhysRevB.76.134512.

(62) Schilling, A.; Cantoni, M.; Guo, J. D.; Ott, H. R. Superconductivity above 130 K in the Hg–Ba–Ca–Cu–O System. *Nature* **1993**, *363* (6424), 56–58. https://doi.org/10.1038/363056a0.

(63) Richardson, C. F.; Ashcroft, N. W. High Temperature Superconductivity in Metallic Hydrogen: Electron-Electron Enhancements. *Phys. Rev. Lett.* **1997**, *78* (1), 118–121. https://doi.org/10.1103/PhysRevLett.78.118.

(64) Barbee, T. W.; García, A.; Cohen, M. L. First-Principles Prediction of High-Temperature Superconductivity in Metallic Hydrogen. *Nature* **1989**, *340* (6232), 369. https://doi.org/10.1038/340369a0.

# Supporting Information

# On Distribution of Superconductivity in Metal Hydrides

*Dmitrii V. Semenok,* [1,*] *Ivan A. Kruglov,* [2,3] *Igor A. Savkin,*[4] *Alexander G. Kvashnin* [1,2,*] *and Artem R. Oganov* [1,2,5]

[1] Skolkovo Institute of Science and Technology, Skolkovo Innovation Center 121205, 3 Nobel Street, Moscow, Russia

[2] Moscow Institute of Physics and Technology, 141700, 9 Institutsky lane, Dolgoprudny, Russia

[3] Dukhov Research Institute of Automatics (VNIIA), Moscow 127055, Russia

[4] Research Computer Center of Lomonosov Moscow State University, Moscow, Russia

[5] International Center for Materials Discovery, Northwestern Polytechnical University, Xi'an, 710072, China

**Corresponding authors**

*Alexander G. Kvashnin: A.Kvashnin@skoltech.ru

*Dmitrii V. Semenok: dmitrii.semenok@skoltech.ru

## Table of Contents

## Crystal structures of metal polyhydrides

**Table S1.** Crystal structures of predicted novel metal polyhydrides

| Phase | Pressure, GPa | Volume per primitive cell, Å$^3$ | Lattice parameters | Coordinates | | | |
|---|---|---|---|---|---|---|---|
| *Cm*-$KH_9$ | 50 | 41.73 | a = 6.50Å α = 90°<br>b = 3.98 Å β = 123.43°<br>c = 3.86Å γ = 90° | K1 | -0.206 | 0.000 | 0.459 |
| | | | | H1 | 0.470 | 0.295 | 0.067 |
| | | | | H2 | -0.134 | 0.296 | 0.060 |
| | | | | H3 | 0.174 | 0.000 | -0.317 |
| | | | | H4 | 0.463 | 0.000 | -0.460 |
| | | | | H5 | 0.175 | 0.000 | 0.086 |
| | | | | H6 | 0.167 | 0.000 | 0.480 |
| | | | | H7 | -0.428 | 0.000 | -0.238 |
| *Imm2*-$KH_{11}$ | 50 | 47.01 | a = 4.72Å α = 90°<br>b = 5.30 Å β = 90°<br>c = 3.76Å γ = 90° | K1 | 0.000 | 0.500 | 0.048 |
| | | | | H1 | -0.348 | 0.219 | 0.068 |
| | | | | H2 | 0.273 | 0.355 | 0.490 |
| | | | | H3 | 0.000 | 0.000 | 0.233 |
| | | | | H4 | 0.000 | 0.000 | 0.435 |
| | | | | H5 | 0.000 | 0.000 | -0.191 |
| *Immm*-$KH_{10}$ | 150 | 30.17 | a = 3.99Å α = 90°<br>b = 5.19Å β = 90°<br>c = 2.90Å γ = 90° | K1 | 0.500 | 0.000 | 0.500 |
| | | | | H1 | 0.000 | 0.089 | 0.500 |
| | | | | H2 | 0.321 | 0.219 | 0.000 |
| | | | | H3 | 0.167 | -0.123 | 0.000 |
| *Immm*-$KH_{12}$ | 150 | 33.97 | a = 4.99 Å α = 90°<br>b = 4.34 Å β =90°<br>c = 3.13 Å γ = 90° | K1 | 0.500 | 0.000 | 0.500 |
| | | | | H1 | 0.361 | 0.279 | 0.000 |
| | | | | H2 | -0.081 | 0.000 | 0.239 |
| | | | | H3 | -0.263 | 0.144 | 0.000 |
| *P-1*-$CaH_{14}$ | 100 | 42.25 | a = 3.12 Å α = 78.20°<br>b = 3.37 Å β =73.61°<br>c = 4.30 Å γ = 79.45° | Ca1 | 0.500 | 0.000 | 0.5 |
| | | | | H1 | 0.050 | 0.639 | 0.566 |
| | | | | H2 | -0.031 | 0.187 | 0.815 |
| | | | | H3 | 0.431 | 0.534 | 0.865 |
| | | | | H4 | 0.590 | -0.073 | -0.047 |
| | | | | H5 | 0.559 | 0.439 | 0.701 |
| | | | | H6 | 0.044 | 0.243 | 0.261 |
| | | | | H8 | -0.038 | 0.294 | -0.040 |
| *C2/m*-$CaH_{12}$ | 150 | 33.18 | a = 5.92 Å α = 90°<br>b = 3.02 Å β = 111.87°<br>c = 3.99 Å γ = 90° | Ca1 | 0.000 | 0.500 | 0.000 |
| | | | | H1 | 0.762 | 0.000 | 0.848 |
| | | | | H2 | 0.565 | 0.000 | 0.563 |
| | | | | H3 | 0.892 | 0.000 | 0.190 |
| | | | | H4 | 0.023 | 0.000 | 0.333 |
| | | | | H5 | 0.259 | 0.273 | 0.408 |
| *C2*-$CaH_{18}$ | 150 | 44.18 | a = 8.47 Å α = 90°<br>b = 2.98 Å β = 114.54°<br>c = 3.84 Å γ = 90° | Ca1 | 0.000 | 0.877 | 0.000 |
| | | | | H1 | 0.549 | 0.234 | 0.528 |
| | | | | H2 | 0.147 | 0.383 | 0.179 |
| | | | | H3 | -0.093 | 0.378 | 0.682 |
| | | | | H4 | 0.684 | 0.384 | 0.489 |
| | | | | H5 | 0.271 | 0.184 | 0.160 |
| | | | | H6 | -0.066 | 0.318 | 0.256 |
| | | | | H7 | 0.120 | 0.190 | 0.654 |
| | | | | H8 | 0.735 | 0.384 | 0.352 |
| | | | | H9 | 0.713 | 0.159 | 0.028 |
| *C2*-$CaH_{18}$ | 200 | 39.91 | a = 8.11 Å α = 90°<br>b = 2.89 Å β= 113.34° | Ca1 | 0.000 | 0.819 | 0.500 |
| | | | | H1 | -0.095 | 0.320 | 0.173 |
| | | | c = 3.69 Å γ = 90° | H2 | 0.728 | 0.107 | 0.339 |
| | | | | H3 | 0.737 | 0.320 | 0.851 |
| | | | | H4 | 0.317 | 0.318 | 0.006 |
| | | | | H5 | 0.067 | 0.278 | 0.246 |
| | | | | H6 | 0.553 | 0.16 | 0.033 |
| | | | | H7 | 0.277 | 0.085 | 0.450 |
| | | | | H8 | 0.117 | 0.115 | 0.156 |
| | | | | H9 | 0.848 | 0.323 | 0.318 |
| $P\bar{1}$-$ZrH_{16}$ | 150 | 42.66 | a = 3.04 Å α = 115.5°<br>b = 4.06 Å β= 110.9°<br>c = 4.11 Å γ = 75.7° | Zr1 | 0.000 | 0.500 | 0.000 |
| | | | | H1 | 0.276 | 0.030 | -0.196 |
| | | | | H2 | 0.046 | 0.243 | -0.485 |
| | | | | H3 | 0.148 | 0.287 | 0.377 |
| | | | | H4 | -0.371 | 0.378 | 0.199 |
| | | | | H5 | -0.424 | 0.151 | -0.267 |
| | | | | H6 | -0.490 | -0.338 | -0.341 |
| | | | | H7 | 0.307 | -0.143 | 0.430 |
| | | | | H8 | 0.171 | 0.012 | -0.042 |
| *C2/m*-$HfH_{14}$ | 300 | 0.46 | a = 4.64 Å α = 90°<br>b = 3.04Å β= 99.92°<br>c = 4.37 Å γ = 90° | Hf1 | 0.000 | 0.500 | 0.500 |
| | | | | H1 | 0.494 | 0.250 | -0.136 |
| | | | | H2 | 0.176 | 0.213 | 0.240 |
| | | | | H3 | 0.308 | 0.000 | 0.051 |
| | | | | H4 | -0.161 | 0.000 | 0.066 |
| | | | | H5 | -0.154 | 0.000 | 0.385 |
| C2/m-$TiH_{14}$ | 200 | 32.01 | a = 4.03 Å α = 90°<br>b = 6.34Å β= 141.34°<br>c = 4.00 Å γ = 90 | Ti1 | 0.000 | 0.000 | 0.500 |
| | | | | H1 | 0.350 | 0.241 | 0.424 |
| | | | | H2 | -0.073 | 0.129 | 0.074 |
| | | | | H3 | -0.389 | 0.131 | -0.082 |
| | | | | H4 | -0.354 | 0.000 | -0.424 |
| *C2/c*-$PrH_7$ | 50 | 68.59 | a = 3.69 Å α = 90°<br>b = 6.44 Å β= 91.58°<br>c = 5.76 Å γ = 90° | Pr1 | 0.000 | 0.164 | 0.250 |
| | | | | H1 | 0.000 | 0.497 | 0.250 |
| | | | | H2 | -0.283 | 0.092 | -0.451 |
| | | | | H3 | -0.334 | 0.391 | 0.094 |
| | | | | H4 | 0.467 | 0.154 | 0.053 |
| $P6_3mc$-$PrH_8$ | 150 | 54.10 | a = 3.50 Å α = 90°<br>b = 3.50 Å β= 90°<br>c = 5.17 Å γ = 120° | Pr1 | 0.333 | 0.667 | -0.061 |
| | | | | H1 | 0.176 | -0.176 | 0.266 |
| | | | | H2 | 0.333 | 0.667 | -0.463 |
| | | | | H3 | 0.000 | 0.000 | 0.436 |
| | | | | H4 | -0.155 | 0.155 | 0.121 |
| *F-43m*-$PrH_9$ | 150 | 28.74 | a = 4.86 Å α = 90°<br>b = 4.86 Å β= 90°<br>c = 4.86 Å γ = 90° | Pr1 | 0.750 | 0.750 | 0.750 |
| | | | | H1 | 0.132 | 0.13 | 0.132 |
| | | | | H2 | 0.387 | 0.387 | 0.387 |
| | | | | H3 | 0.000 | 0.000 | 0.000 |
| *Fm-3m*-$NdH_8$ | 150 | 53.85 | a = 4.75 Å α = 90°<br>b = 4.75 Å β= 90°<br>c = 4.75 Å γ = 90° | Nd1 | 0.500 | 0.500 | 0.500 |
| | | | | H1 | 0.132 | 0.132 | 0.132 |
| *Cc*-$HoH_7$ | 150 | 52.05 | a = 3.27 Å α = 90°<br>b = 5.75 Å β= 89.35°<br>c = 5.52 Å γ = 90° | Ho1 | -0.380 | 0.166 | -0.129 |
| | | | | H1 | 0.101 | 0.001 | 0.356 |
| | | | | H2 | -0.385 | 0.330 | -0.431 |
| | | | | H3 | -0.121 | 0.415 | 0.065 |
| | | | | H4 | 0.173 | 0.115 | -0.352 |
| | | | | H5 | 0.319 | 0.405 | 0.055 |
| | | | | H6 | 0.077 | 0.343 | -0.242 |
| | | | | H7 | -0.049 | 0.085 | 0.152 |
| *I4/mmm*-$HoH_4$ | 150 | 40.71 | a = 2.77 Å α = 90°<br>b = 2.77 Å β= 90°<br>c = 5.30 Å γ = 90° | Ho1 | 0.000 | 0.000 | 0.500 |
| | | | | H1 | 0.000 | 0.000 | -0.133 |
| | | | | H2 | 0.000 | 0.500 | 0.250 |
| *R-3m*-$ErH_6$ | 150 | 43.44 | a = 5.11 Å α = 90°<br>b = 5.11 Å β= 90° | Er1 | 0.000 | 0.000 | 0.000 |
| | | | | H1 | -0.278 | 0.000 | 0.500 |
| | | | c = 2.90 Å γ = 120° | | | | |
| *P-62m*-$ErH_{15}$ | 150 | 38.40 | a = 3.99 Å α = 90°<br>b = 3.99 Å β= 90°<br>c = 2.78 Å γ = 120° | Er1<br>H1<br>H2<br>H3 | 0.000<br>-0.117<br>-0.415<br>-0.340 | 0.000<br>0.441<br>-0.258<br>0.000 | 0.000<br>0.000<br>0.500<br>0.500 |
| *P4/nmm*-$TmH_5$ | 150 | 37.78 | a = 3.49 Å α = 90°<br>b = 3.49 Å β= 90°<br>c = 3.37 Å γ = 90° | Tm1<br>H1<br>H2 | 0.250<br>0.250<br>0.750 | 0.250<br>-0.483<br>0.250 | -0.281<br>0.192<br>0.500 |
| *Fm-3m*-$TmH_8$ | 150 | 25.46 | a = 4.66 Å α = 90°<br>b = 4.66 Å β= 90°<br>c = 4.66 Å γ = 90° | Tm1<br>H1 | 0.000<br>0.369 | 0.000<br>0.369 | 0.000<br>0.369 |
| *C222*-$LuH_{12}$ | 150 | 35.45 | a = 4.868 Å α = 90°<br>b = 2.823 Å β= 90°<br>c = 5.158 Å γ = 90° | Lu1<br>H1<br>H2<br>H3 | 0.000<br>0.433<br>-0.164<br>0.337 | 0.000<br>0.433<br>0.498<br>0.022 | 0.000<br>0.357<br>0.189<br>-0.427 |
| *Cm*-$LuH_{13}$ | 150 | 37.03 | a = 3.161 Å α = 90°<br>b = 5.511 Å β= 105.29°<br>c = 4.407 Å γ = 90° | Lu1<br>H1<br>H2<br>H3<br>H4<br>H5<br>H6<br>H7<br>H8 | 0.445<br>-0.226<br>-0.38<br>-0.439<br>0.152<br>-0.344<br>-0.289<br>0.429<br>-0.249 | 0.000<br>0.000<br>0.274<br>0.000<br>0.000<br>0.371<br>0.270<br>0.333<br>0.145 | 0.375<br>0.068<br>0.061<br>-0.068<br>-0.306<br>-0.052<br>-0.34<br>0.343<br>-0.256 |
| $P6_3/mmc$-$PaH_7$ | 50 | 70.83 | a = 3.798 Å α = 90°<br>b = 3.798 Å β= 90°<br>c = 5.665 Å γ = 120° | Pr1<br>H1<br>H2 | 0.333<br>0.176<br>0.000 | 0.666<br>0.353<br>0.000 | 0.750<br>0.074<br>0.250 |
| *Fm-3m*-$PaH_8$ | 150 | 54.99 | a = 3.503 Å α = 90°<br>b = 3.503 Å β= 90°<br>c = 5.175 Å γ = 120° | Pr1<br>H1<br>H2<br>H3<br>H4 | 0.333<br>0.176<br>0.333<br>0.000<br>-0.155 | 0.666<br>-0.177<br>0.666<br>0.000<br>0.155 | -0.061<br>0.266<br>-0.463<br>0.436<br>0.121 |
| *F-43m*-$PaH_9$ | 150 | 30.48 | a = 4.959 Å α = 90°<br>b = 4.959 Å β= 90°<br>c = 4.959 Å γ = 90° | Pr1<br>H1<br>H2<br>H3 | 0.750<br>0.387<br>0.130<br>0.000 | 0.750<br>0.387<br>0.130<br>0.000 | 0.750<br>0.387<br>0.131<br>0.000 |
| $P6_3/mmc$-$NpH_7$ | 50 | 64.49 | a = 3.719 Å α = 90°<br>b = 3.719 Å β= 90°<br>c = 5.384 Å γ = 120° | Np1<br>H1<br>H2 | 0.333<br>0.182<br>0.000 | 0.666<br>0.364<br>0.000 | 0.750<br>0.080<br>0.250 |
| *Fm-3m*-$NpH_8$ | 150 | 27.08 | a = 4.767 Å α = 90°<br>b = 4.767 Å β= 90°<br>c = 4.767 Å γ = 90° | Np1<br>H1 | 0.000<br>0.364 | 0.000<br>0.364 | 0.000<br>0.364 |
| *C2/m*-$AmH_5$ | 50 | 52.91 | a = 6.23 Å α = 90°<br>b = 3.503 Å β= 122.93°<br>c = 5.78 Å γ = 90° | Am1<br>H1<br>H2<br>H3 | -0.314<br>0.454<br>0.352<br>0.063 | 0.000<br>0.271<br>0.000<br>0.000 | -0.288<br>0.078<br>-0.302<br>0.423 |
| $Fm\bar{3}m$-$AmH_8$ | 150 | 26.01 | a = 4.704 Å α = 90°<br>b = 4.704 Å β= 90°<br>c = 4.704 Å γ = 90° | Am1<br>H1 | 0.000<br>-0.361 | 0.000<br>-0.364 | 0.000<br>-0.361 |
| *C2/m*-$CmH_5$ | 50 | 52.38 | a = 9.71 Å α = 90°<br>b = 3.46 Å β= 32.68°<br>c = 5.75 Å γ = 90° | Cm1<br>H1<br>H2<br>H3<br>H4 | 0.309<br>-0.047<br>-0.354<br>-0.063<br>0.277 | 0.000<br>0.229<br>0.000<br>0.000<br>0.000 | -0.331<br>0.171<br>0.014<br>-0.298<br>0.359 |
| $P6_3/mmc$-$CmH_8$ | 150 | 51.53 | a = 3.32 Å α = 90°<br>b = 3.32 Å β= 90°<br>c = 5.37 Å γ = 120° | Cm1<br>H1<br>H2 | 0.333<br>-0.170<br>0.333 | 0.667<br>-0.341<br>0.667 | 0.750<br>-0.063<br>0.250 |
| | | | | H3 | 0.000 | 0.000 | 0.250 |
| *C2*-$CmH_{20}$ | 150 | 47.17 | a = 5.78 Å α = 90°<br>b = 3.37 Å β= 90.19°<br>c = 4.82 Å γ = 90° | Cm1 | 0.000 | -0.003 | 0.500 |
| | | | | H1 | -0.248 | 0.262 | -0.303 |
| | | | | H2 | -0.169 | 0.494 | 0.435 |
| | | | | H3 | -0.167 | 0.392 | 0.012 |
| | | | | H4 | 0.337 | 0.483 | 0.183 |
| | | | | H5 | 0.354 | 0.114 | 0.022 |
| | | | | H6 | 0.011 | 0.234 | -0.164 |
| | | | | H7 | 0.130 | 0.030 | 0.158 |
| | | | | H8 | -0.014 | 0.495 | -0.309 |
| | | | | H9 | -0.462 | 0.249 | 0.174 |
| | | | | H10 | -0.252 | 0.235 | 0.319 |
| *P4/nmm*-$FrH_7$ | 50 | 89.64 | a = 4.77 Å α = 90°<br>b = 4.77 Å β= 90°<br>c = 3.94 Å γ = 90° | Fr1 | 0.750 | 0.250 | 0.500 |
| | | | | H1 | 0.250 | 0.250 | 0.4112 |
| | | | | H2 | -0.445 | -0.445 | 0.0292 |
| | | | | H3 | 0.250 | 0.250 | 0.2119 |
| | | | | H4 | 0.250 | 0.250 | -0.2408 |
| *C2*-$RaH_{12}$ | 50 | 58.81 | a = 7.51 Å α = 90°<br>b = 3.61 Å β= 139.69°<br>c = 6.73 Å γ = 90° | Ra1 | 0.000 | 0.377 | 0.000 |
| | | | | H1 | -0.119 | 0.391 | 0.337 |
| | | | | H2 | 0.196 | 0.380 | -0.168 |
| | | | | H3 | 0.497 | 0.366 | 0.439 |
| | | | | H4 | -0.401 | 0.381 | -0.152 |
| | | | | H5 | 0.159 | 0.169 | 0.457 |
| | | | | H6 | -0.304 | 0.249 | 0.427 |
| *Cmcm*-$RaH_6$ | 50 | 82.62 | a = 3.35Å α = 90°<br>b = 8.01 Å β= 90°<br>c = 6.14 Å γ = 90° | Ra1 | 0.000 | -0.081 | 0.250 |
| | | | | H1 | 0.000 | 0.220 | 0.092 |
| | | | | H2 | 0.000 | 0.373 | -0.316 |
| | | | | H3 | 0.000 | 0.500 | 0.000 |
| | | | | H4 | 0.000 | 0.213 | 0.250 |
| *Cmmm*-$RaH_{10}$ | 150 | 36.51 | a = 3.08 Å α = 90°<br>b = 7.96 Å β= 90°<br>c = 2.97 Å γ = 90° | Ra1 | 0.500 | 0.000 | 0.500 |
| | | | | H1 | 0.000 | -0.089 | 0.000 |
| | | | | H2 | 0.000 | -0.208 | 0.374 |
| | | | | H3 | -0.214 | 0.307 | 0.000 |
| *C2/m*-$RaH_{12}$ | 150 | 40.82 | a = 6.07 Å α = 90°<br>b = 4.12 Å β= 116.24°<br>c = 3.63 Å γ = 90° | Ra1 | 0.000 | 0.500 | 0.500 |
| | | | | H1 | 0.376 | 0.304 | -0.062 |
| | | | | H2 | 0.003 | 0.000 | -0.244 |
| | | | | H3 | -0.213 | 0.352 | -0.169 |
| | | | | H4 | 0.128 | 0.000 | 0.474 |

## Equations for calculating $T_C$ and related parameters

To calculate $T_C$, the "full" Allen-Dynes formula was used for calculating $T_C$ in the following form [1]:

$$T_C = \omega_{log}\frac{f_1 f_2}{1.2}\exp\left(\frac{-1.04(1+\lambda)}{\lambda-\mu^*-0.62\lambda\mu^*}\right) \tag{S1}$$

with

$$f_1 f_2 = \sqrt[3]{1+\left(\frac{\lambda}{2.46(1+3.8\mu^*)}\right)^{\frac{3}{2}}}\cdot\left(1-\frac{\lambda^2(1-\omega_2/\omega_{log})}{\lambda^2+3.312(1+6.3\mu^*)^2}\right), \tag{S2}$$

while $f_1 f_2 = 1$ in the modified McMillan formula having a similar form. The EPC constant λ, logarithmic average frequency $\omega_{log}$ and mean square frequency $\omega_2$ were calculated as:

$$\lambda = \int_{\omega_{min}}^{\omega_{max}} \frac{2\cdot a^2F(\omega)}{\omega}d\omega \tag{S3}$$

and

$$\omega_{log} = \exp\left(\frac{2}{\lambda}\int_{\omega_{min}}^{\omega_{max}}\frac{d\omega}{\omega}\alpha^2F(\omega)\ln(\omega)\right), \omega_2 = \sqrt{\frac{1}{\lambda}\int_0^{\omega_{max}}\left[\frac{2a^2F(\omega)}{\omega}\right]\omega^2 d\omega} \tag{S4}$$

and $\mu^*$ is the Coulomb pseudopotential, for which we used the widely accepted lower bound value of 0.10.

To calculate the isotopic coefficient $\beta$ the Allen-Dynes interpolation formulas were used:

$$\beta_{McM} = -\frac{d\ln T_C}{d\ln M} = \frac{1}{2}\left[1-\frac{1.04(1+\lambda)(1+0.62\lambda)}{[\lambda-\mu^*(1+0.62\lambda)]^2}\mu^{*2}\right] \tag{S5}$$

$$\beta_{AD} = \beta_{McM} - \frac{2.34\mu^{*2}\lambda^{3/2}}{(2.46+9.25\mu^*)\cdot((2.46+9.25\mu^*)^{3/2}+\lambda^{3/2})} - \frac{130.4\cdot\mu^{*2}\lambda^2(1+6.3\mu^*)\left(1-\frac{\omega_{\log}}{\omega_2}\right)\frac{\omega_{\log}}{\omega_2}}{\left(8.28+104\mu^*+329\mu^{*2}+2.5\cdot\lambda^2\frac{\omega_{\log}}{\omega_2}\right)\cdot\left(8.28+104\mu^*+329\mu^{*2}+2.5\cdot\lambda^2\left(\frac{\omega_{\log}}{\omega_2}\right)^2\right)} \tag{S6}$$

where the last two correction terms are usually small (~0.01). The Sommerfeld constant was found as

$$\gamma = \frac{2}{3}\pi^2 k_B^2 N(0)(1+\lambda), \tag{S7}$$

and was used to estimate the upper critical magnetic field. The superconductive gap was estimated using the well-known semi-empirical equation, working satisfactorily for $T_C/\omega_{\log} < 0.25$ (Equation 4.1 in Ref. [2]):

$$\frac{2\Delta(0)}{k_B T_C} = 3.53\left[1+12.5\left(\frac{T_C}{\omega_{\log}}\right)^2\ln\left(\frac{\omega_{\log}}{2T_C}\right)\right]. \tag{S8}$$

The semi-empirical equations of the BCS theory (Equations 5.11 and 5.9 in Ref. [2]) were used to estimate the upper critical magnetic field and the jump in specific heat:

$$\frac{\gamma T_C^2}{H_C^2(0)} = 0.168\left[1-12.2\left(\frac{T_C}{\omega_{\log}}\right)^2\ln\left(\frac{\omega_{\log}}{3T_C}\right)\right] \tag{S9}$$

and

$$\frac{\Delta C(T_C)}{\gamma T_C} = 1.43\left[1+53\left(\frac{T_C}{\omega_{\log}}\right)^2\ln\left(\frac{\omega_{\log}}{3T_C}\right)\right], \tag{S10}$$

we estimated the upper critical magnetic field and the specific heat jump.

## Thermodynamic and structural data for predicted novel metal polyhydrides

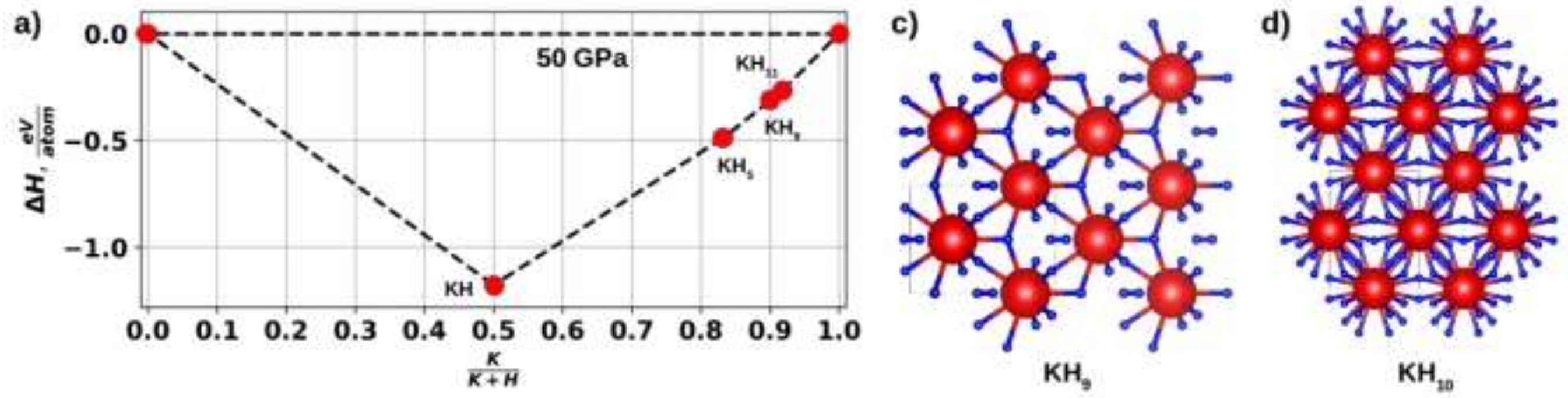

**Figure S1.** The convex hulls and structures of the higher potassium polyhydrides.

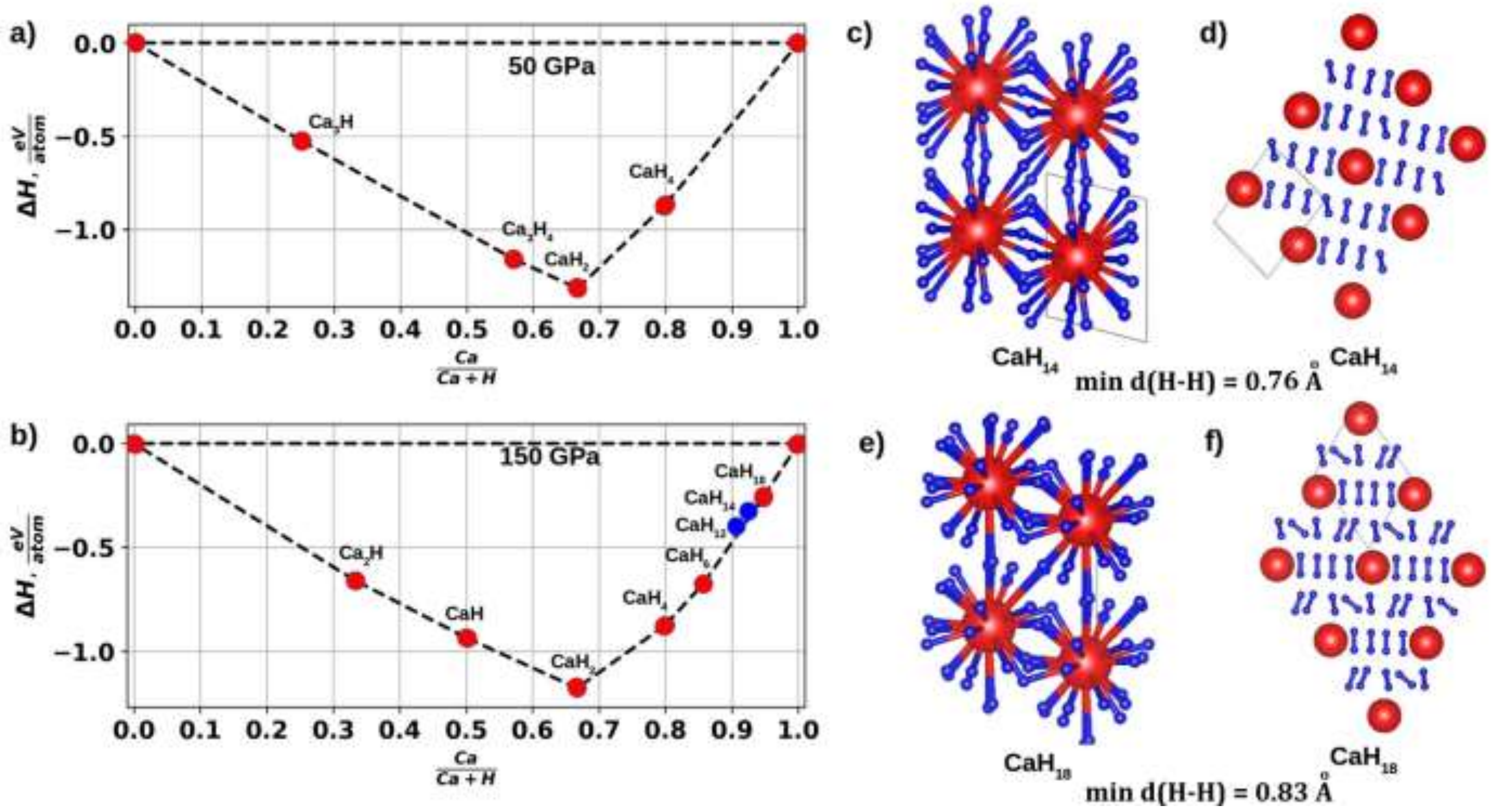

**Figure S2.** The convex hulls and structures of the higher calcium polyhydrides.

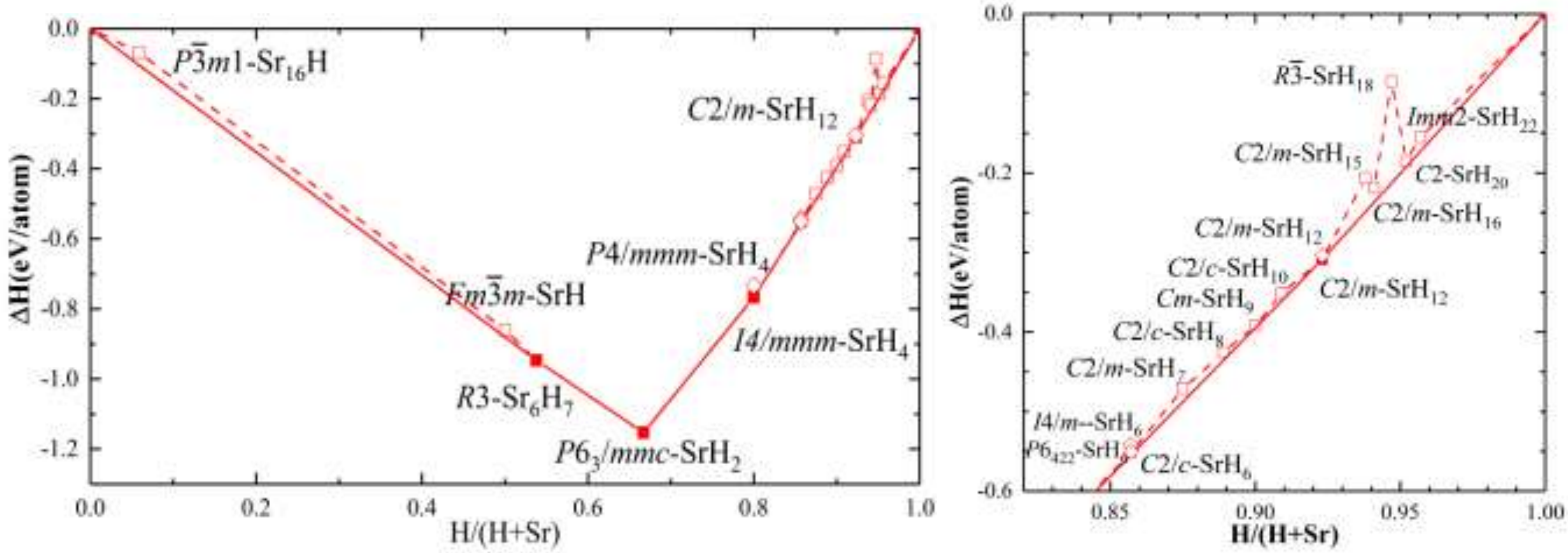

**Figure S3.** The convex hulls of the higher strontium polyhydrides at 50 GPa.

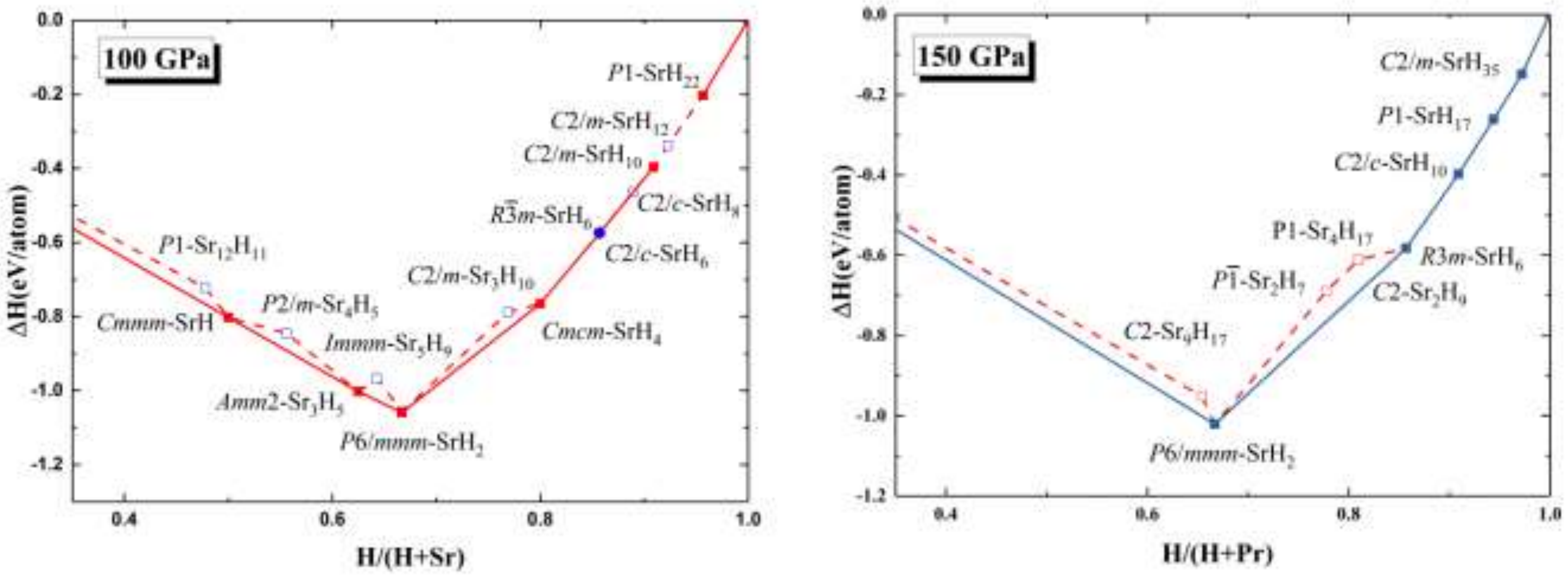

**Figure S4.** The convex hulls of the higher strontium polyhydrides at 100 and 150 GPa.

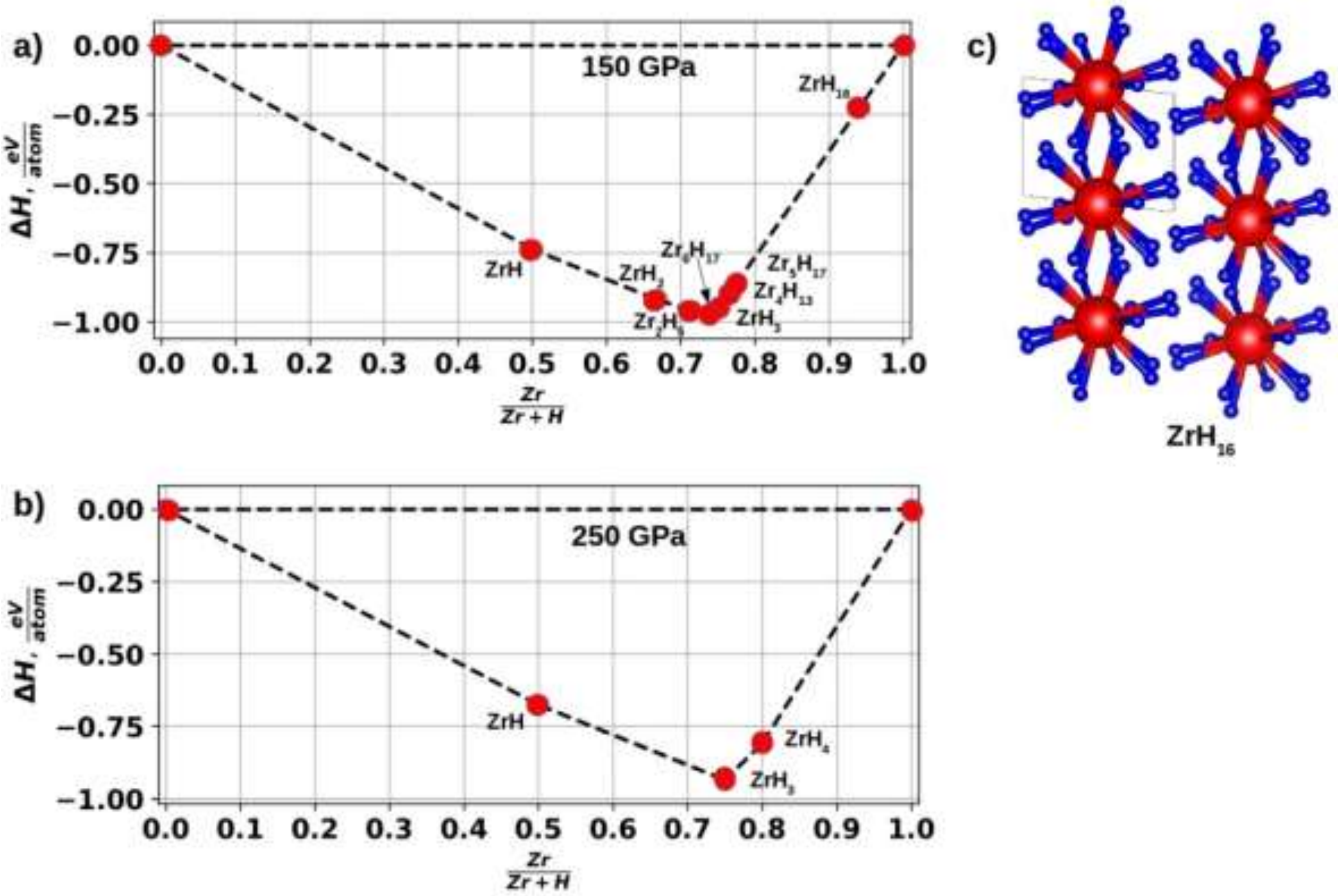

**Figure S5.** The convex hulls and structures of the higher zirconium polyhydrides.

**Table S2.** The predicted stable phases in the Zr-H and Hf-H systems.

| Pressure, GPa | Stable phases | Pressure, GPa | Stable phases |
|---|---|---|---|
| 150 | *Immm*-ZrH<br>*Pm*-$Zr_6H_{17}$<br>*P4₂/nmc*-$Zr_2H_5$<br>*Cmma*-$ZrH_2$<br>*P1*-$ZrH_3$<br>*P*$\bar{1}$-$ZrH_{16}$ | 100 | *Pn*$\bar{3}$*m*-$HfH_3$<br>*Amm2*-$Hf_3H_{13}$<br>*I4/mmm*-$HfH_2$<br>*Fmmm*-HfH |
| | | 300 | *C2/m*-$HfH_{14}$<br>*Cmc2₁*-$HfH_6$<br>*Fddd*-$HfH_4$<br>*R*$\bar{3}$*m*-HfH<br>*Pn*$\bar{3}$*m*-$HfH_3$ |

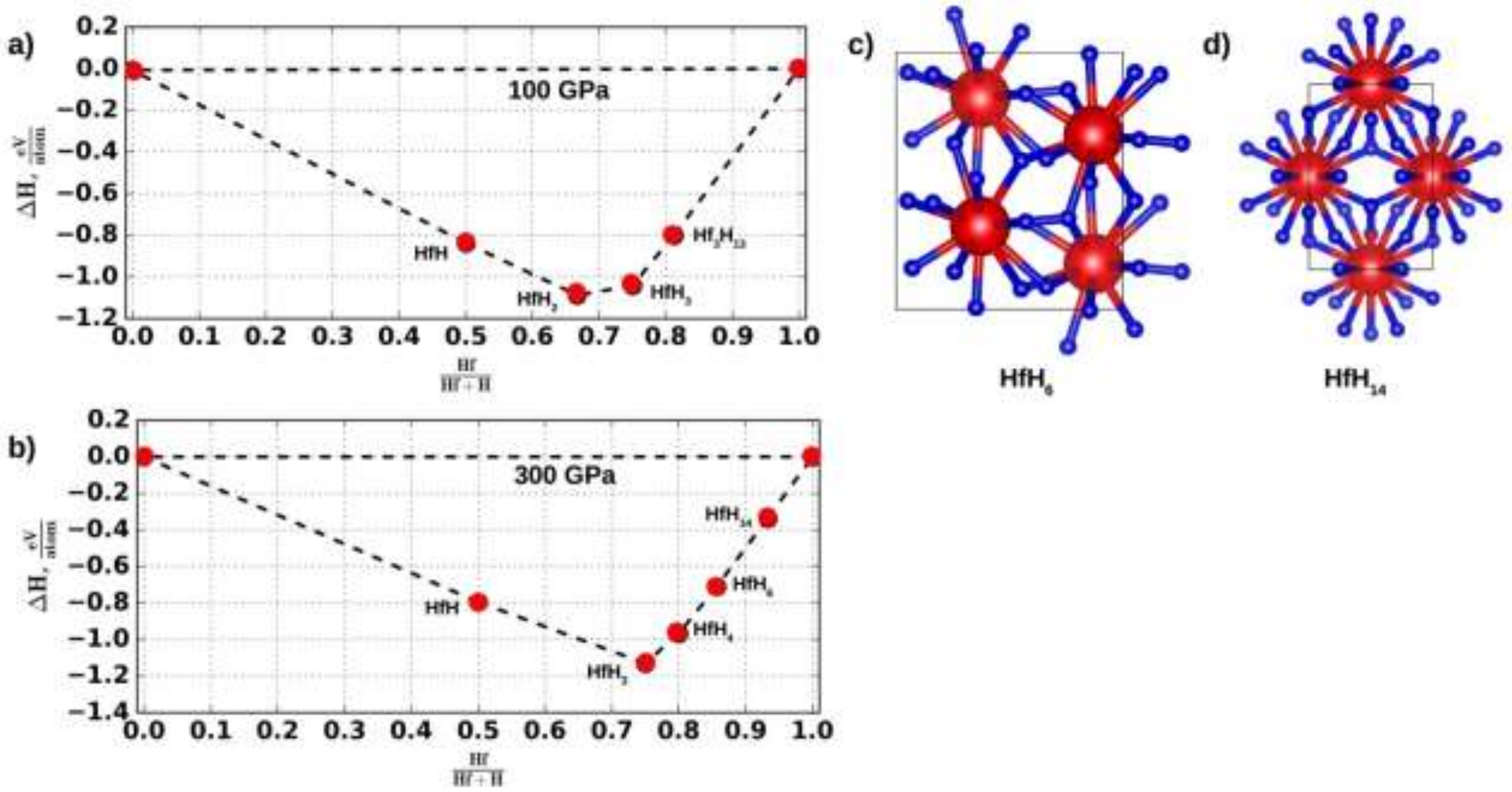

**Figure S6.** The convex hulls and structures of the higher hafnium polyhydrides.

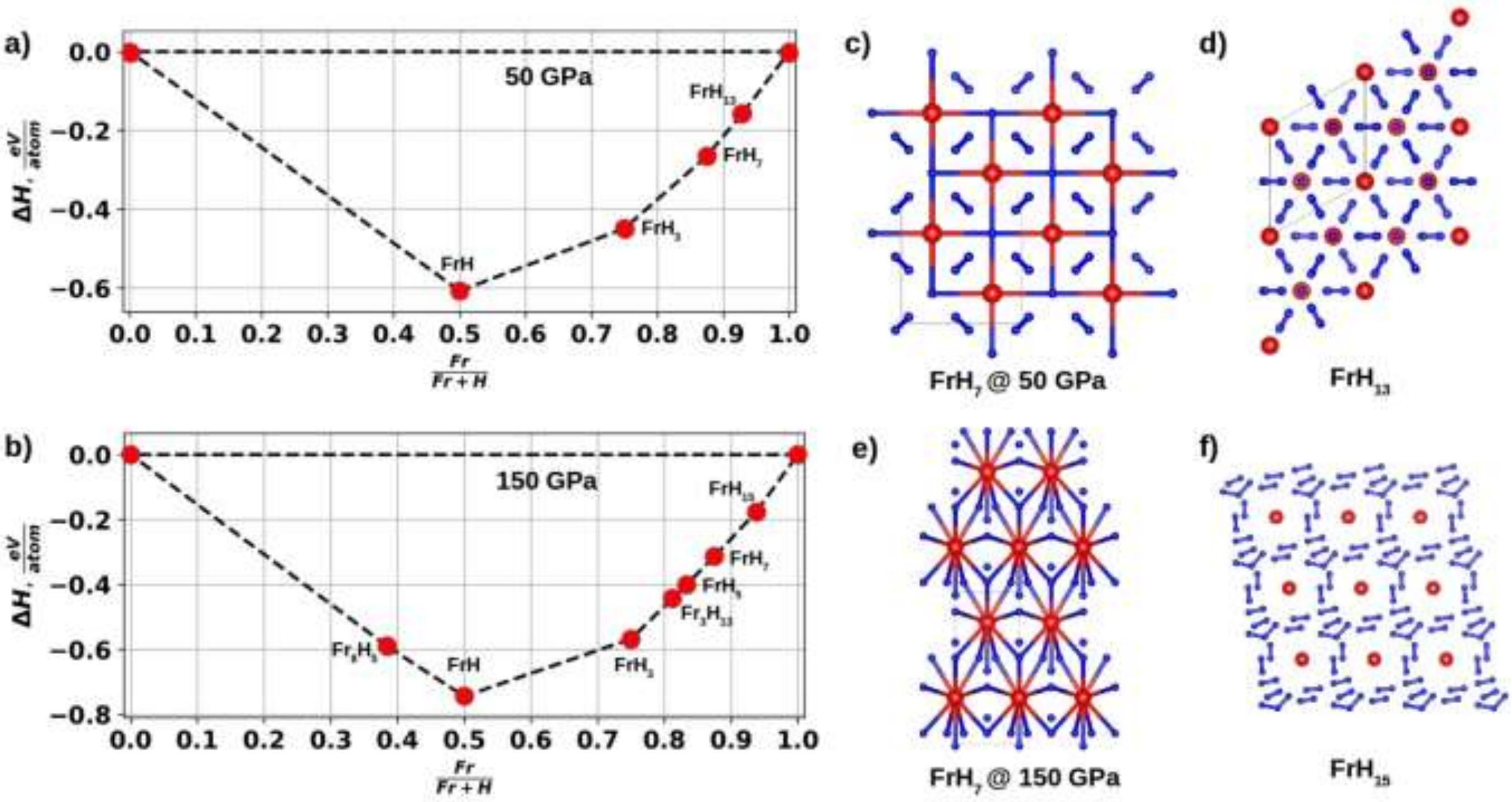

**Figure S7.** The convex hulls and structures of the higher francium polyhydrides.

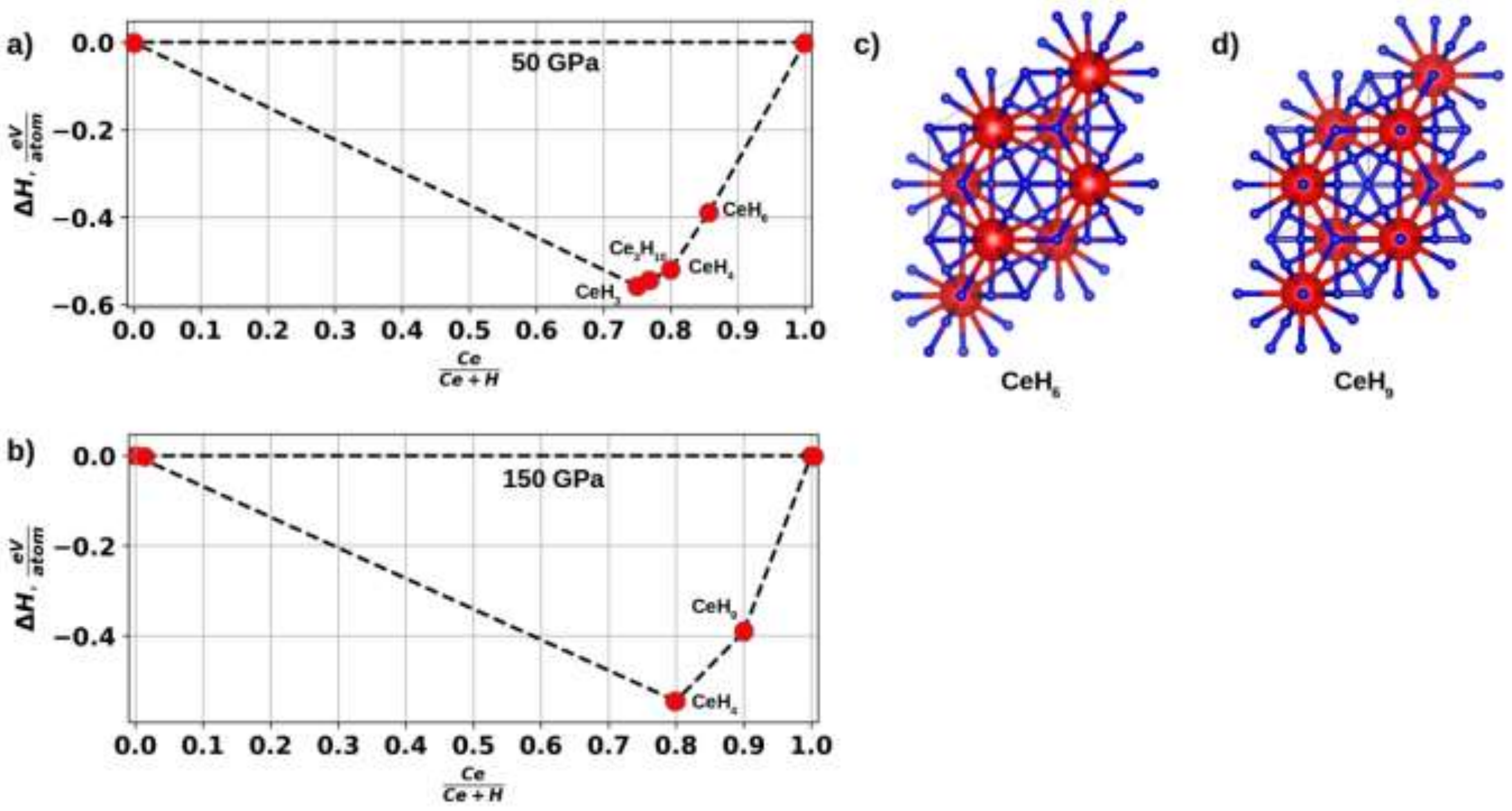

**Figure S8.** The convex hulls and structures of the higher cerium polyhydrides.

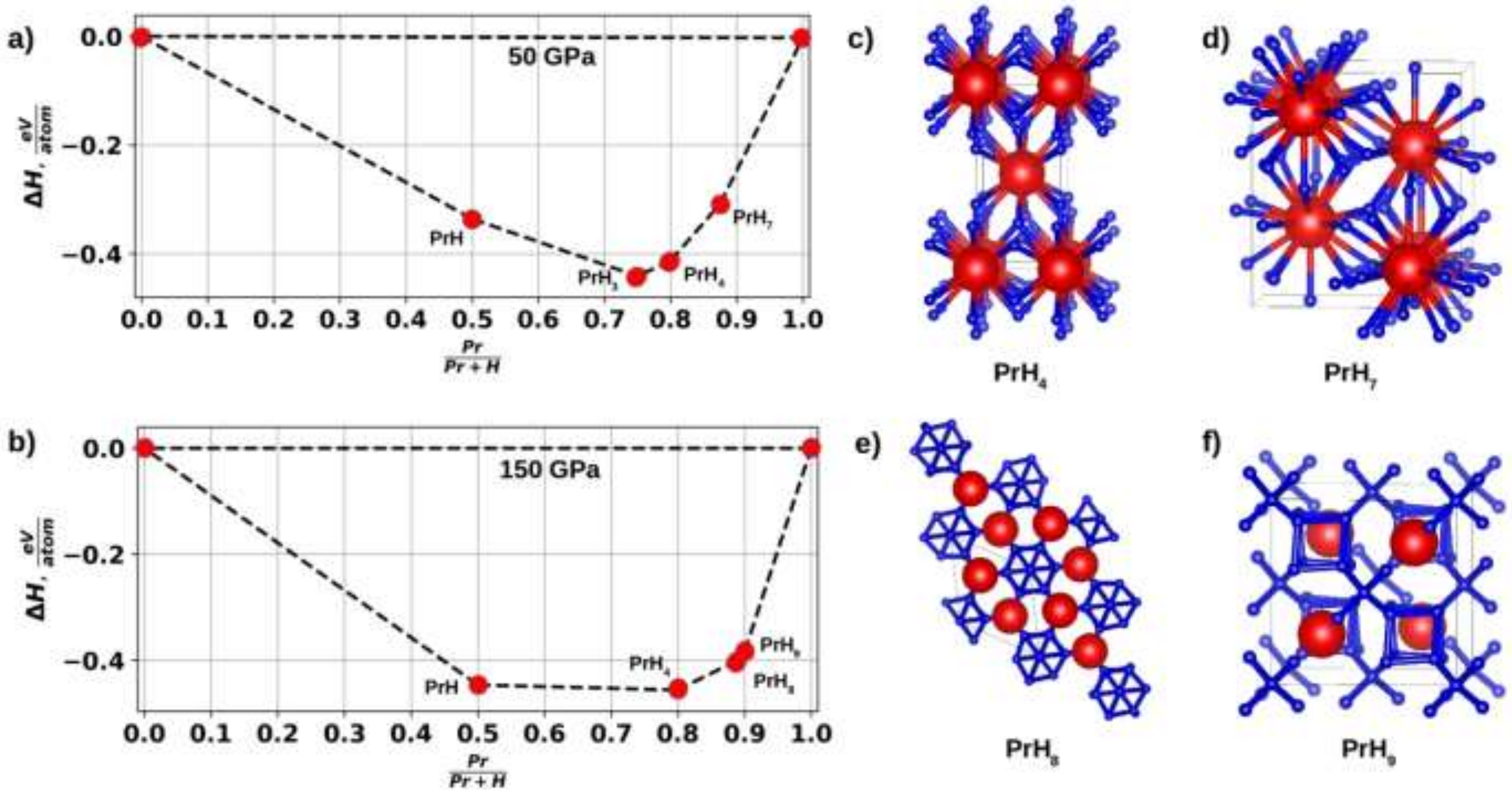

**Figure S9.** The convex hulls and structures of the higher praseodymium polyhydrides.

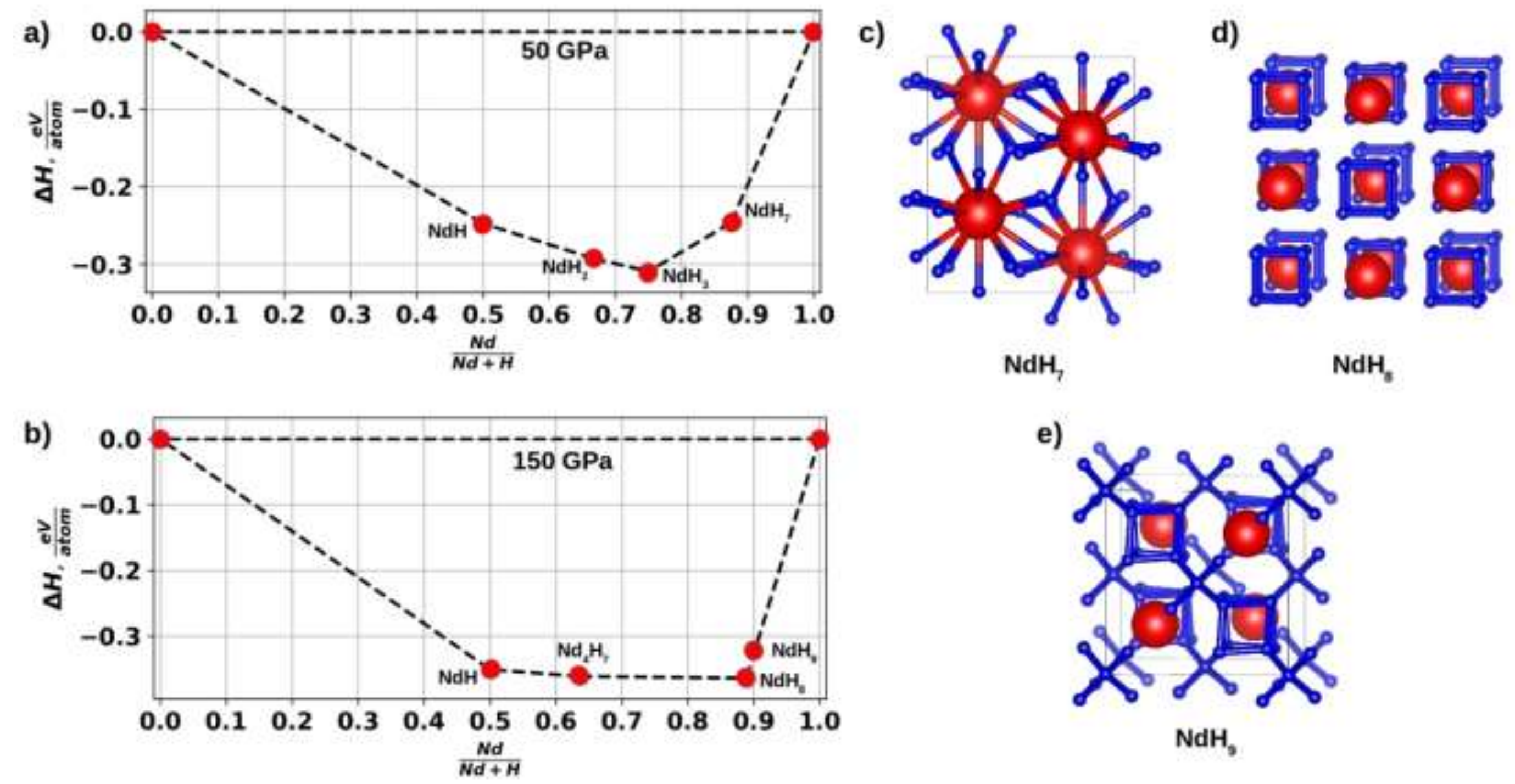

**Figure S10.** The convex hulls and structures of the higher neodymium polyhydrides.

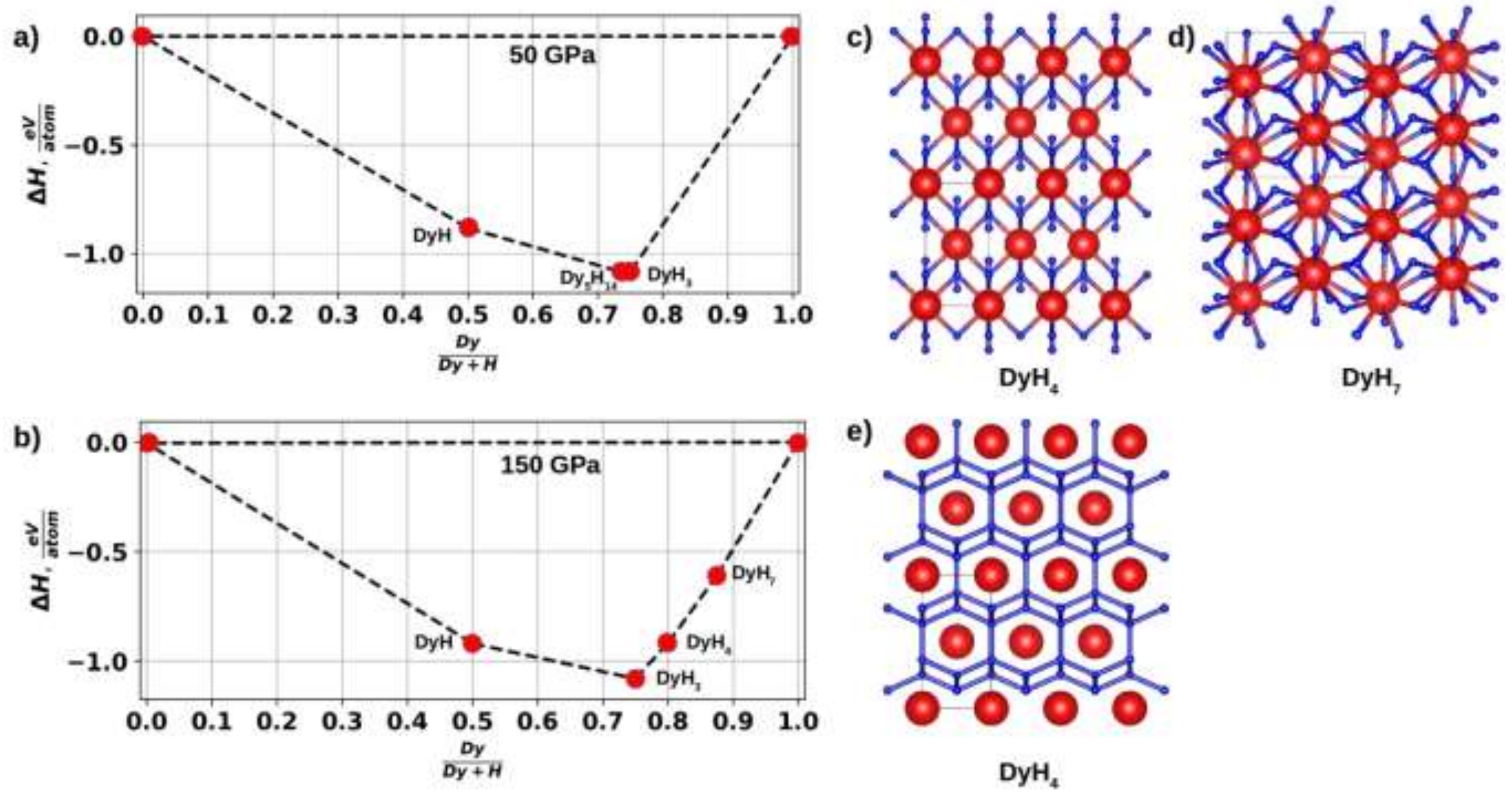

**Figure S11.** The convex hulls and structures of the higher dysprosium polyhydrides. The hydrogen sublattice in $DyH_4$ with d(H-H) = 1.5 Å is shown in panel e).

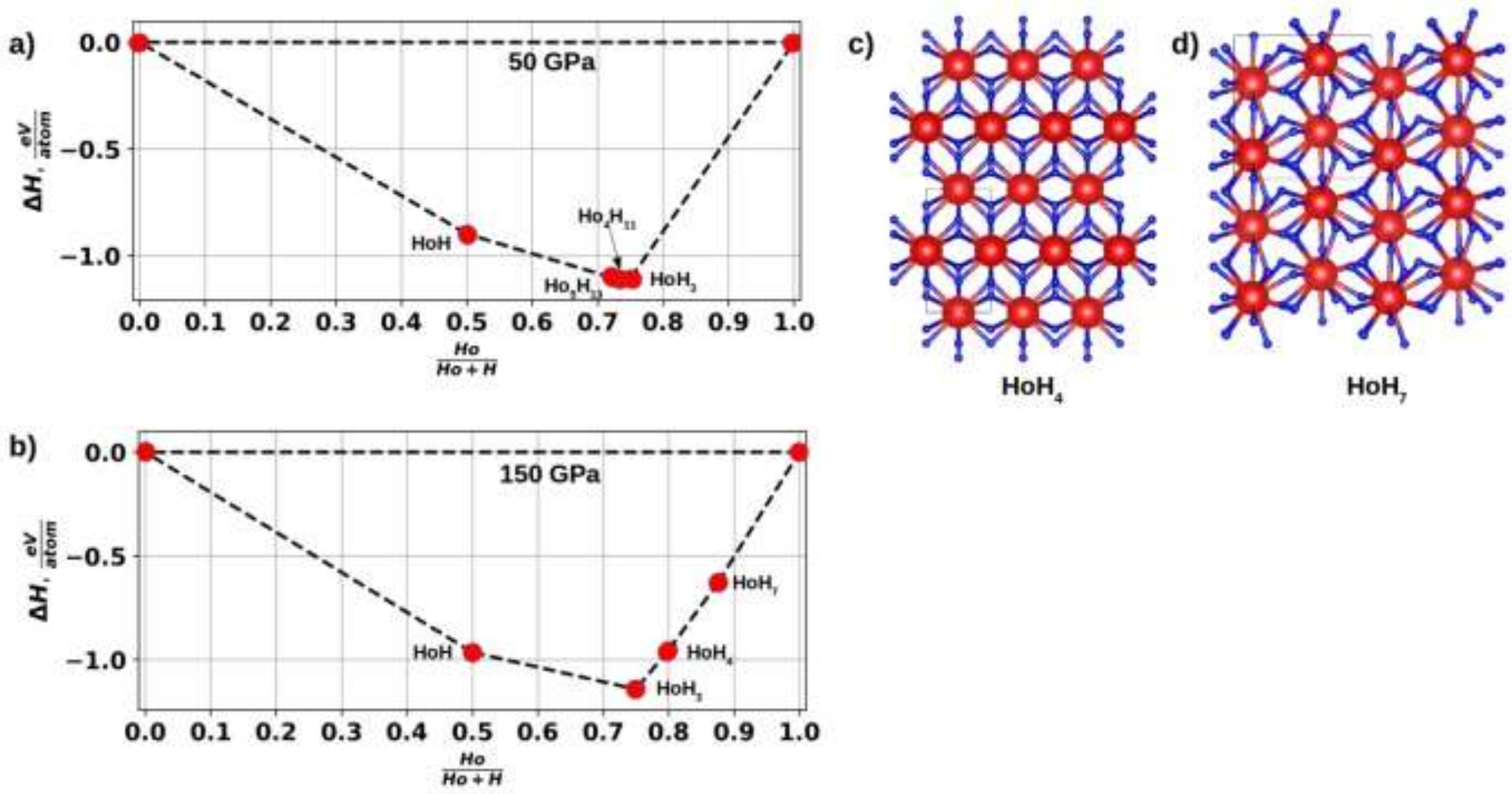

**Figure S12.** The convex hulls and structures of the higher holmium polyhydrides.

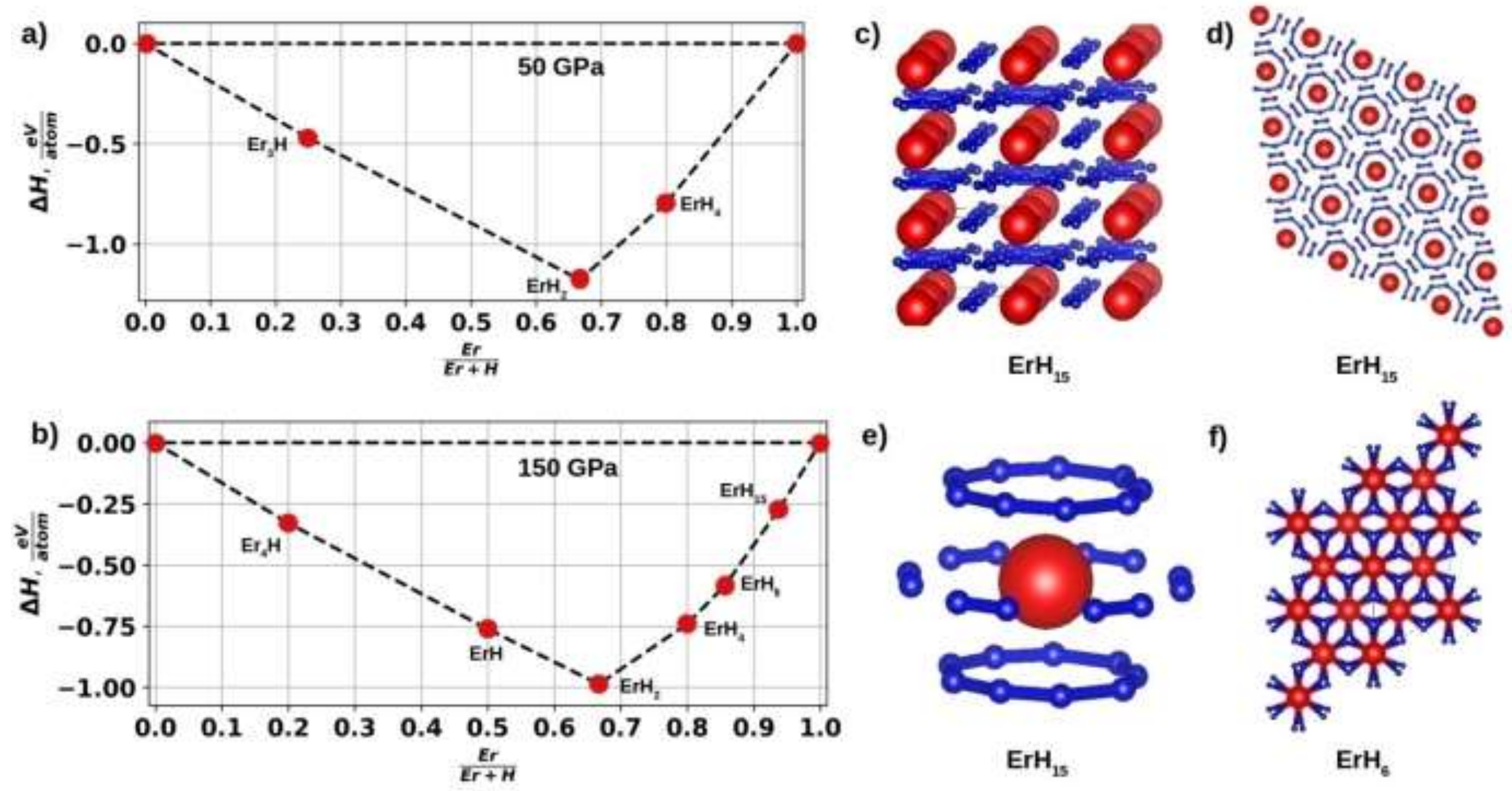

**Figure S13.** The convex hulls and structures of the higher erbium polyhydrides; c) d) The structures of $ErH_{15}$ have a layered character. e) The coordination sphere of the Er atoms includes six $H_2$ molecules and two $H_9$ nanogons. f) The structure of $ErH_6$.

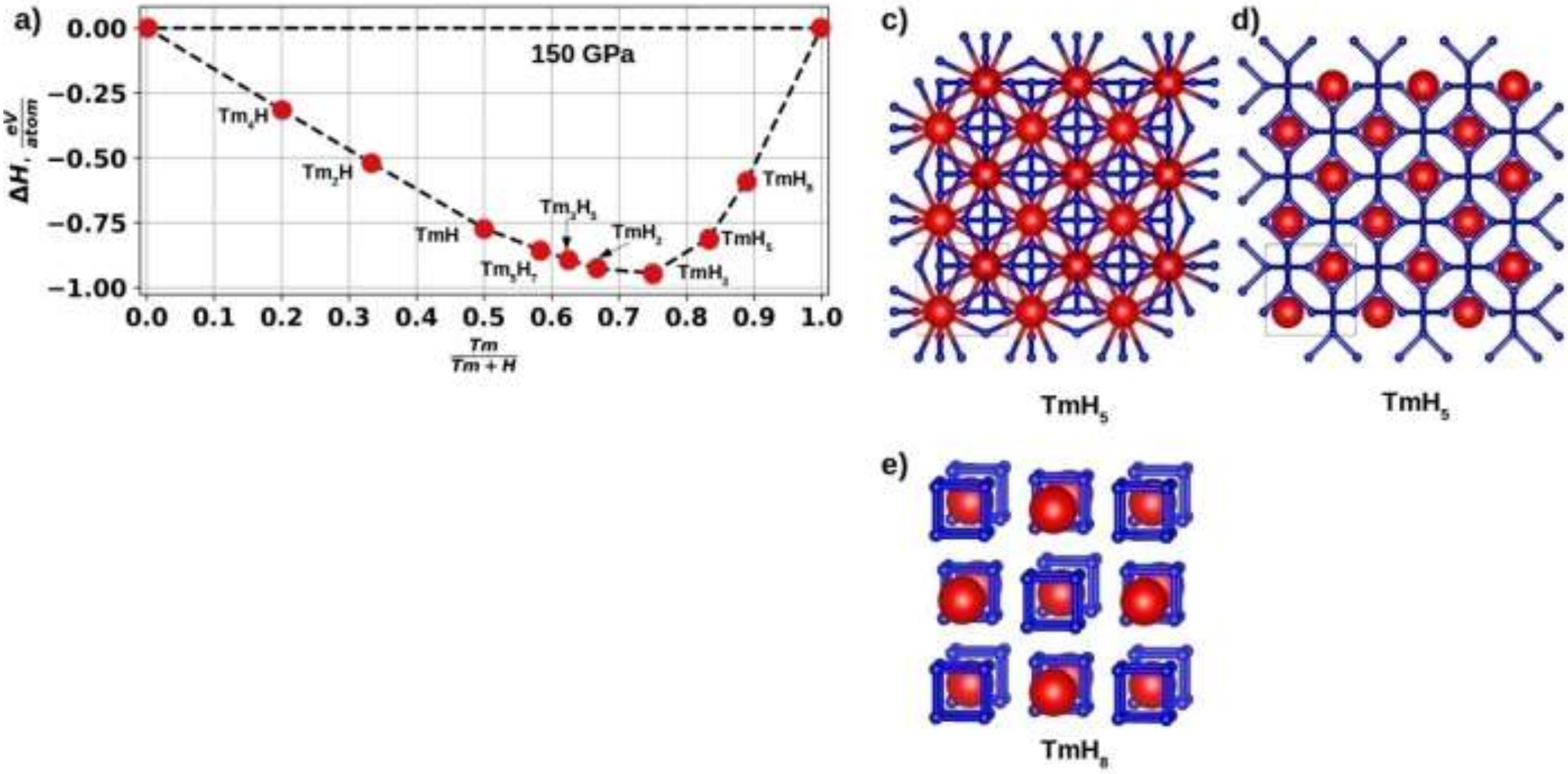

**Figure S14.** The convex hulls and structures of the higher thulium polyhydrides.

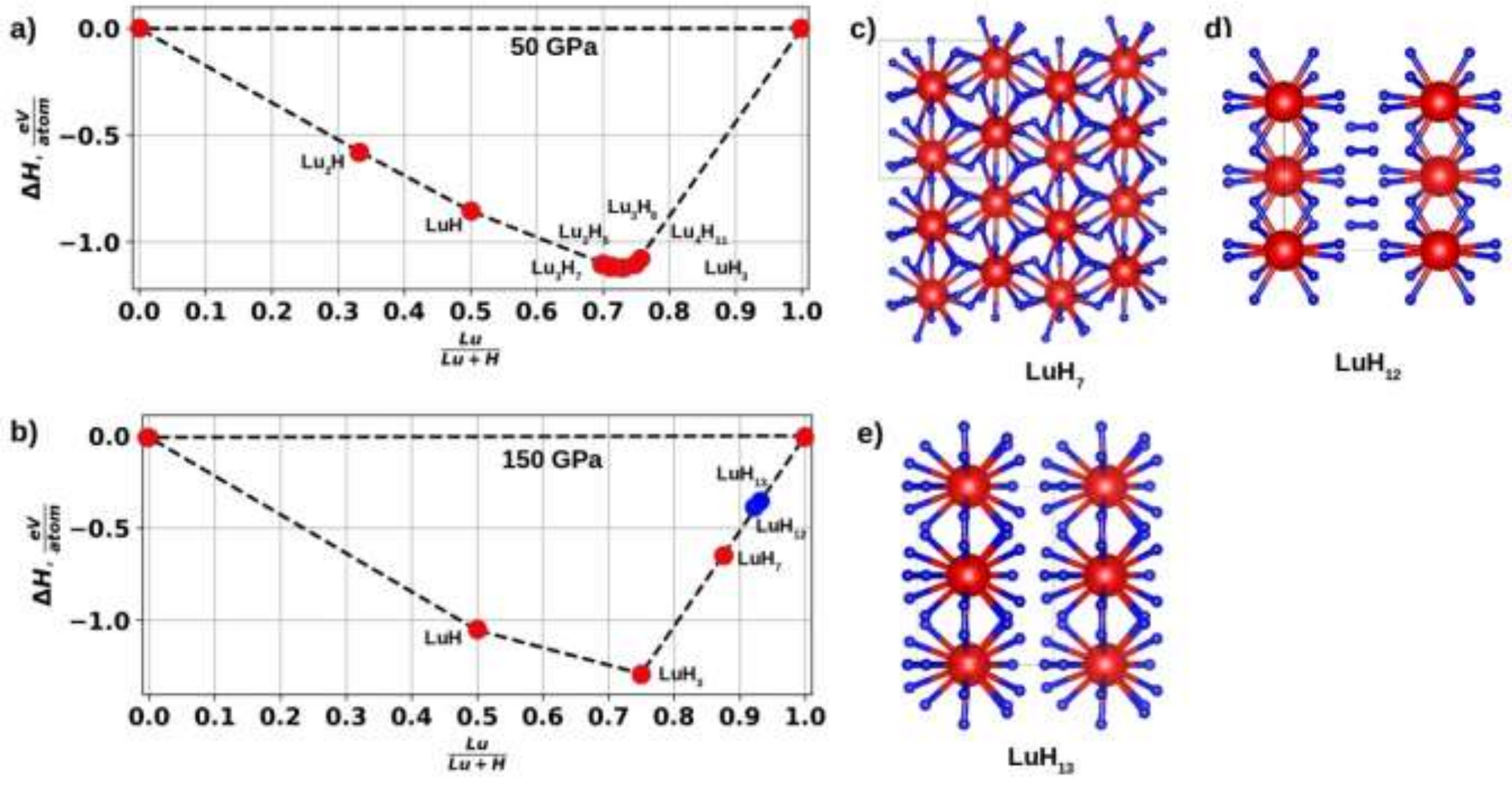

**Figure S15.** The convex hulls and structures of the higher lutetium polyhydrides, including two metastable structures, $LuH_{12}$, and $LuH_{13}$.

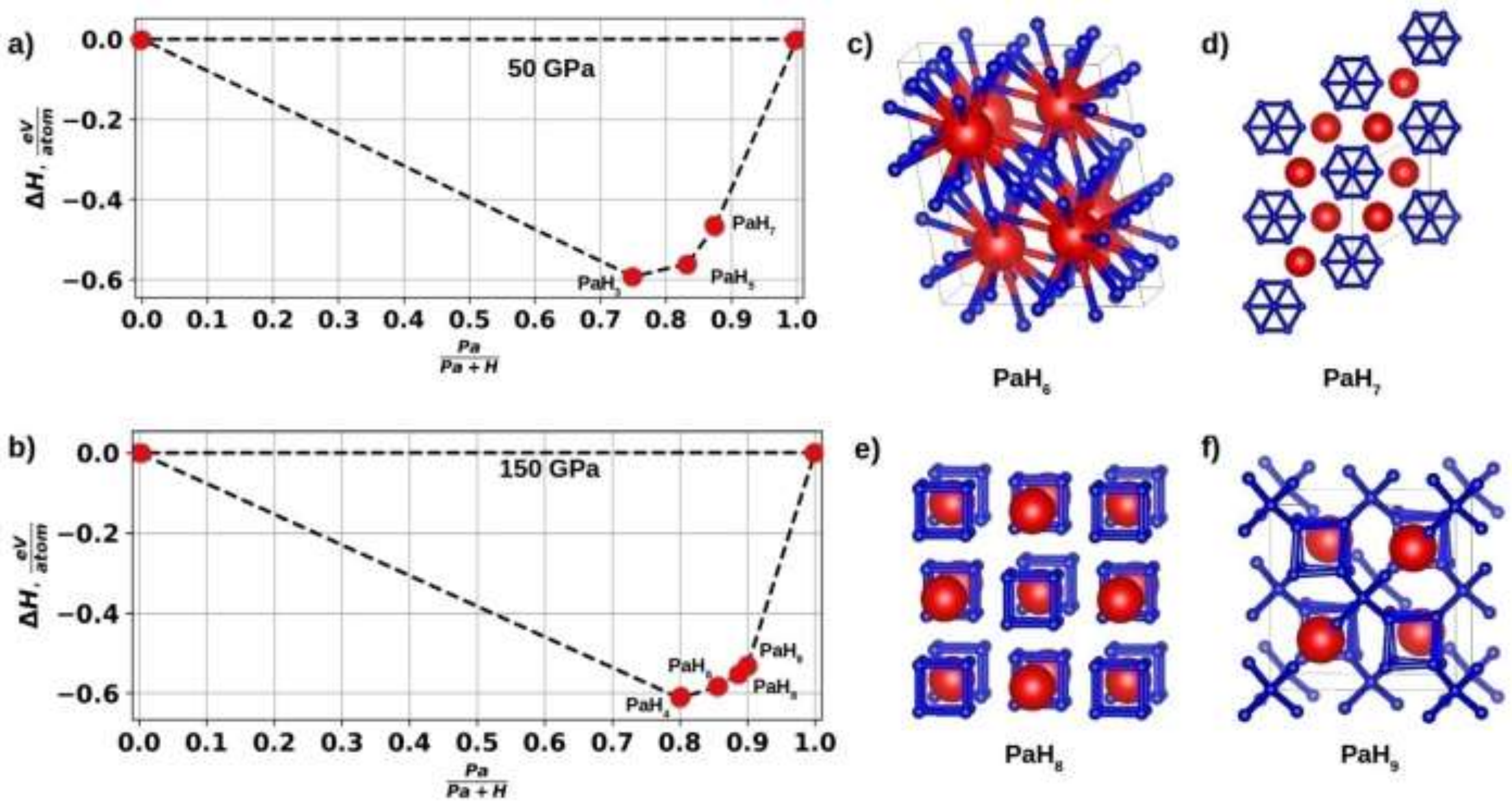

**Figure S16.** The convex hulls and structures of the higher protactinium polyhydrides.

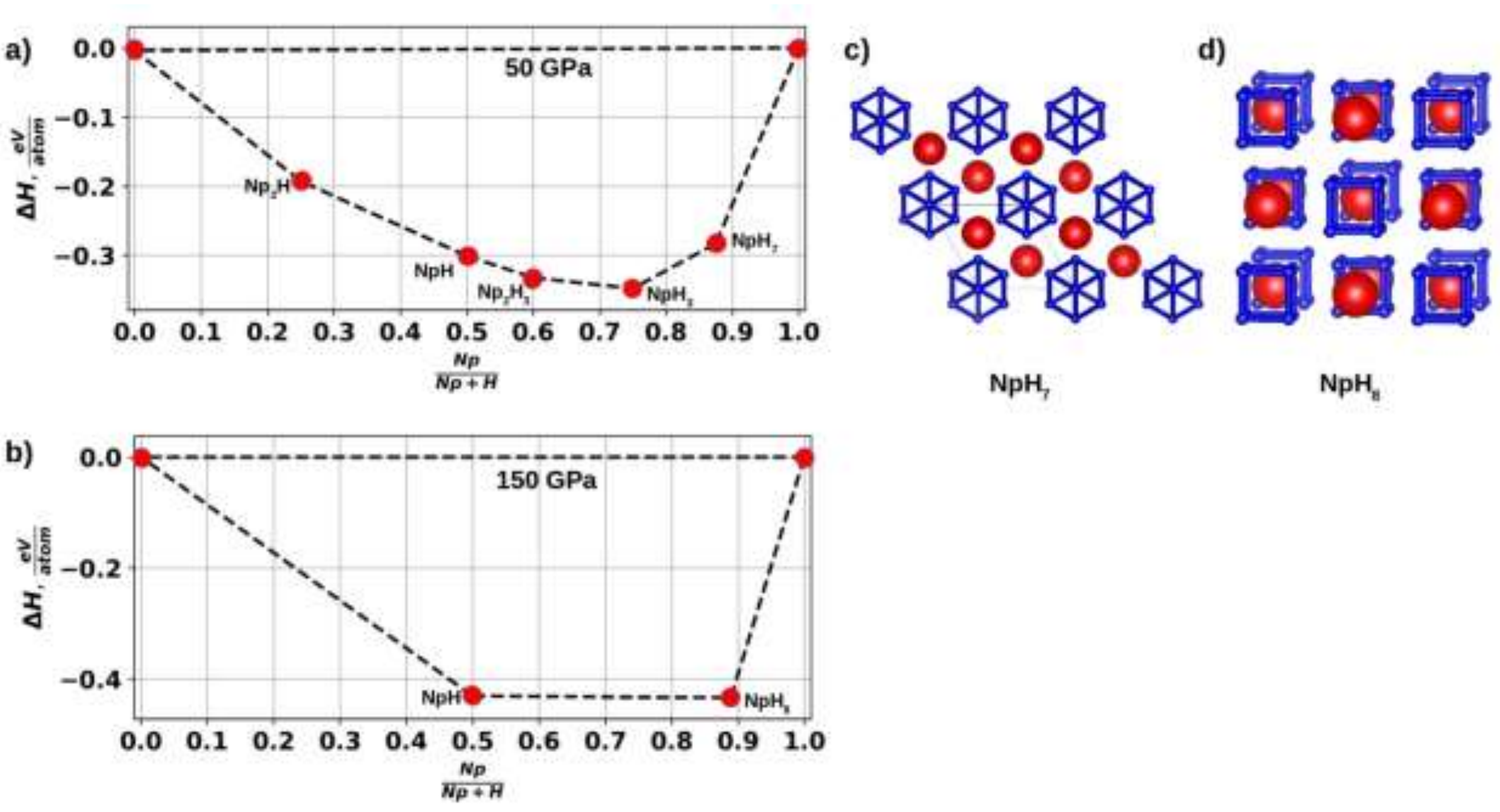

**Figure S17.** The convex hulls and structures of the higher neptunium polyhydrides.

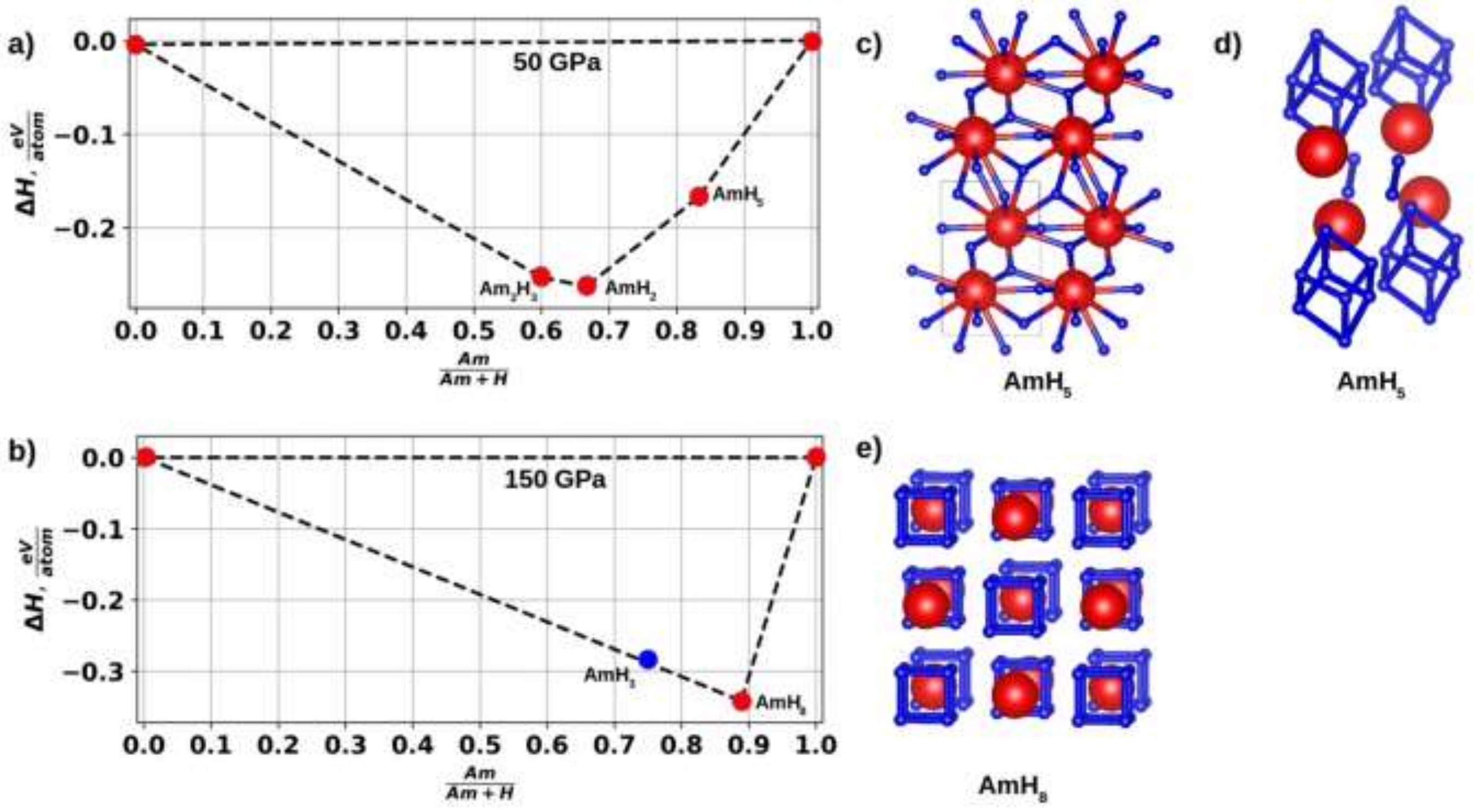

**Figure S18.** The convex hulls and structures of the higher americium polyhydrides.

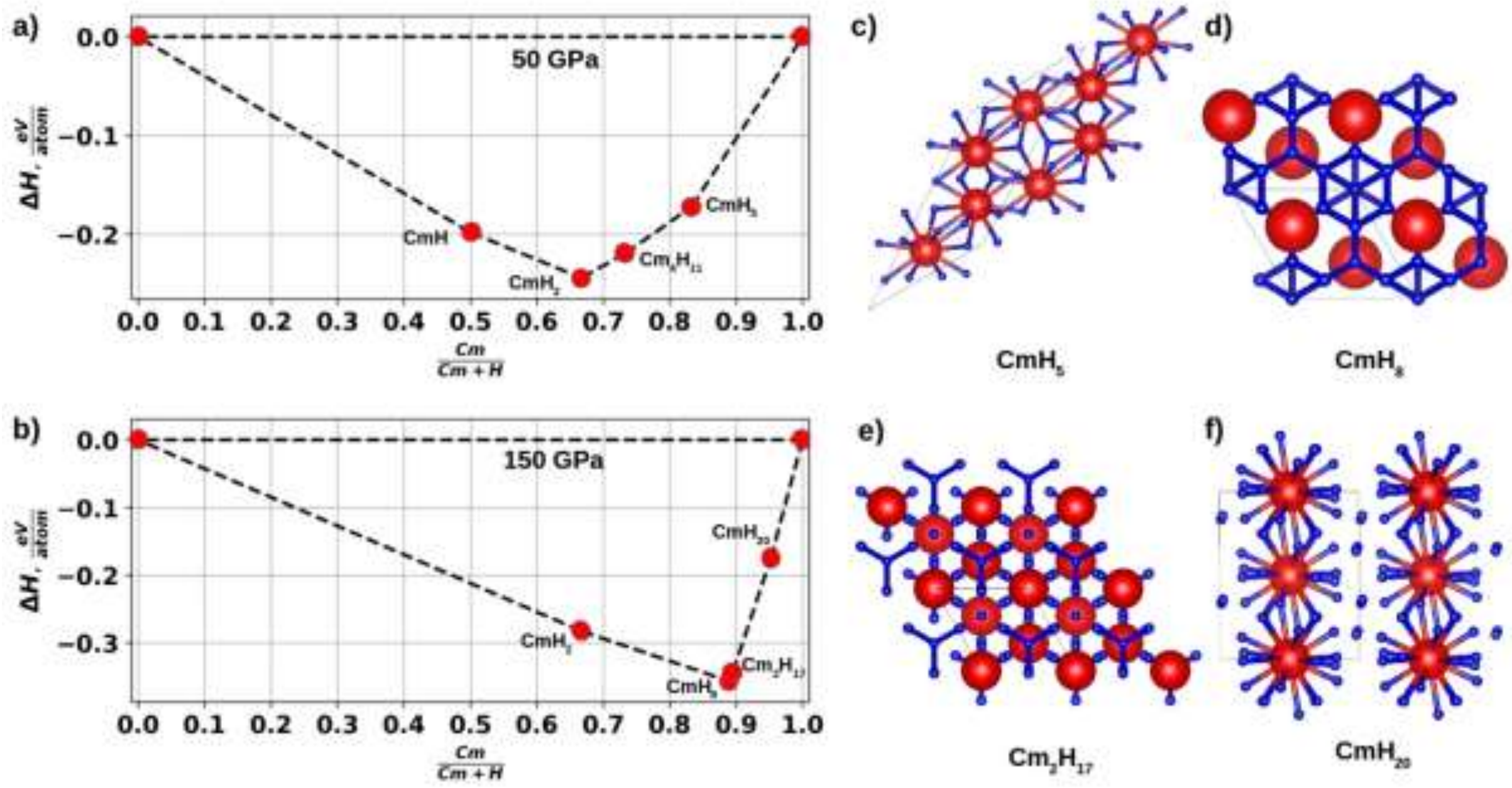

**Figure S19.** The convex hulls and structures of the higher curium polyhydrides.

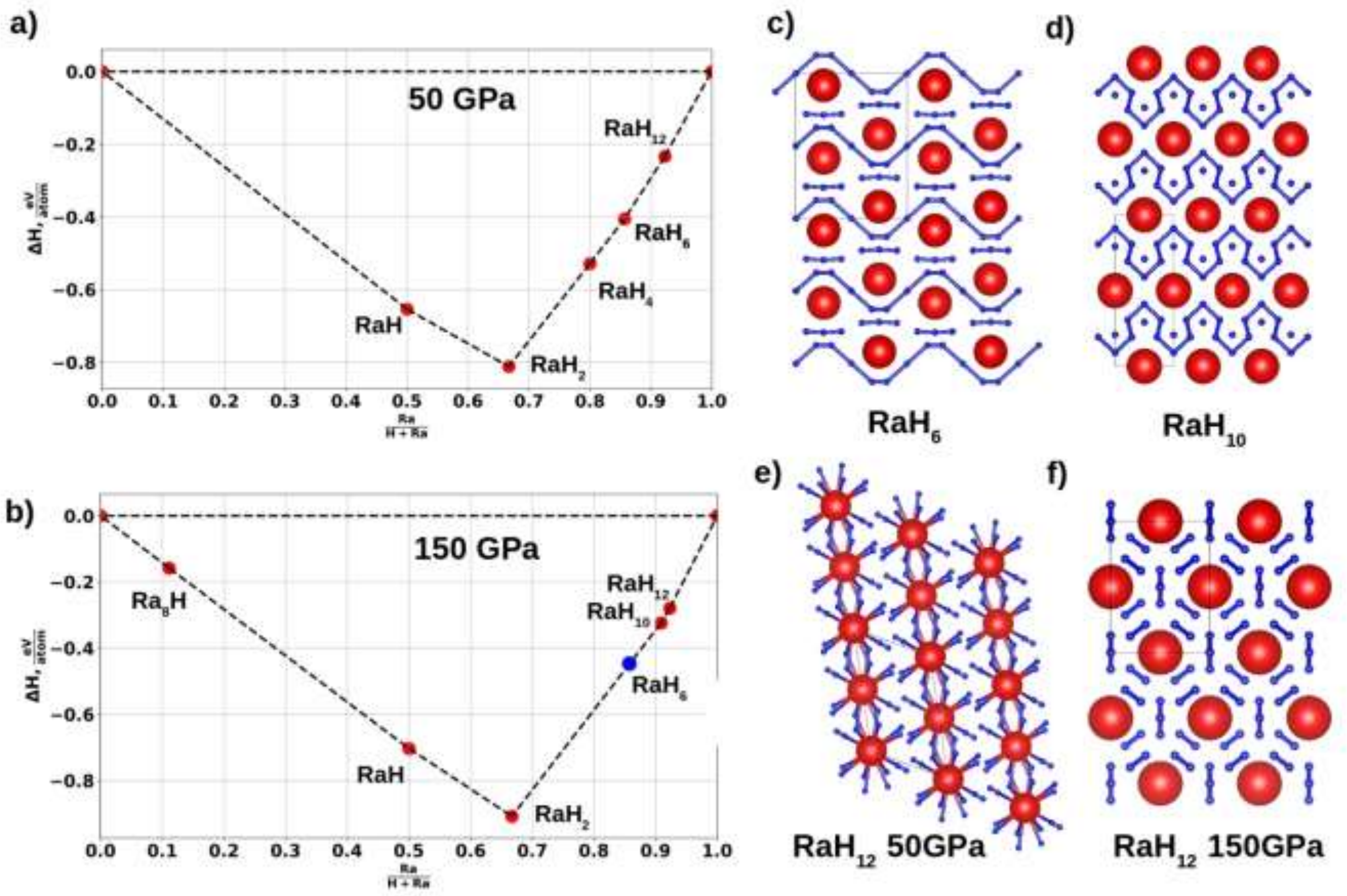

**Figure S20.** The convex hulls and structures of the higher radium polyhydrides.

## The predicted superconducting properties of the hydrides of lanthanides and actinides

**Table S3**. The predicted superconducting properties of the hydrides of lanthanides and actinides. The $T_C$ values are given for μ* = 0.1

| System/Phases | Polyhydrides | $P$, GPa | λ | $\omega_{log}$, K | $T_C$ (Allen-Dynes), K |
|---|---|---|---|---|---|
| **Pr-H**<br>$Pm\bar{3}m$-PrH<br>$Immm$-Pr$_3$H$_{10}$<br>$Pm\bar{3}n$-PrH$_3$<br>$Fmmm$-PrH$_4$<br>$C2/c$-PrH$_7$ | $C2/c$-PrH$_7$ | 50 | 0.72 | 703 | 26.8 |
| **Pr-H**<br>$C2/c$-PrH$_7$<br>$Fm\bar{3}m$-PrH$_8$<br>$F\bar{4}3m$-PrH$_9$ | $Fm\bar{3}m$-PrH$_8$ | 150 | 0.65 | 1078 | 31.3 |
| | $F\bar{4}3m$-PrH$_9$ | | 0.49 | 762 | 8.7 |
| **Nd-H**<br>$Pm\bar{3}m$ -NdH<br>$Amm2$-Nd$_4$H$_7$<br>$Fm\bar{3}m$ -NdH$_8$<br>$F\bar{4}3m$-NdH$_9$ | $Fm\bar{3}m$-NdH$_8$ | 150 | 0.44 | 876 | 6.1 |
| **Ho-H**<br>$Fm\bar{3}m$-HoH<br>$I4/mmm$-HoH$_4$<br>$I4/mmm$-HoH$_3$<br>$Cc$-HoH$_7$ | Cc-HoH$_7$ | 150 | 0.78 | 477 | 22.6 |
| | $I4/mmm$-HoH$_4$ | | 0.76 | 823 | 36.7 |
| **Er-H**<br>$R\bar{3}c$-Er$_4$H<br>$Pm\bar{3}m$ -ErH<br>$Imma$-ErH$_2$<br>$Cmcm$-ErH$_4$<br>$R\bar{3}m$-ErH$_6$<br>$P\bar{6}2m$-ErH$_{15}$ | $R\bar{3}m$-ErH$_6$ | 150 | 0.27 | 906 | 0.15 |
| | $P\bar{6}2m$-ErH$_{15}$ | | 0.72 | 794 | 31.5 |
| **Tm-H**<br>$Cmcm$-TmH<br>$P6/mmm$-TmH$_2$<br>$C2/m$-TmH$_3$<br>$I4/mmm$-TmH$_4$<br>$P4/nmm$-TmH$_5$*<br>$Fm\bar{3}m$ -TmH$_8$ | $Fm\bar{3}m$-TmH$_8$ | 150 | 0.22 | 832 | 0.003 |
| **Lu-H**<br>$Fm\bar{3}m$-LuH<br>$Fm\bar{3}m$-LuH$_3$<br>$Cc$-LuH$_7$<br>$C222$-LuH$_{12}$<br>$Cm$-LuH$_{13}$* | $C222$-LuH$_{12}$ | 150 | 0.46 | 840 | 6.7 |
| **Pa-H**<br>$Pmma$-PaH$_3$<br>$P6_3mc$-PaH$_5$<br>$P6_3/mmc$ PaH$_7$ | $P6_3/mmc$-PaH$_7$ | 50 | 1.30 | 428 | 47.3 |
| **Pa-H**<br>$Fmmm$-PaH$_4$<br>$C2/m$-PaH$_6$<br>$Fm\bar{3}m$ -PaH$_8$<br>$F\bar{4}3m$ -PaH$_9$ | $Fm\bar{3}m$-PaH$_8$ | 150 | 0.87 | 565.2 | 33.1 |
| | $F\bar{4}3m$-PaH$_9$ | | 1.21: | 627.2 | 62.7 |
| **Np-H**<br>$R\bar{3}c$-Np$_3$H<br>$P1$-NpH | $P6_3/mmc$-NpH$_7$ | 50 | 0.50 | 747.1 | 9.6 |

| System/Phases | Polyhydrides | *P*, GPa | λ | $\omega_{log}$, K | $T_C$ (Allen-Dynes), K |
|---|---|---|---|---|---|
| *C2/c*-$Np_2H_3$<br>*Cmcm*-$NpH_3$<br>*P6₃/mmc*-$NpH_7$ | | | | | |
| **Np-H**<br>$P6_3/mmc$-NpH<br>$Fm\bar{3}m$-$NpH_8$ | $Fm\bar{3}m$-$NpH_8$ | 150 | 0.29 | 1149 | 0.41 |
| **Am-H**<br>*C2/m*-$Am_2H_3$<br>*C2/c*-$AmH_2$<br>*C2/m*-$AmH_5$ | *C2/m*-$AmH_5$ | 50 | 0.31 | 477 | 0.31 |
| **Am-H**<br>$Fm\bar{3}m$-$AmH_8$ | $Fm\bar{3}m$-$AmH_8$ | 150 | 0.28 | 1027 | 0.28 |
| **Cm-H**<br>*Pnma*-$CmH_2$<br>$P6_3/mmc$-$CmH_8$<br>*R3m*-$Cm_2H_{17}$<br>*C2*-$CmH_{20}$* | $P6_3/mmc$-$CmH_8$ | 150 | 0.34 | 725 | 0.94 |

*Semiconductors

**Table S4.** The structural hallmarks of the higher actinide hydrides stable at 150 GPa*.

| **Ac, $7s^26d^1$** | **Th, $6s^26d^2$** | **Pa, $7s^26d^15f^2$** | **U, $7s^26d^15f^3$** | **Np, $7s^26d^15f^4$** | **Am, $7s^25f^7$** | **Cm, $7s^26d^15f^7$** |
|---|---|---|---|---|---|---|
| $P6/mmm$ -$AcH_2$<br>$Cmcm$ -$AcH_3$<br>$Cmmm$-$Ac_3H_{10}$<br>$P\bar{1}$-$AcH_5$<br>$C2/m$-$AcH_8$<br>$R\bar{3}m$-$AcH_{10}$<br>$P\bar{6}m2$-$AcH_{16}$<br>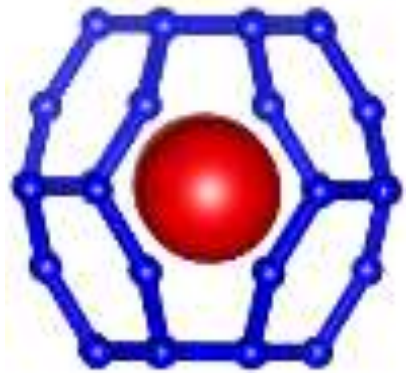<br>**$AcH_{10}$**<br>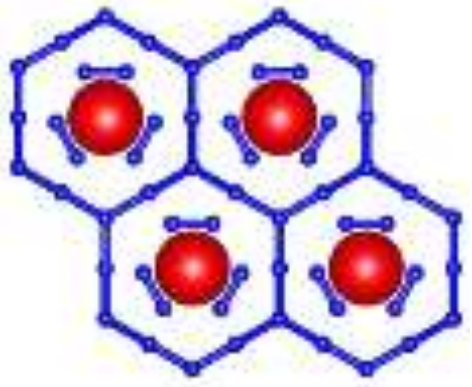<br>**$AcH_{16}$** | $I4/mmm$-$ThH_4$<br>$P2_1/c$-$ThH_7$<br>$Fm\bar{3}m$-$ThH_{10}$<br>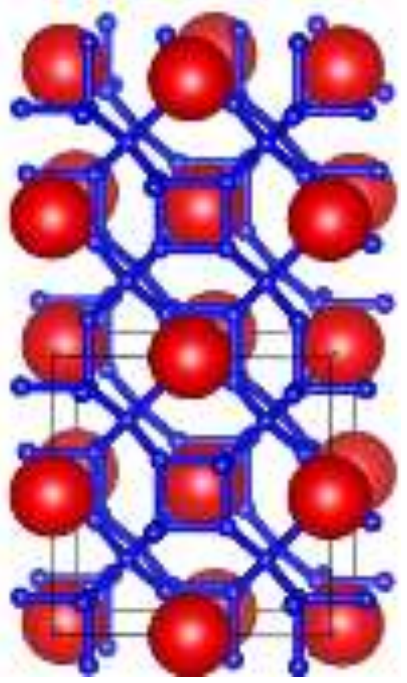<br>**$ThH_{10}$**<br>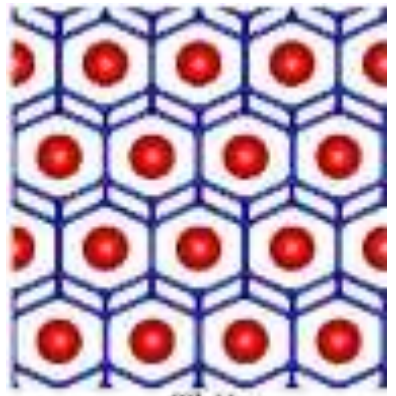<br>$ThH_4$ | $Fmmm$-$PaH_4$<br>$C2/m$-$PaH_6$<br>$Fm\bar{3}m$-$PaH_8$<br>$F\bar{4}3m$-$PaH_9$<br>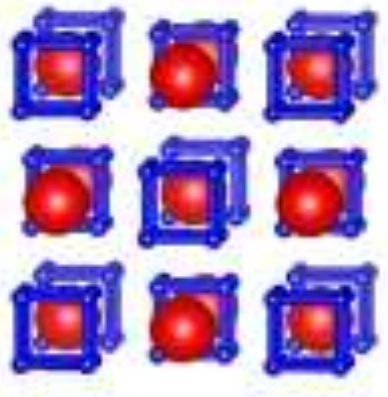<br>$PaH_8$<br>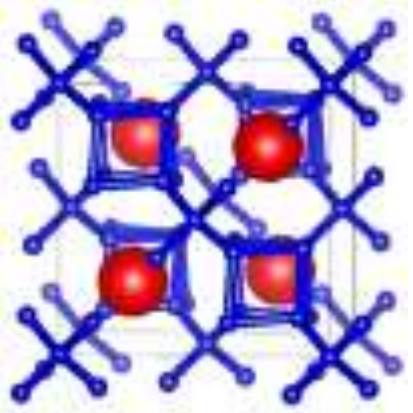<br>$PaH_9$ | $P6_3/mmc$-UH<br>$Fm\bar{3}m$-$UH_8$<br>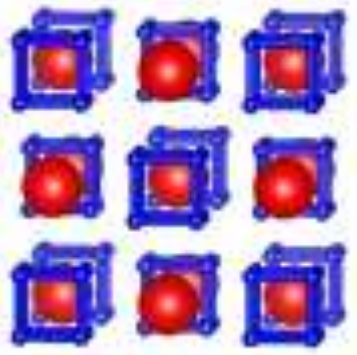<br>**$UH_8$** | $P6_3/mmc$-NpH<br>$Fm\bar{3}m$-$NpH_8$<br>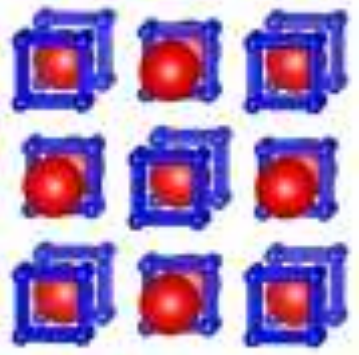<br>**$NpH_8$** | $Fm\bar{3}m$-$AmH_8$<br>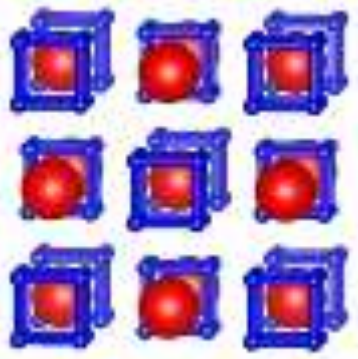<br>**$AmH_8$** | $Pnma$-$CmH_2$<br>$P6_3/mmc$-$CmH_8$<br>$R3m$-$Cm_2H_{17}$<br>$C2$-$CmH_{20}$<br>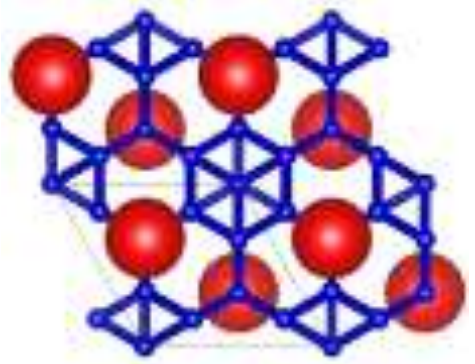<br>$CmH_8$<br>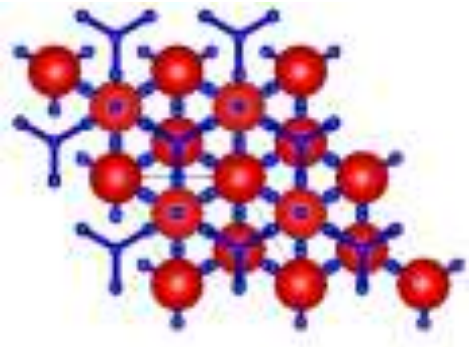<br>$Cm_2H_{17}$ |

*) The influence of the relativistic effects on the outer electron shells of lanthanides and actinides should be taken into account. The relativistic effects leading to a further decrease in the atomic radius and localization of *s*-electrons turn out to be similar to the effects of external pressure. This is one of the reasons why the $XH_6$-$XH_8$ compounds become stable much earlier, at 50 GPa, while the higher hydrides $XH_{12}$-$XH_{15}$ (and up to $XH_{20}$) reach stability at 150 GPa. This effect is more pronounced in some lighter actinides (e.g., Th, U, and Pu), accompanied by a loss of metallic properties of these hydrides.

The size of hydride forming atom is also very important. Large atoms can easily stabilize clathrate hydrogen shell while atoms of smaller radius can stabilize the same shell only when additional pressure is applied.

**Table S5.** Structural hallmarks of higher lanthanide hydrides stable at 150 GPa.

| **La, $6s^25d^1$** | **Ce, $6s^24f^2$ a)** | **Pr, $6s^24f^3$** | **Nd, $6s^24f^4$** | **Dy, $6s^24f^{10}$** | **Ho, $6s^24f^{11}$** | **Er, $6s^24f^{12}$** | **Tm, $6s^24f^{13}$** | **Lu, $6s^24f^{14}5d^1$** |
|---|---|---|---|---|---|---|---|---|
| *P6/mmm* -$LaH_2$<br>*P1*-$LaH_3$<br>*Cmmm*-$La_3H_{10}$<br>*I4/mmm*-$LaH_4$<br>$P\bar{1}$-$LaH_5$<br>$R\bar{3}m$-$LaH_{10}$<br>$P\bar{6}m2$-$LaH_{16}$<br>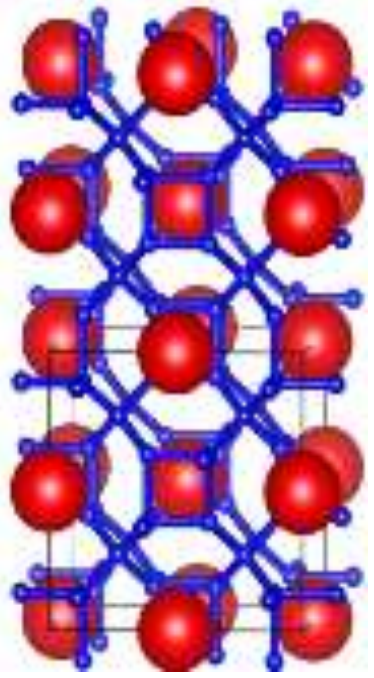<br>**$LaH_{10}$** | *I4/mmm*-$CeH_4$<br>$P6_3/mmc$-$CeH_9$<br>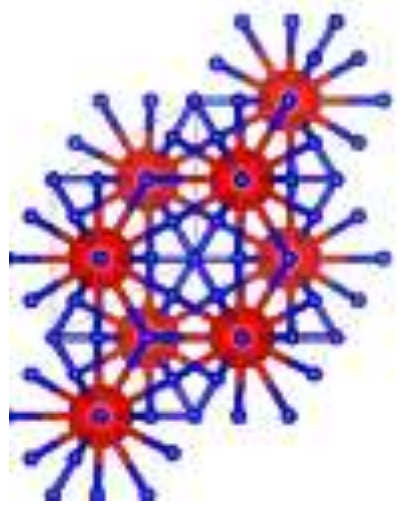<br>**$CeH_9$** | *C2/c*-$PrH_7$<br>$Fm\bar{3}m$-$PrH_8$<br>$F\bar{4}3m$-$PrH_9$<br>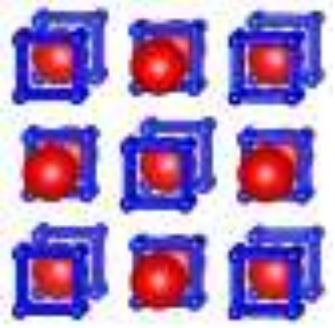<br>**$PrH_8$**<br>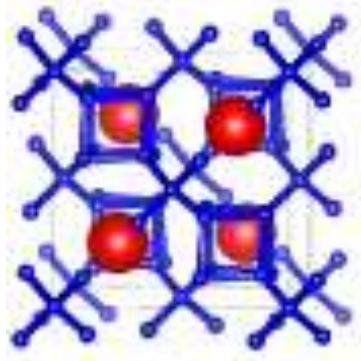<br>$PrH_9$ | $Pm\bar{3}m$ -NdH<br>*Amm2*-$Nd_4H_7$<br>$Fm\bar{3}m$ -$NdH_8$<br>$F\bar{4}3m$-$NdH_9$<br>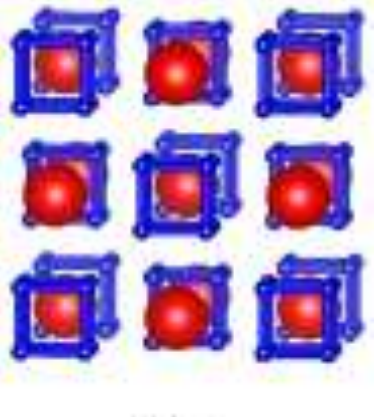<br>$NdH_8$<br>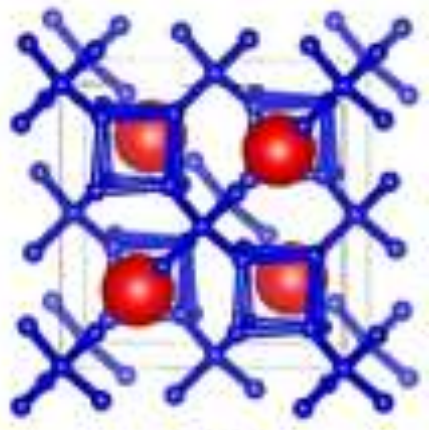<br>$NdH_9$ | $Fm\bar{3}m$-DyH<br>*Pnma*-$DyH_3$<br>*I4/mmm*-$DyH_4$<br>*Cc*-$DyH_7$<br>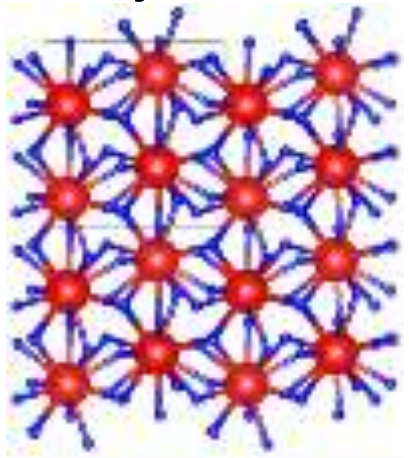<br>$DyH_7$ | $Fm\bar{3}m$-HoH<br>*I4/mmm*-$HoH_3$<br>*I4/mmm*-$HoH_4$<br>*Cc*-$HoH_7$<br>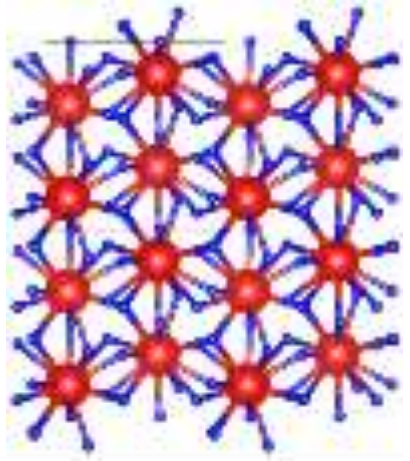<br>$HoH_7$ | *R3-c*-$Er_4H$<br>$Pm\bar{3}m$-ErH<br>*Imma*-$ErH_2$<br>*Cmcm*-$ErH_4$<br>$R\bar{3}m$-$ErH_6$<br>$P\bar{6}m2$-$ErH_{15}$<br>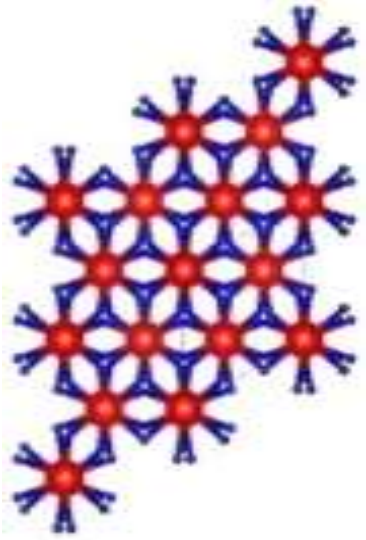<br>$ErH_6$<br>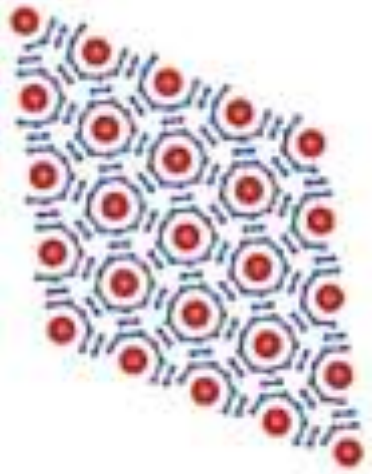<br>$ErH_{15}$ | *Cmcm*-TmH<br>*P6/mmm*-$TmH_2$<br>*C2/m*-$TmH_3$<br>*I4/mmm*-$TmH_4$<br>*P4/nmm*-$TmH_5$<br>$Fm\bar{3}m$-$TmH_8$<br>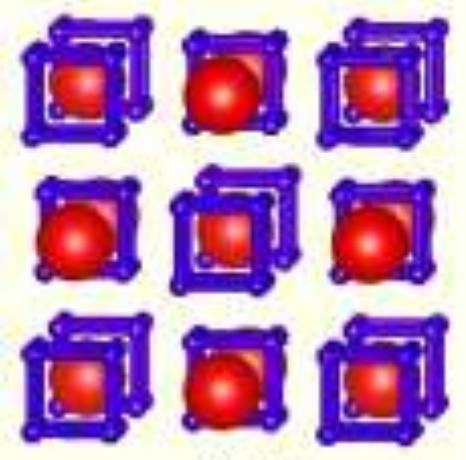<br>$TmH_8$<br>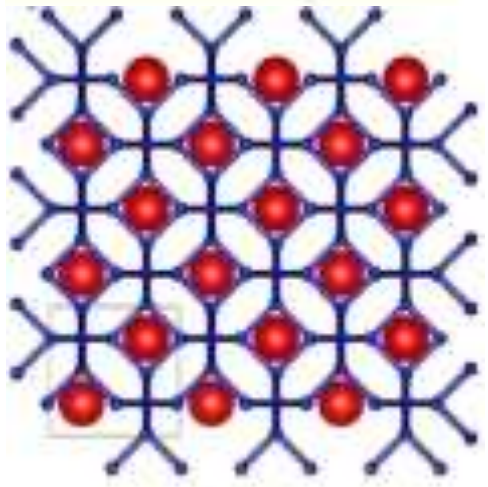<br>$TmH_5$ | $Fm\bar{3}m$-LuH<br>$Fm\bar{3}m$-$LuH_3$<br>*Cc*-$LuH_7$<br>*C222*-$LuH_{12}$<br>*Cm*-$LuH_{13}$<br>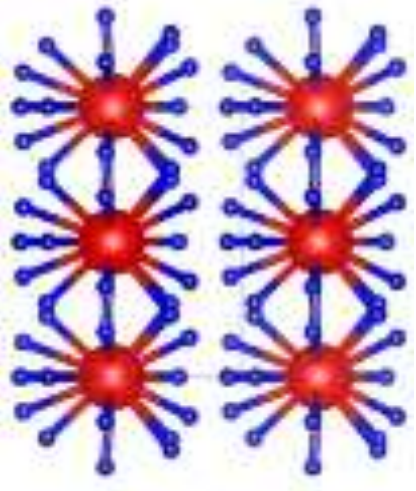<br>$LuH_{13}$<br>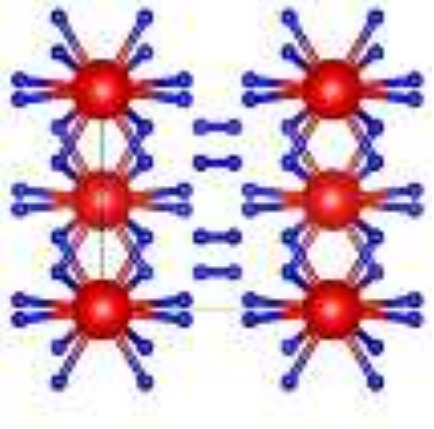<br>$LuH_{12}$ |

**a)** Cerium hydrides $P6_3mc$-$CeH_6$ and $P6_3/mmc$-$CeH_9$ are similar to scandium hydrides ($P6_3mmc$-$ScH_6$, $I4_1md$-$ScH_9$) but differ from the hydrides of the neighboring La and Pr. This may be because the *d*- and *f*-orbitals appear for the first time only in Sc and Ce, respectively.

## The Eliashberg functions of the most interesting metal hydrides superconductors

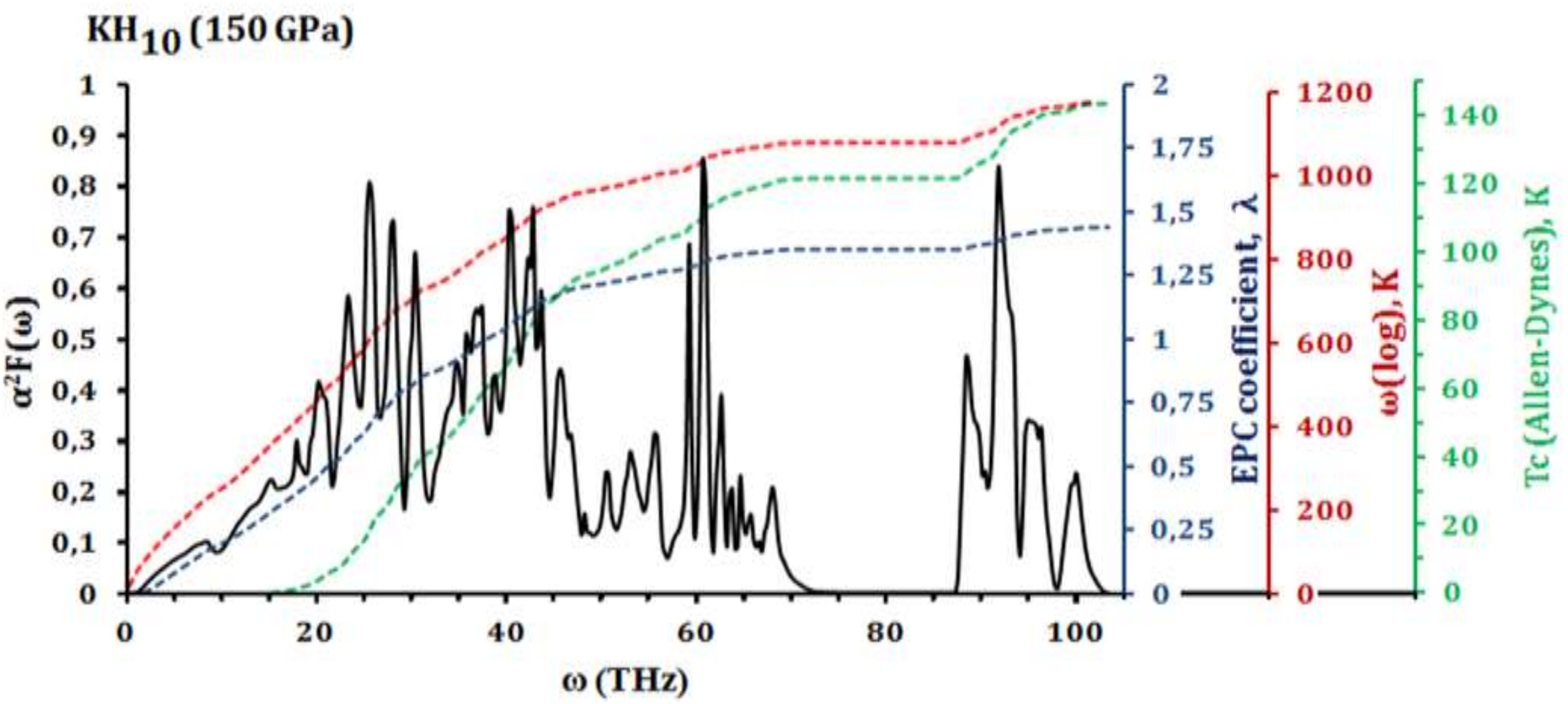

**Figure S21.** The Eliashberg function of *Immm*-$KH_{10}$ at 150 GPa.

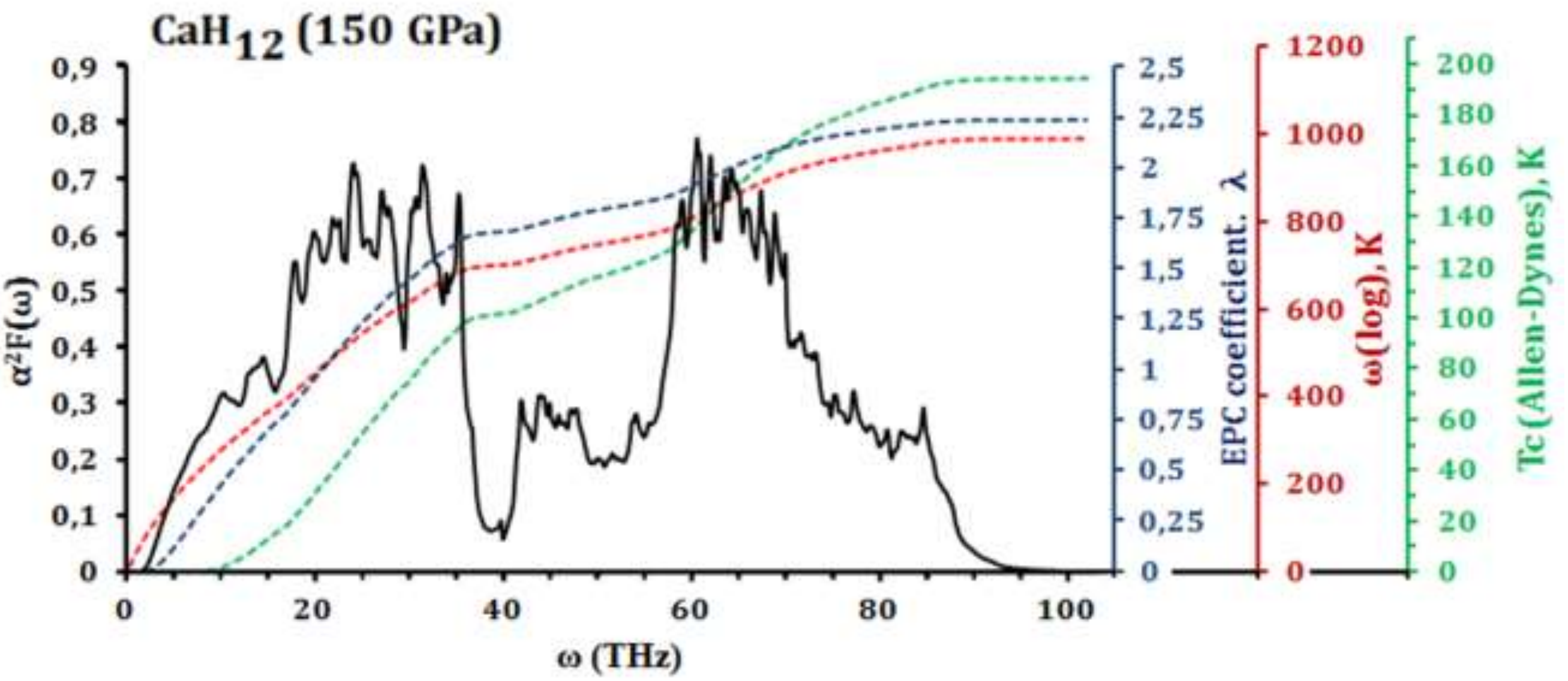

**Figure S22.** The Eliashberg function of *C2/m*-$CaH_{12}$ at 150 GPa.

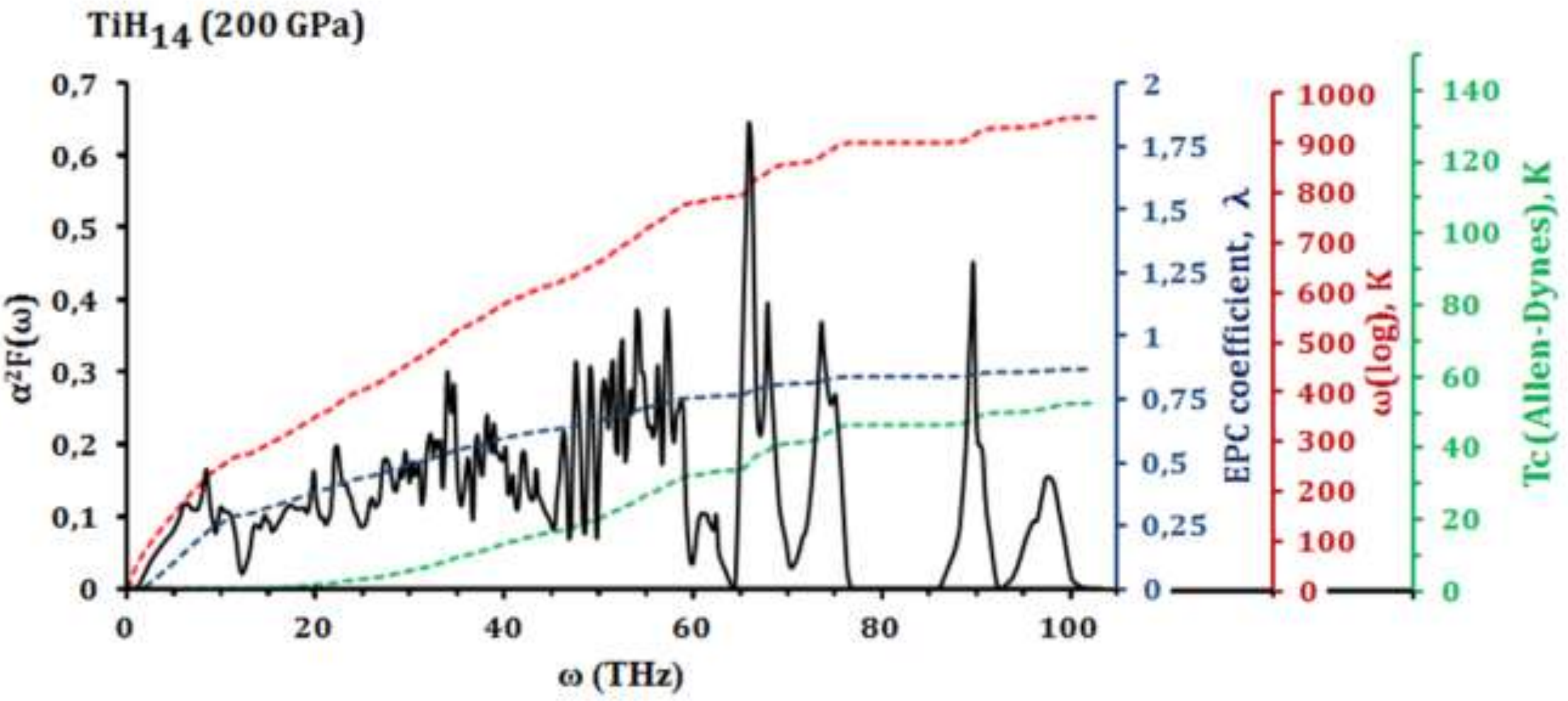

**Figure S23.** The Eliashberg function of *C*2/*m*-$TiH_{14}$ at 200 GPa.

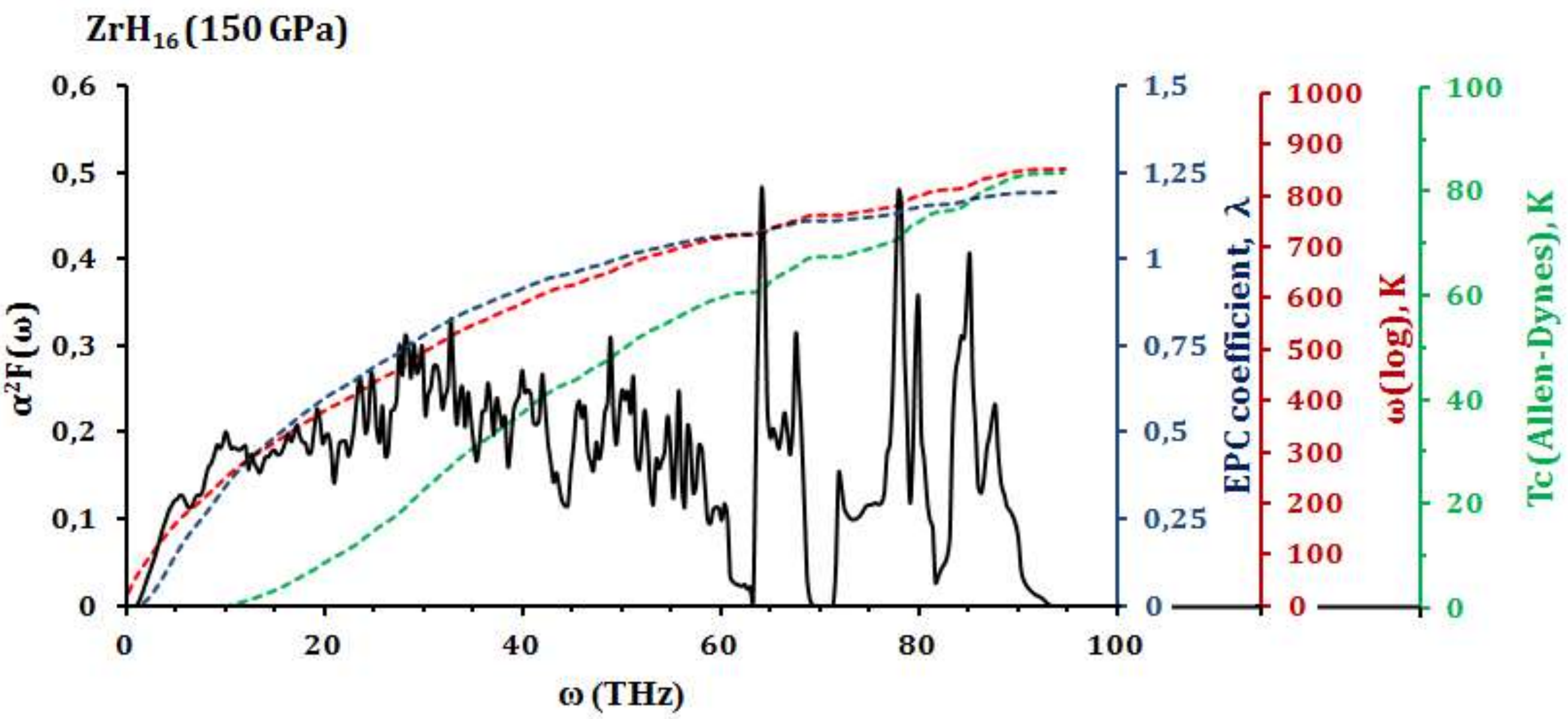

**Figure S24.** The Eliashberg function of $P\bar{1}$-$ZrH_{16}$ at 150 GPa.

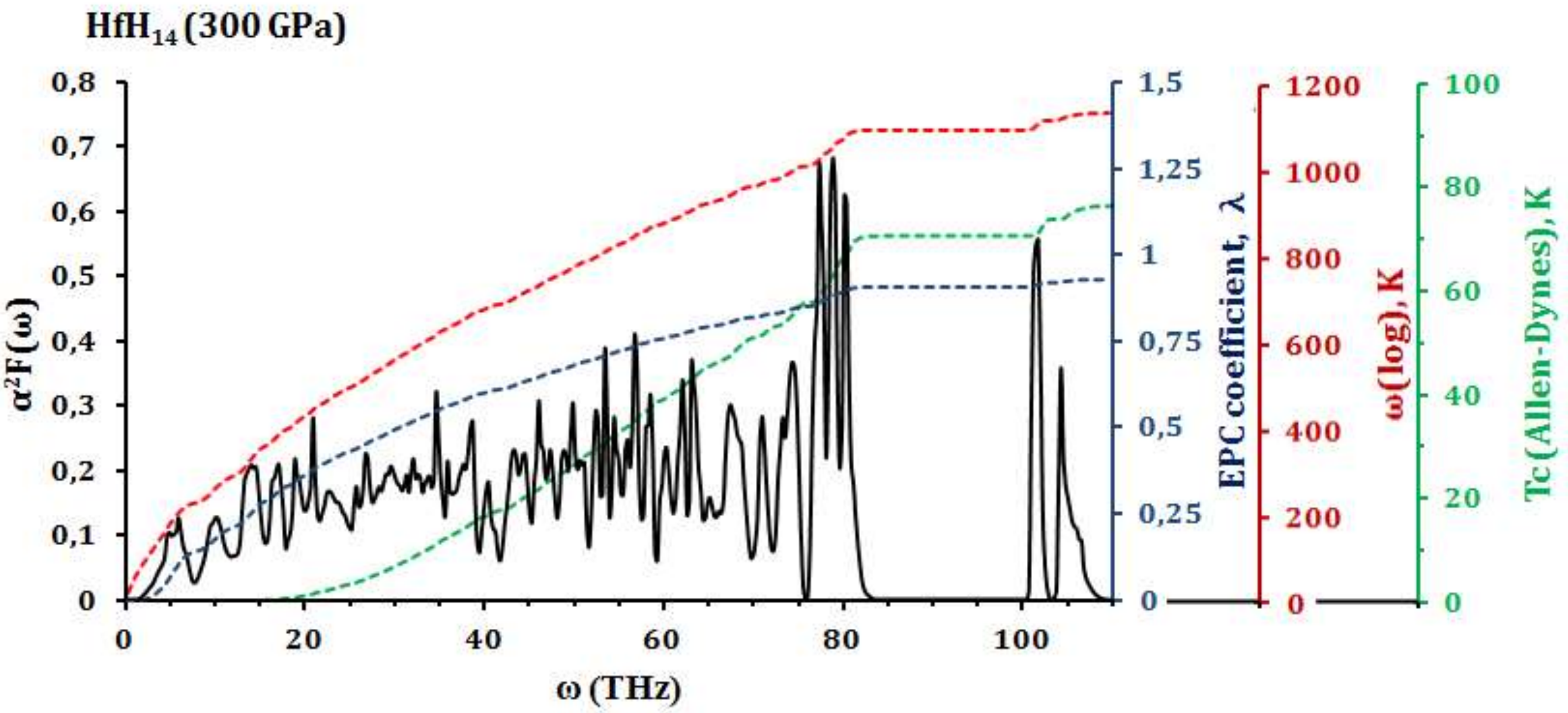

**Figure S25.** The Eliashberg function of *C2/m*-$HfH_{14}$ at 300 GPa.

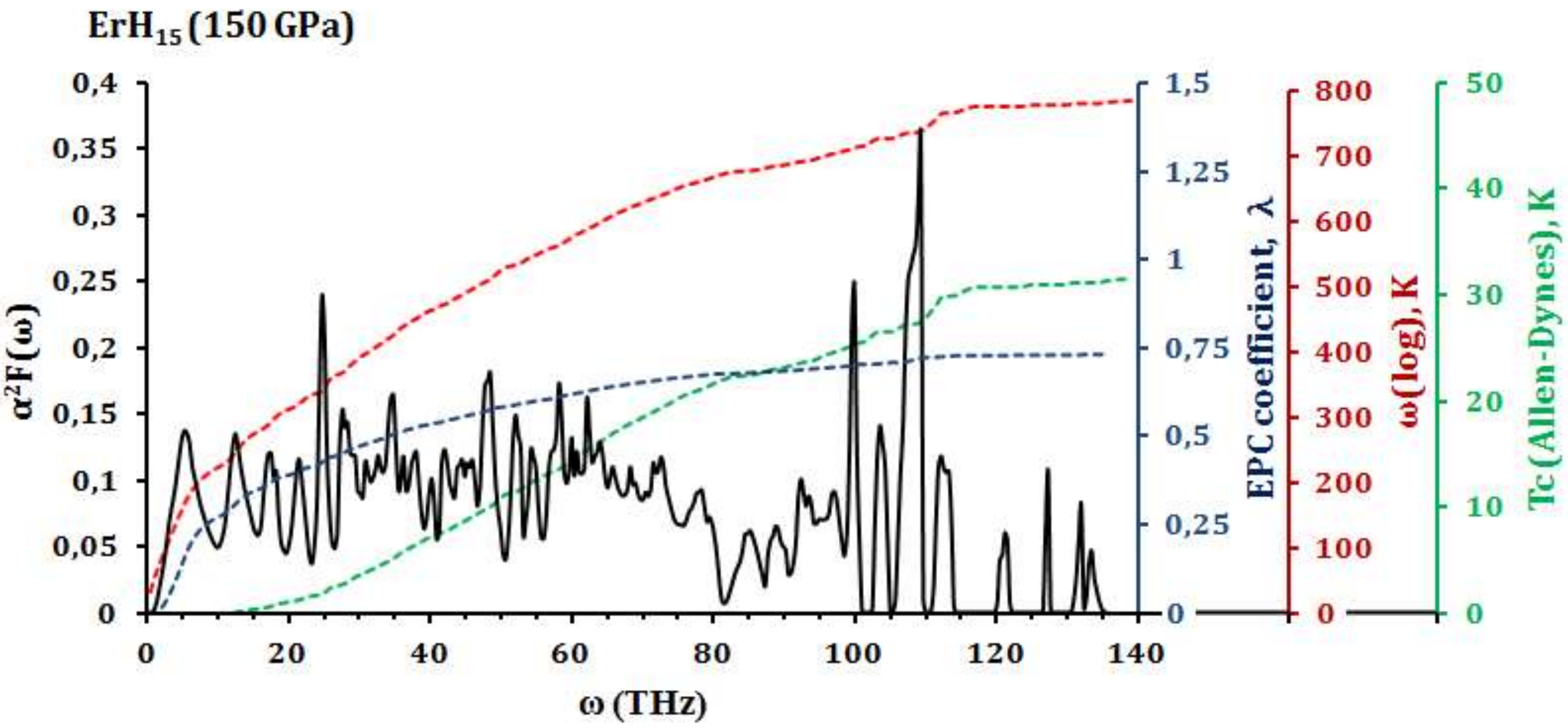

**Figure S26.** The Eliashberg function of $P\bar{6}m2$-$ErH_{15}$ at 150 GPa.

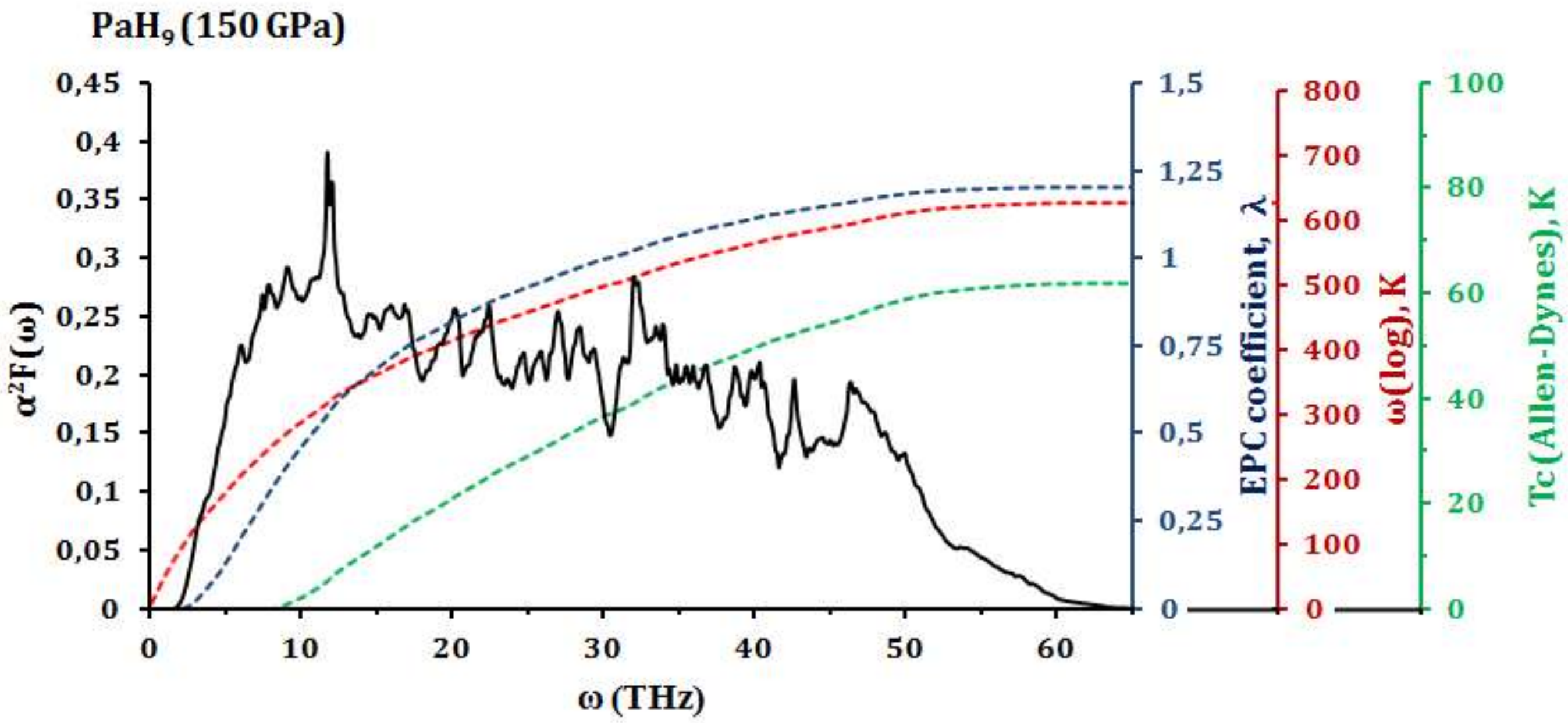

**Figure S27.** The Eliashberg function of $F\bar{4}3m$-$PaH_9$ at 150 GPa.

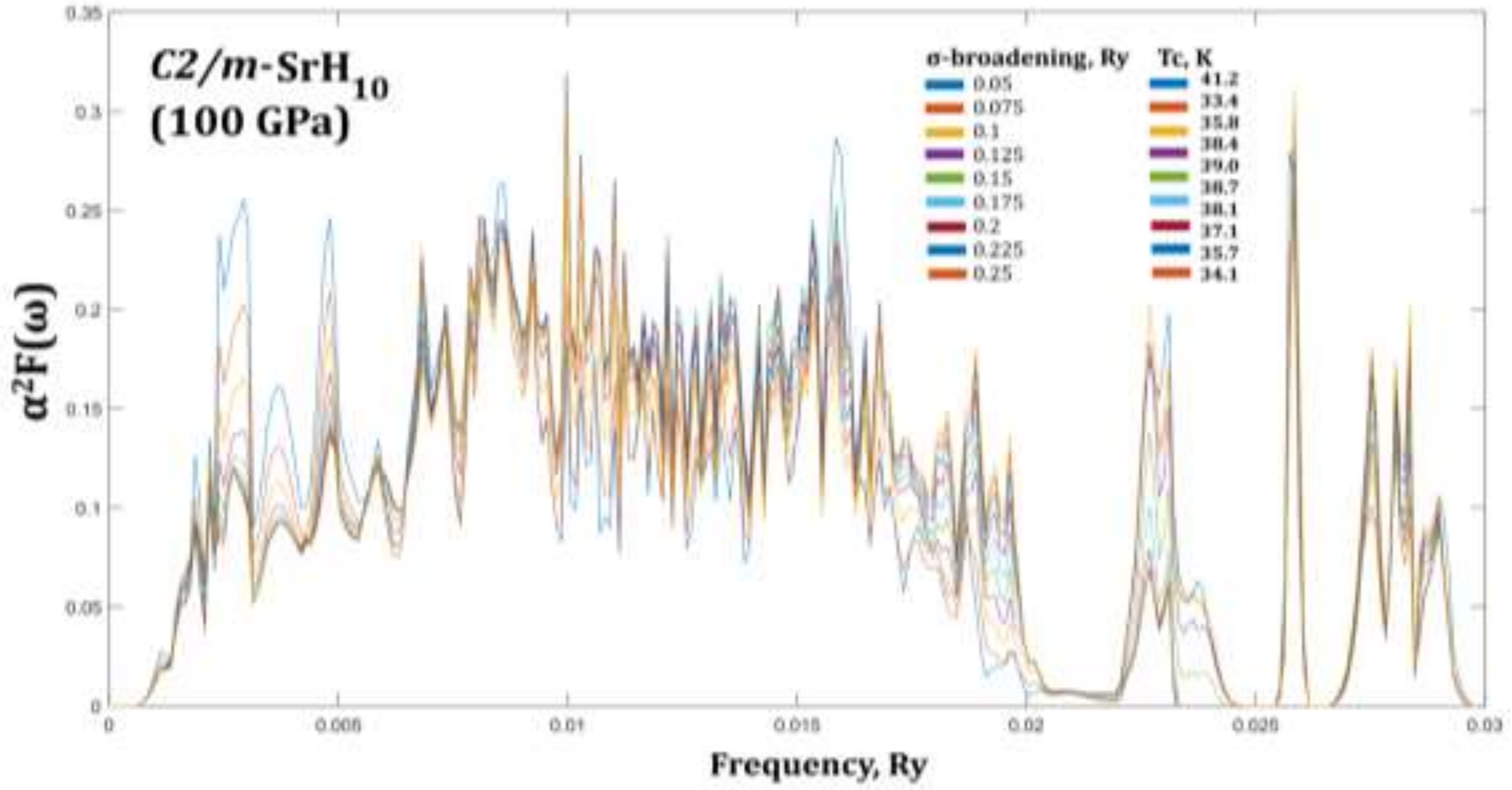

**Figure S28.** A series of the Eliashberg functions of *C2/m*-$SrH_{10}$ at 100 GPa with different σ-broadening.

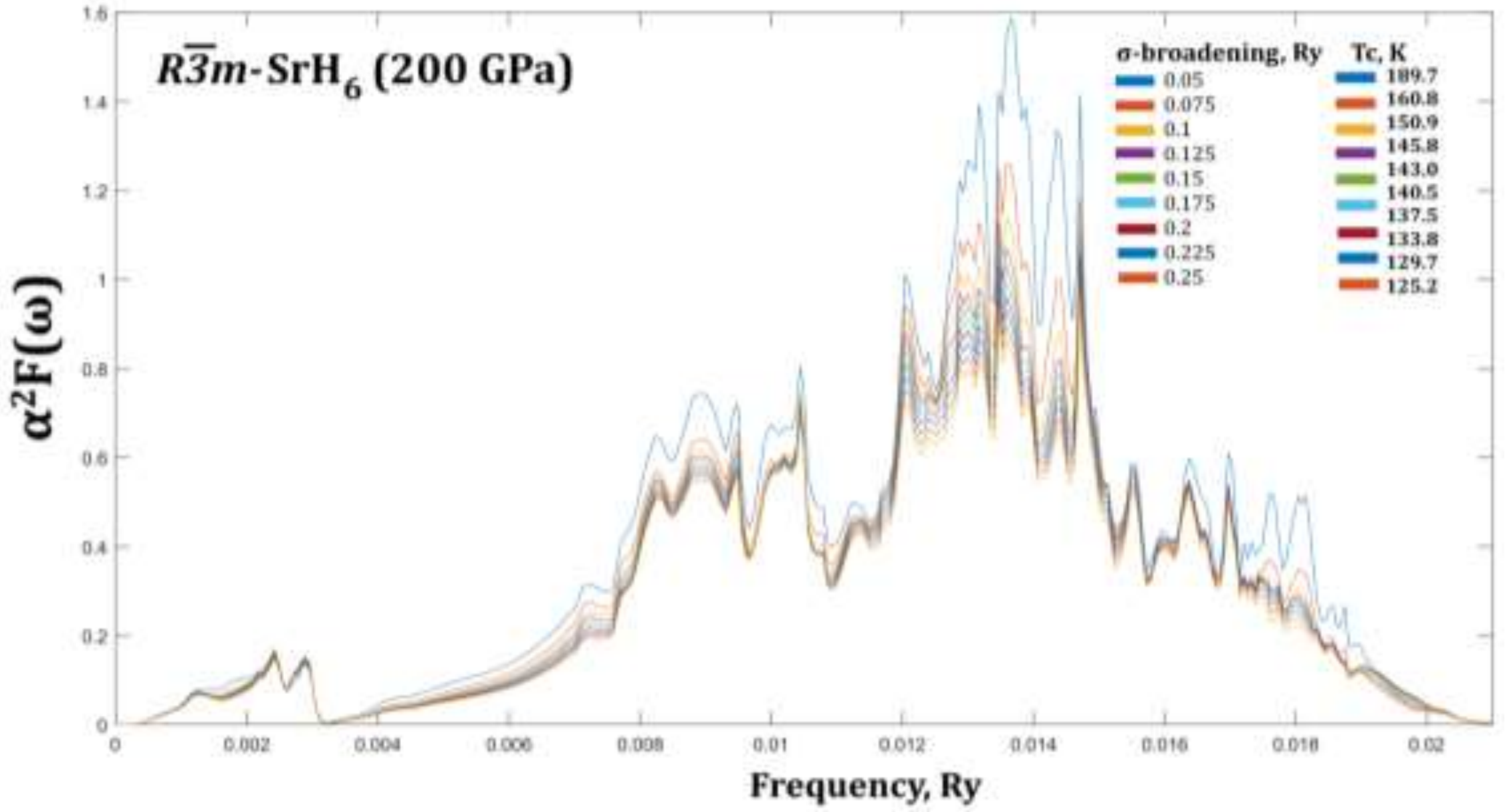

**Figure S29.** A series of the Eliashberg functions of *R3m*-$SrH_6$ at 200 GPa with different σ-broadenings.

## Statistics of the superconducting hydrides

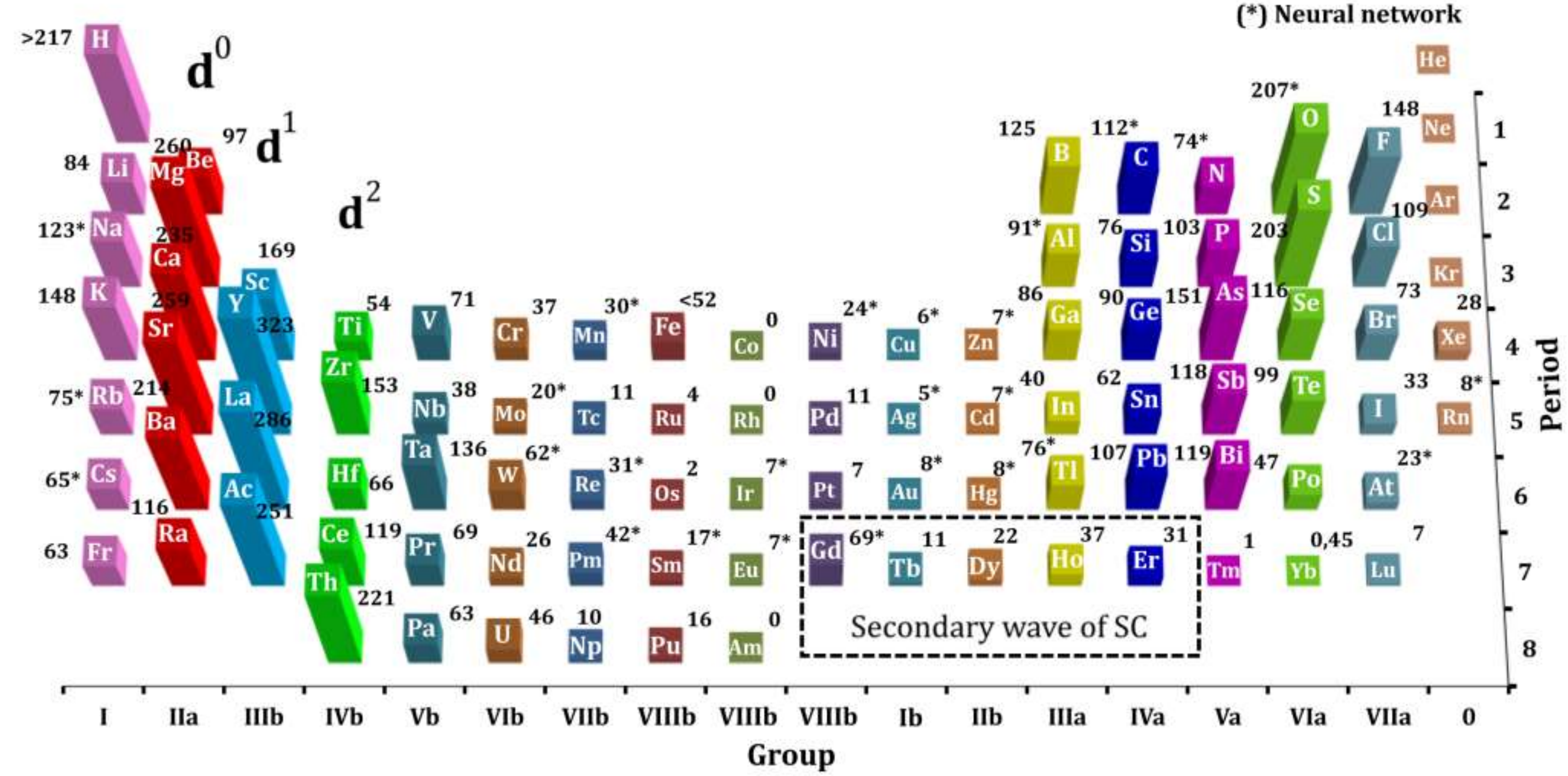

**Figure S30.** Mendeleev's Table with the already studied binary hydrides $XH_n$. The maximum values of $T_C$ were calculated using *ab initio* methods or predicted by a neural network (marked by *).

The distribution shown in Figure S30 was constructed using the following conditions. For metallic hydrogen, the minimum published value of $T_C = 217$ K (at ~500 GPa) found in Ref. [3] was taken. For the unstudied hydrides of alkali metals at high pressures (Na, Rb, Cs), the max$T_C$ values were estimated by the constructed neural network (see Methods section). New compounds in the Fr-H, Np-H, Pu-H, Am-H, and Ra-H systems were predicted in this study For the Mg-H system, the value of max$T_C$ from Ref. [5] for $MgH_6$ at 300 GPa was chosen among several recent estimations published in Refs. [4,5]. For the Zr-H system, the max$T_C$ value was taken from Ref. [6]. For the Th-H system, the maximum $T_C$ of $ThH_{10}$ calculated using the Allen-Dynes formula was taken from Ref. [7], the value is close to the experimental one from Ref. [8]. For the Ta-H system, the maximum value of $T_C$ was taken from Ref. [9], for $TaH_6$ despite the recent experimental results showing an inability to synthesize higher tantalum hydrides [10]. Results for the Pr-H system were taken from the new experimental study [11]. The Pa-H system was studied theoretically in this work, while recent research [12] indicates a slightly higher $T_C$ (79 K at 10 GPa). For the W-H system, max$T_C = 62$ K corresponds to $WH_5$, which agrees well with the recently published value of 61 K at 280 GPa [13]. [13] For the Fe-H system, the upper limit of $T_C$ was taken from Ref. [14]. The maximum $T_C$ value for the Pt-H system was taken from the recent experimental work [15]. For the Ba-H system, the data were based on ab initio calculations of the experimentally synthesized $BaH_{12}$ phase having $Cmc2_1$ space group symmetry with $T_C = 214$ K at 135 GPa [16]. All other data were either taken from the literature cited in the main text or obtained from a trained neural network.

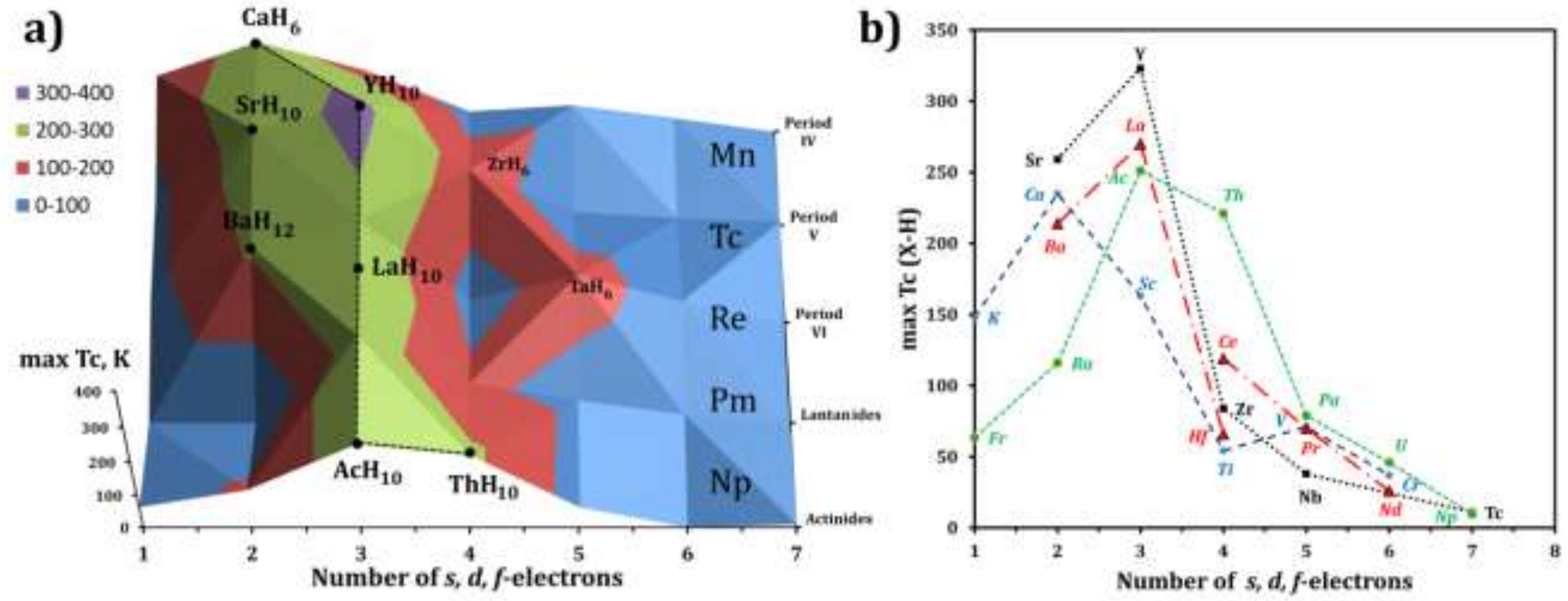

**Figure S31.** a) Distribution of max$T_C$ in the left part of Mendeleev's table; b) max$T_C$ as a function of the number of electrons for all studied metal hydrides.

The analysis of the dependence of max$T_C$ on the number of electrons in outer orbitals (Figure S30) shows that almost all the high-$T_C$ superconducting (HTSC) hydrides are concentrated along the $d^0$- and $d^1$-belts. The other metals hold little promise, if any, in terms of high-temperature superconducting hydrides. For instance, search for stable phases in the Hg-H and Cu-H systems, showed total absence of any polyhydrides.

**Table S6.** Statistical analysis of the influence of the number of *s*-, *d*-, and *f*-electrons on the max$T_C$ function of metal hydrides.

| Element | $N_s$ | $N_d$ | $N_f$ | Total (d+f) | max $T_C$, K | Eq. S11, K | Eq. S12, K |
|---|---|---|---|---|---|---|---|
| **K** | 1 | 0 | 0 | 0 | 148 | 197.6 | 208.85 |
| **Ca** | 2 | 0 | 0 | 0 | 235 | 235.2 | 261.1 |
| **Sc** | 2 | 1 | 0 | 1 | 169 | 194 | 199.74 |
| **Ti** | 2 | 2 | 0 | 2 | 54 | 152.8 | 147.66 |
| **Y** | 2 | 1 | 0 | 1 | 323 | 194 | 199.74 |
| **Zr** | 2 | 2 | 0 | 2 | 88 | 152.8 | 147.66 |
| **Nb** | 1 | 4 | 0 | 4 | 38 | 32.8 | 19.09 |
| **La** | 2 | 1 | 0 | 1 | 286 | 194 | 199.74 |
| **Hf** | 2 | 2 | 0 | 2 | 66 | 152.8 | 147.66 |
| **Ce** | 2 | 1 | 1 | 2 | 117 | 138.5 | 122.46 |
| **Pr** | 2 | 0 | 3 | 3 | 31 | 68.7 | 35.5 |
| **Nd** | 2 | 0 | 4 | 4 | 6 | 13.2 | 2.7 |
| **Ac** | 2 | 1 | 0 | 1 | 251 | 194 | 199.74 |
| **Th** | 2 | 2 | 0 | 2 | 221 | 152.8 | 147.66 |
| **Pa** | 2 | 1 | 2 | 3 | 63 | 83 | 66.38 |
| **U** | 2 | 1 | 3 | 4 | 46 | 27.5 | 31.5 |
| **Np** | 2 | 1 | 4 | 5 | 10 | -28 | 17.82 |
| **V** | 2 | 3 | 0 | 3 | 71 | 111.6 | 104.86 |
| **Cr** | 1 | 5 | 0 | 5 | 37 | -8.4 | -5.15 |
| **Tc** | 2 | 5 | 0 | 5 | 11 | 29.2 | 47.1 |
| **Ta** | 2 | 3 | 0 | 3 | 136 | 111.6 | 104.86 |
| **Correlation with maxTc** | 0.173 | -0.334 | -0.504 | -0.78 | - | 0.807 | 0.8245 |

The following equations were used to approximate the data from Table S6:

$$T_C = 160 + 37.6N_s - 41.2N_d - 55.5N_f \qquad (S11)$$

$$T_C = 156.6 + 52.25N_s - 66N_d - 107N_f + 4.64N_d^{\ 2} + 10.6N_f^{\ 2} + 19.2N_dN_f \qquad (S12)$$

The replacement of a linear equation by a quadratic form (Eq. S12) does not lead to a significant increase in the accuracy. The error is caused by the influence of period number, but in this work we focus only on the group number.

The correlation with max$T_C$ was calculated using the standard formula:

$$r = \frac{\sum (X(i) - \overline{X}) \cdot (\max T_C(i) - \overline{\max T_C})}{\sqrt{\sum (X(i) - \overline{X})^2 \cdot \sum (\max T_C(i) - \overline{\max T_C})^2}}$$

The statistical analysis shows that the effect of *s*-electrons on superconductivity is positive but small. However, only three examples of $s^1$-$s^2$ comparisons were included in the statistics. As a result, the importance of *s*-orbitals can be clarified only after the detailed investigation of all alkaline and alkaline earth metal hydrides. The influence of *d*- and *f*-electrons on superconductivity is negative, being more distinct for *f*-electrons. The explicit negative correlation (-0.75) between max$T_C$ and the total number of electrons confirms our ideas about the influence of filling of the outer electron shells on superconductivity.

## Electronic density of states at the Fermi level

**Table S7.** Electronic density of states at the Fermi level in polyhydrides used to plot the graph in Figure 7a.

| Polyhydride | $N_H$ | $N_e$ | Pressure, GPa | $T_C$, K | *N(E$_F$)*, states/Ry/Å$^3$ |
|---|---|---|---|---|---|
| $CaH_6$ | 6 | 2 | 150 | 225 | 0.51 |
| $H_3S$ | 3 | 6 | 165 | 204 | 0.30 |
| $YH_{10}$ | 10 | 3 | 250 | 326 | 0.50 |
| $LaH_{10}$ | 10 | 3 | 250 | 286 | 0.35 |
| $ThH_{10}$ | 10 | 3 | 250 | 221 | 0.23 |
| $AcH_{10}$ | 10 | 3 | 250 | 251 | 0.36 |
| $KH_{10}$ | 10 | 1 | 150 | 148 | 0.18 |
| $KH_{12}$ | 12 | 1 | 150 | 117 | 0.16 |
| $PrH_8$ | 8 | 5 | 150 | 31 | 1.75 |
| $PrH_7$ | 7 | 5 | 50 | 27 | 6.43 |
| $NdH_8$ | 8 | 6 | 50 | 6 | 2.38 |
| $NdH_7$ | 7 | 6 | 50 | 6 | 3.06 |
| $PuH_8$ | 8 | 8 | 50 | 8.2 | 10.7 |
| $UH_8$ | 8 | 6 | 50 | 27.5 | 1.58 |
| $FeH_6$ | 6 | 8 | 150 | 43 | 0.35 |
| $FeH_5$ | 5 | 8 | 150 | 46 | 0.11 |
| $ErH_{15}$ | 15 | 14 | 150 | 31.5 | 0.09 |
| $TmH_8$ | 8 | 15 | 150 | 1 | 0.84 |
| $CeH_9$ | 9 | 4 | 150 | 117 | 0.30 |
| $TiH_{14}$ | 14 | 4 | 200 | 54 | 0.36 |
| $CmH_5$ | 5 | 10 | 50 | 1 | 4.34 |

**Table S8.** Parameters of superconductivity of metal hydrides used for building the distribution diagrams in Figure 8. The critical temperatures were calculated using the Allen-Dynes equation; γ is the Sommerfeld constant.

| Polyhydride | Pressure, GPa | λ | $\omega_{log}$, K | $T_C$, K | *N(E$_F$)*, states/eV/f.u. | *γ, mJ/mol·K$^2$* | *$\mu_0 H_C(0)$, T* |
|---|---|---|---|---|---|---|---|
| $KH_6$ | 166 | 0.9 | 1165 | 70 | 0.5 | 4.5 | 12 |
| *Im3m*-$CaH_6$ | 150 | 2.7 | - | 235 | 0.84 | 14.6 | 80 |
| *hcp*-$ScH_9$ | 300 | 1.94 | 1156 | 163 | 0.9 | 12.5 | 50 |
| $TiH_2$ | 0 | 0.84 | 127 | 6.7 | 2.59 | 22.4 | 2.5 |
| $VH_8$ | 200 | 1.13 | 876 | 71.4 | 1.05 | 10.5 | 19 |
| $CrH_3$ | 81 | 0.95 | 568 | 37.1 | - | - | - |
| *Fddd*-$TaH_6$ | 300 | 1.56 | 1151 | 136 | 0.95 | 11.4 | 39 |
| $YH_{10}$ | 300 | 2.6 | 1282 | 323 | 0.92 | 15.6 | 111 |
| *Im3m*-$H_3S$ | 200 | 2.19 | 1335 | ~ 200 | 0.63 | 9.5 | 54 |
| *fcc*-$LaH_{10}$ | 210 | 3.41 | 848 | 250-286 | 0.76 | 15 | 81 |
| *$P2_1/c$*-$ZrH_6$ | 295 | 1.7 | 914 | 153 | 0.75 | 9.5 | 41.7 |
| *I4/mmm*-$NbH_4$ | 300 | 0.82 | 938 | 38 | 8.1 | 69.4 | 25 |
| *F43m*-$PrH_9$ | 120 | 0.8 | 1387 | 69 | 1.1 | 9.3 | 17 |
| *hcp*-$PrH_9$ | 120 | 0.8 | 1100 | 55 | 1.35 | 11.4 | 15 |
| *I4/mmm*-$TcH_2$ | 200 | 0.52 | 736 | 11 | 0.69 | 4.94 | 1.9 |
| ZrH | 120 | 0.71 | 295 | 10 | 1.0 | 8 | 2.2 |
| *I4/mmm*-$BaH_{12}$ | 135 | 2.66 | 927 | 214 | 0.3 | 5.2 | 43 |
| *C2/m*-$HfH_{14}$ | 300 | 0.93 | 1138 | 76 | 0.375 | 3.4 | 11.3 |
| *hcp*-$CeH_9$ | 200 | 2.30 | 740 | 117 | 0.71 | 11 | 34 |

| Polyhydride | Pressure, GPa | λ | $\omega_{log}$, K | $T_C$, K | $N(E_F)$, states/eV/f.u. | $\gamma$, $mJ/mol{\cdot}K^2$ | $\mu_0 H_C(0)$, T |
|---|---|---|---|---|---|---|---|
| *C2/c*-$CeH_9$ | 100 | 1.48 | 662 | 75 | 0.73 | 8.5 | 18.5 |
| *fcc*-$ThH_{10}$ | 100 | 2.19 | 1043 | 221 | 0.61 | 11.1 | 71 |
| *R3m*-$AcH_{10}$ | 250 | 3.46 | 711 | 251 | 0.79 | 16.6 | 75 |
| *Immm*-$KH_{10}$ | 150 | 1.45 | 1175 | 148 | 0.4 | 4.6 | 27 |
| *hcp*-$NdH_9$ | 120 | 0.67 | 769 | 25 | 7.67 | 60.3 | 15.2 |
| *C2/m*-$TiH_{14}$ | 200 | 0.82 | 950 | 54 | 0.87 | 7.45 | 12 |
| $ErH_{15}$ | 150 | 0.72 | 794 | 31.5 | 0.28 | 2.27 | 3.7 |
| *Fm3m*-$TmH_8$ | 150 | 0.22 | 832 | 0 | 1.58 | 9.1 | 0 |
| $CmH_5$ | 50 | 0.34 | 725 | 0.94 | 8.16 | 51.5 | 0.55 |
| *Fm3m*-$UH_8$ | 50 | 0.73 | 838 | 27.5 | 0.57 | 4.6 | 4.6 |
| *hcp*-$UH_9$ | 300 | 0.67 | 933 | 36 | 0.63 | 5 | 6.3 |
| *hcp*-$UH_7$ | 20 | 0.83 | 874 | 47.6 | 1.03 | 8.9 | 11.3 |

# References

(1) Allen, P. B.; Dynes, R. C. Transition Temperature of Strong-Coupled Superconductors Reanalyzed. *Phys. Rev. B* **1975**, *12* (3), 905–922. https://doi.org/10.1103/PhysRevB.12.905.

(2) Carbotte, J. P. Properties of Boson-Exchange Superconductors. *Rev. Mod. Phys.* **1990**, *62* (4), 1027–1157. https://doi.org/10.1103/RevModPhys.62.1027.

(3) Kudryashov, N. A.; Kutukov, A. A.; Mazur, E. A. Critical Temperature of Metallic Hydrogen at a Pressure of 500 GPa. *JETP Lett.* **2016**, *104* (7), 460–465. https://doi.org/10.1134/S0021364016190061.

(4) Szczęśniak, R.; Durajski, A. P. Superconductivity Well above Room Temperature in Compressed MgH6. *Front. Phys.* **2016**, *11* (6), 117406. https://doi.org/10.1007/s11467-016-0578-1.

(5) Feng, X.; Zhang, J.; Gao, G.; Liu, H.; Wang, H. Compressed Sodalite-like MgH6 as a Potential High-Temperature Superconductor. *RSC Adv.* **2015**, *5* (73), 59292–59296. https://doi.org/10.1039/C5RA11459D.

(6) Abe, K. High-Pressure Properties of Dense Metallic Zirconium Hydrides Studied by Ab Initio Calculations. *Phys. Rev. B* **2018**, *98* (13), 134103. https://doi.org/10.1103/PhysRevB.98.134103.

(7) Kvashnin, A. G.; Semenok, D. V.; Kruglov, I. A.; Oganov, A. R. High-Temperature Superconductivity in Th-H System at Pressure Conditions. *ACS Appl. Mater. Interfaces* **2018**, *10* (50), 43809–43816.

(8) Semenok, D. V.; Kvashnin, A. G.; Ivanova, A. G.; Svitlyk, V.; Troyan, I. A.; Oganov, A. R. Synthesis of ThH4, ThH6, ThH9 and ThH10: A Route to Room-Temperature Superconductivity. *ArXiv190210206 Cond-Mat* **2019**.

(9) Zhuang, Q.; Jin, X.; Cui, T.; Ma, Y.; Lv, Q.; Li, Y.; Zhang, H.; Meng, X.; Bao, K. Pressure-Stabilized Superconductive Ionic Tantalum Hydrides. *Inorg. Chem.* **2017**, *56* (7), 3901–3908. https://doi.org/10.1021/acs.inorgchem.6b02822.

(10) Ying, J.; Li, X.; Greenberg, E.; Prakapenka, V. B.; Liu, H.; Struzhkin, V. V. Synthesis and Stability of Tantalum Hydride at High Pressures. *Phys. Rev. B* **2019**, *99* (22), 224504. https://doi.org/10.1103/PhysRevB.99.224504.

(11) Zhou, D.; Semenok, D.; Duan, D.; Xie, H.; Huang, X.; Chen, W.; Li, X.; Liu, B.; Oganov, A. R.; Cui, T. Superconducting Praseodymium Superhydrides. *ArXiv190406643 Cond-Mat* **2019**.

(12) Xiao, X.; Duan, D.; Xie, H.; Shao, Z.; Li, D.; Tian, F.; Song, H.; Yu, H.; Bao, K.; Cui, T. Structure and Superconductivity of Protactinium Hydrides under High Pressure. *J. Phys. Condens. Matter* **2019**, *31* (31), 315403. https://doi.org/10.1088/1361-648X/ab1d03.

(13) Zheng, S.; Zhang, S.; Sun, Y.; Zhang, J.; Lin, J.; Yang, G.; Bergara, A. Structural and Superconducting Properties of Tungsten Hydrides Under High Pressure. *Front. Phys.* **2018**, *6*. https://doi.org/10.3389/fphy.2018.00101.

(14) Kvashnin, A. G.; Kruglov, I. A.; Semenok, D. V.; Oganov, A. R. Iron Superhydrides FeH5 and FeH6: Stability, Electronic Properties, and Superconductivity. *J. Phys. Chem. C* **2018**, *122* (8), 4731–4736. https://doi.org/10.1021/acs.jpcc.8b01270.
(15) Matsuoka, T.; Hishida, M.; Kuno, K.; Hirao, N.; Ohishi, Y.; Sasaki, S.; Takahama, K.; Shimizu, K. Superconductivity of Platinum Hydride. *Phys. Rev. B* **2019**, *99* (14), 144511. https://doi.org/10.1103/PhysRevB.99.144511.
(16) Personal Information from Huang’s Group, Jilin University.